\documentclass[11pt]{article}
\usepackage{amsfonts}
\usepackage{amsmath}
\usepackage{amssymb}
\usepackage{bbold}
\usepackage[margin=2.2cm]{geometry}
\usepackage{longtable}
\usepackage{lscape,graphicx} 
\usepackage{graphics}
\usepackage{color}
\usepackage{cite}
\usepackage{array}
\usepackage{makecell}

\usepackage{sidecap}
\usepackage{caption}
\usepackage{subcaption}

\newcommand{\ba}{\begin{array}{c}}
\newcommand{\ea}{\end{array}}

\newcommand{\be}{\beta}

\def\be{\begin{equation}}
\def\ee{\end{equation}}
\def\beq{\begin{equation}}
\def\eeq{\end{equation}}
\def\bc{\begin{center}}
\def\ec{\end{center}}
\def\bea{\begin{eqnarray}}
\def\eea{\end{eqnarray}}

\definecolor{darkgreen}{rgb}{0.1,0.7,0.3}

\begin{document}
\begin{titlepage}
\vspace*{-1cm}
\phantom{hep-ph/***}
\flushright
\hfil{IFIC/21-21}
\hfil{FTUV-21-0611.3946}


\vskip 1.5cm
\begin{center}
\mathversion{bold}
{\LARGE\bf 
Flavour and CP symmetries in the inverse seesaw}\\[3mm]
\mathversion{normal}
\vskip .3cm
\end{center}
\vskip 0.5  cm
\begin{center}
{\large C.~Hagedorn}$^{1,2}$,
{\large J.~Kriewald}$^{3}$,
{\large J.~Orloff}$^{3}$,
{\large A.~M.~Teixeira}$^{3}$
\\
\vskip .7cm
{\footnotesize
$^{1}$ Instituto de F\'isica Corpuscular, Universidad de Valencia and CSIC,
Edificio Institutos Investigaci\'on, Catedr\'atico Jos\'e Beltr\'an 2, 46980 Paterna, Spain\\[0.3cm]
$^{2}$ Istituto Nazionale di Fisica Nucleare, Sezione di Padova, Via F. Marzolo 8, 35131 Padua, Italy\\[0.3cm]
$^{3}$ Laboratoire de Physique de Clermont (UMR 6533), CNRS/IN2P3, Univ. Clermont Auvergne, 4 Av. Blaise Pascal, 63178 Aubi\`ere Cedex, France
\vskip .5cm
}
\end{center}
\vskip 1cm
\begin{abstract}
We consider an inverse seesaw mechanism of neutrino mass generation in
which the Standard Model is extended by $3+3$ (heavy) sterile
states, and endowed with a flavour symmetry $G_f$, $G_f=\Delta (3 \, n^2)$ or
$G_f=\Delta (6 \, n^2)$, and a CP symmetry. 
These symmetries are broken in a peculiar way, so that in the charged
lepton sector a residual symmetry $G_\ell$ is preserved, while the
neutral fermion sector remains invariant 
under the residual symmetry $G_\nu=Z_2 \times CP$. We study the concrete setup, where
the Majorana mass term for three of the sterile states
conserves $G_\nu$, while the remaining mass terms (i.e.
couplings of left-handed leptons and heavy sterile states, 
as well as the Dirac-type couplings among the latter) do not break the flavour or CP symmetry. 
We perform a comprehensive analysis of lepton mixing for different
classes of residual symmetries, giving examples for each of these, and
study in detail 
the impact of the additional sterile states on the 
predictions for lepton mixing. We further 
confront our results with those obtained in the 
model-independent scenario, in which   
the light neutrino mass matrix leaves the residual symmetry $G_\nu$
intact. 
We consider the phenomenological impact of the inverse seesaw mechanism endowed
with flavour and CP symmetries, in 
particular concerning 
effects of non-unitarity of the lepton mixing matrix (which strongly constrain the parameter space
of the scenario), prospects for neutrinoless double beta decay and 
for charged lepton flavour  violating processes. 
\end{abstract}
\end{titlepage}
\setcounter{footnote}{0}

\section{Introduction}
\label{intro}
The Standard Model (SM) of particle physics can successfully explain a plethora of experimental observations. Yet, the existence of three generations of SM fermions,
the origin of neutrino masses, the features of lepton and quark mixing, as well as the striking differences between these remain open issues. 
Symmetries acting on flavour space can address the first and the third point~\cite{Gfreviews,Gfreview_math}, while different types of new particles can be added to the SM
in order to generate at least two non-vanishing neutrino masses~\cite{type1seesaw,type2seesaw,type3seesaw,Mohapatra:1986bd,GonzalezGarcia:1988rw,Mohapatra:1986aw,Bernabeu:1987gr,radnureview}. 

In the present study, we choose a non-abelian discrete 
symmetry $G_f$ 
combined with a CP symmetry, both acting non-trivially on flavour space. This combination has proven to be highly constraining~\cite{GfCPearly,S4CPgeneral,GfCPothers} since, as long as $G_f$ and CP are broken to different residual symmetries $G_\ell$ among charged leptons and $G_\nu=Z_2 \times CP$ among the neutral states,
the Pontecorvo-Maki-Nakagawa-Sakata (PMNS) mixing matrix
depends on a single free parameter.
We select $G_f$ to be a member of the series of groups $\Delta (3 \, n^2)$~\cite{Delta3n2} and $\Delta (6 \, n^2)$~\cite{Delta6n2}, $n$ integer, because these have shown to lead to several interesting mixing patterns~\cite{HMM,DeltaCPothers,D6n2CPZ2Z2,smallDeltaCPmodels}. Four of these, called Case 1), Case 2), Case 3 a) and Case 3 b.1), have been identified in~\cite{HMM}.

Among the different realisations of the Weinberg
operator (including the well-known type-I, type-II and type-III
seesaw mechanisms - as well as their variants), the so-called inverse seesaw (ISS) mechanism~\cite{Mohapatra:1986bd,GonzalezGarcia:1988rw,Mohapatra:1986aw,Bernabeu:1987gr} emerges as another interesting possibility. 
In particular, the ISS mechanism offers a direct connection between the smallness of neutrino masses and the breaking of lepton number (LN) conservation: when compared to the
canonical type-I seesaw, a potentially tiny
LN violating (LNV) dimensionful coupling $\mu_S$ 
provides an additional
source of suppression for the light neutrino masses, while being technically natural in the sense of 't Hooft~\cite{tHooft:1979rat} (in the limit in which the LNV couplings vanish, LN conservation is restored as an accidental symmetry of the ISS Lagrangian).
The ISS mechanism thus allows to accommodate light neutrino masses for natural values of the Dirac neutrino Yukawa couplings
($\sim\mathcal{O}(1)$) at comparatively low scales (TeV or below).

In addition to being a theoretically well-motivated framework, the ISS mechanism can have an important phenomenological impact:
as a consequence of the sizeable mixing between active neutrinos and the comparatively light additional sterile states (possibly within collider reach),
extensive contributions to numerous observables can occur. 
Among the latter, one can mention several 
charged lepton flavour violation (cLFV) processes~\cite{Abada:2014vea,Arganda:2014dta,Abada:2014kba,Abada:2014cca,Arganda:2015naa,Abada:2015oba,DeRomeri:2016gum}, CP violating observables such
as the electric dipole moment (EDM) of the electron~\cite{Abada:2016awd}, or neutrinoless double beta ($0\nu\beta\beta$) decays~\cite{Abada:2014nwa,Abada:2018qok}. The impact of the ISS mechanism regarding the Higgs sector (for instance concerning the one-loop effects of the heavy sterile states on the triple Higgs coupling) has been also explored, and found to be non-negligible (see, for instance~\cite{Baglio:2016bop}).

Flavour (and CP) symmetries have been studied in association with several scenarios of neutrino mass generation, see, e.g.,~\cite{Gfmodels,smallDeltaCPmodels,ISSsymmetries}.

In this study, we endow an ISS framework with 
a flavour symmetry $G_f$ and a CP symmetry. We focus on the so-called $(3,3)$ ISS framework, in which the SM field content is extended by $3+3$ heavy sterile states, $N_i$ and $S_j$.
We note that different realisations of the ISS mechanism with flavour (and CP) symmetries have been considered in the literature, see, e.g.,~\cite{ISSsymmetries}.
The main features of the present ISS framework are the following: left-handed (LH) lepton doublets, and the sterile states $N_i$ and $S_j$ all transform as irreducible triplets of $G_f$,
while right-handed (RH) charged leptons are assigned to singlets, so that the three different charged lepton masses can be easily accommodated. While the source of
breaking of $G_f$ and CP to the residual symmetry $G_\ell$ 
is unique in the charged lepton sector (corresponding to the  charged lepton mass terms), the breaking to $G_\nu$ among the neutral states can be 
realised in different ways. Indeed, we can consider three minimal options, depending on which of the neutral fermion mass terms 
 encodes the symmetry breaking. In this study, we use
an option (henceforth called ``option 1''), in which only the Majorana mass matrix $\mu_S$ 
breaks $G_f$ and CP to $G_\nu$. In this way, $\mu_S$ is the unique source of lepton flavour
and LN violation in the neutral sector. 
Similar to what is found for the charged lepton masses, light neutrino masses are not constrained
in this scenario, and 
their mass spectrum can follow either a normal ordering (NO) or an inverted ordering (IO). 
The mass spectrum of the heavy sterile states is instead strongly restricted, since they combine to form three approximately degenerate pseudo-Dirac pairs (to a very high degree).

We show analytically and numerically that the impact of these heavy sterile states on lepton mixing (i.e., results for lepton mixing angles, predictions for CP phases as well as (approximate) sum rules)
is always small, with relative deviations below $1\%$ from the results previously obtained in the model-independent scenario~\cite{HMM}.
This is a consequence of effects arising due to deviations from unitarity of the PMNS mixing matrix,\footnote{In SM extensions including enlarged 
lepton sectors, in which the new states have non-vanishing mixing to the active neutrinos, the PMNS mixing matrix (corresponding to the LH mixing encoded in the upper left three-by-three block of the full lepton mixing matrix) is in general non-unitary.} 
which are subject to stringent experimental limits.
The matrix encoding these effects is of a peculiar form in our scenario, being both flavour-diagonal and flavour-universal. 
Due to their pseudo-Dirac nature, the heavy states' contribution to $0\nu\beta\beta$ decay is always strongly suppressed. 
As we will discuss, and in stark contrast to typical ISS models, new contributions to cLFV are also negligible.
Our scenario thus complies with all experimental limits for masses of the heavy sterile states
as low as $500$~GeV and Dirac neutrino Yukawa couplings of order $0.1$, and successfully reproduces the results for lepton mixing obtained in the model-independent scenario.

The remainder of the paper is organised as follows: in
section~\ref{sec2} we present the chosen approach to lepton mixing, first in
the model-independent scenario, and then 
in the $(3,3)$ ISS framework. 
Section 3 is devoted to a brief summary of the main results for lepton
mixing in the model-independent scenario. The impact of the heavy
sterile states of the $(3,3)$ ISS framework on lepton mixing is analytically
evaluated in section~\ref{sec4}. The results of the numerical study
are discussed in depth in section~\ref{sec5}, 
using an explicit example for each of the different cases, Case 1) through Case 3 b.1), and emphasising the impact of the deviations from unitarity of the PMNS mixing matrix. 
Sections~\ref{sec6} and~\ref{sec:clfv} are devoted to the results concerning $0\nu\beta\beta$ decays, and prospects for cLFV, respectively.  
We briefly summarise and give an outlook in section~\ref{summ}.
Additional information and complementary discussions are collected in several appendices.
 
\section{Approach to lepton mixing}
\label{sec2}

We assume the existence of a flavour symmetry $G_f=\Delta (3 \, n^2)$ or $G_f=\Delta (6 \, n^2)$
and a $Z_3$ symmetry $Z_3^{(\mathrm{aux})}$, as well as a CP symmetry in the theory.\footnote{Since $\Delta (3 \, n^2)$ is a subgroup of $\Delta (6 \, n^2)$, it is sufficient to focus on the latter in the analysis.} These are broken (without specifying the breaking mechanism) to a residual $Z_3$ symmetry $G_\ell$, corresponding to the diagonal subgroup
of a $Z_3$ group contained in $G_f$ and $Z_3^{(\mathrm{aux})}$,\footnote{In the original study~\cite{HMM}, the residual symmetry $G_\ell$ was assumed to be fully contained in $G_f$. This was possible, since in~\cite{HMM}
the focus has been on the mass matrix combination $m_\ell^{\phantom{\dagger}} \, m_\ell^\dagger$ and not on the charged lepton mass matrix $m_\ell$ alone. Thus, only the transformation properties of LH lepton doublets were necessary.
However, when considering also $m_\ell$ and, consequently, RH charged leptons, a possibility to distinguish among these is needed. Nevertheless, the results for lepton mixing are not affected by this change.} in the charged lepton sector
and to $G_\nu= Z_2 \times CP$ (with $Z_2$ being a subgroup of $G_f$) among the neutral states. The $Z_2$ symmetry is given by the generator $Z$, denoted as $Z(\mathrm{{\bf r}})$ in the representation $\mathrm{{\bf r}}$.
 The CP symmetry is described by a CP transformation $X$ in flavour space. In the different representations $\mathrm{{\bf r}}$ of $G_f$, $X (\mathrm{{\bf r}})$ corresponds to a unitary matrix fulfilling
\begin{equation}
\label{eq:Xrcond}
X (\mathrm{{\bf r}}) \, X(\mathrm{{\bf r}})^\star = X(\mathrm{{\bf r}})^\star \, X (\mathrm{{\bf r}}) = \mathbb{1}
\end{equation}
so that $X$ is always represented as a symmetric matrix.\footnote{For more details on this choice, see~\cite{S4CPgeneral}.} A consistent definition of a theory with $G_f$ and CP necessitates the fulfilment of the consistency 
condition
\begin{equation}
\label{eq:GfCPconsist}
X (\mathrm{{\bf r}}) \, g (\mathrm{{\bf r}})^\star \, X (\mathrm{{\bf r}})^\star = g^\prime (\mathrm{{\bf r}})
\end{equation}
with $g$ and $g^\prime$ being elements of $G_f$ and $g^{(\prime)} (\mathrm{{\bf r}})$
their representation matrices in the representation $\mathrm{{\bf r}}$.
This condition must be fulfilled for all representations $\mathrm{{\bf r}}$, or at least for the representations used for charged leptons and the neutral states.
Since the product $Z_2 \times CP$ is direct, $Z(\mathrm{{\bf r}})$ and 
$X(\mathrm{{\bf r}})$ commute
\begin{equation}
\label{eq:XZcond}
 X (\mathrm{{\bf r}}) \, Z (\mathrm{{\bf r}})^\star - Z (\mathrm{{\bf r}}) \, X (\mathrm{{\bf r}}) =\mathbb{0}
\end{equation}
for all representations $\mathrm{{\bf r}}$.
The flavour and CP symmetries, together with their residuals, determine the lepton mixing pattern.
Since we follow the approach to lepton mixing presented in~\cite{HMM}, 
we further assume
that the index of $G_f$ is not divisible by three, i.e.~$3 \nmid n$.
All choices of CP symmetries and residual $Z_2$ groups in the sector of the neutral states fulfil the conditions in eqs.~(\ref{eq:Xrcond},\ref{eq:GfCPconsist},\ref{eq:XZcond}).
For convenience, we summarise in appendix \ref{app1} the relevant group theory aspects of $G_f$, i.e.~the generators and their form in the chosen irreducible representations $\mathrm{{\bf r}}$ of $G_f$.
Details about the form of the CP transformation $X (\mathrm{{\bf r}})$ can also be found in appendix~\ref{app1}.

In the following, we first review the implementation of these symmetries and their residuals in the model-independent scenario that has been considered in~\cite{HMM}, and then turn to the $(3,3)$ ISS framework, focusing on one particular implementation, called option 1. We comment on two other minimal options at the end of this section.

\subsection{Model-independent scenario}
\label{sec21}

In the model-independent scenario, we consider the mass terms
\begin{equation}
\label{eq:masseseff}
-\bar{\ell}_{\alpha L} \, (m_\ell)_{\alpha\beta} \, \ell_{\beta R} - \frac 12 \, \overline{\nu}^c_{\alpha L} \, (m_\nu)_{\alpha\beta} \, \nu_{\beta L} + \mathrm{h.c.}
\end{equation}
for charged leptons, $m_\ell$, and for neutrinos, $m_\nu$, and with indices $\alpha, \beta = e, \mu, \tau$. 
While charged leptons acquire their (Dirac) masses from the Yukawa couplings to the Higgs,  
the LNV neutrino mass term can be effectively generated by means of the Weinberg operator, 
\begin{equation}
\label{eq:YukawaWeinberg}
-(y_\ell)_{\alpha\beta} \, \overline{L}_\alpha \, H \, \ell_{\beta R} + 
\frac{1}{\Lambda_{\mathrm{LN}}} \, (y_\nu)_{\alpha\beta} \, \Big( \overline{L}^c_\alpha \, H \Big) \, \Big( L_\beta \, H \Big)  + \mathrm{h.c.}
\end{equation}
with LH lepton doublets defined as $L_\alpha = \Big( \begin{array}{c} \nu_{\alpha L} \\ \ell_{\alpha L} \end{array} \Big) \sim ({\bf 2}, - \frac 12)$, RH charged leptons
$\ell_{\alpha R} \sim ({\bf 1}, -1)$ and the Higgs doublet $H \sim ({\bf 2}, \frac 12)$ under $SU(2)_L \times U(1)_Y$. 
$\Lambda_\mathrm{LN}$ defines the scale at which LN is broken and Majorana neutrino masses are generated.
After electroweak symmetry breaking, $\langle H \rangle = \Big( \begin{array}{c} 0 \\ \frac{v}{\sqrt{2}} \end{array} \Big)$ with $v \approx 246 \, \mathrm{GeV}$, the mass matrices $m_\ell$
and $m_\nu$ are given by  
\begin{equation}
\label{eq:massesEWth}
m_\ell= y_\ell \, \frac{v}{\sqrt{2}} \;\; \mbox{and} \;\; m_\nu= y_\nu \, \frac{v^2}{\Lambda_{\mathrm{LN}}} \; .
\end{equation}
The physical (mass) basis, denoted by $\hat{\phantom{x}}$, is related to the interaction basis by the unitary transformations
\begin{equation}
\label{eq:Umassbasis}
\ell_{L} = U_\ell \, \hat{\ell}_{ L} \; , \;\; \ell_{R} = U_R \, \hat{\ell}_{R} \;\; \mbox{and} \;\; \nu_{ L} = U_\nu \, \hat{\nu}_{ L} \; .
\end{equation}
The mass matrices $m_\ell$ and $m_\nu$ are then diagonalised
 as follows\begin{equation}
\label{eq:massesdiag}
U_\ell^\dagger \, m_\ell \, U_R = m_\ell^\mathrm{diag} = \mathrm{diag}\left(m_e, m_\mu, m_\tau \right)
\;\; \mbox{and} \;\;
U_\nu^T \, m_\nu \, U_\nu = m_\nu^\mathrm{diag} = \mathrm{diag}\left(m_1, m_2, m_3 \right)
\end{equation}
and the (unitary) PMNS mixing matrix\footnote{The conventions of lepton mixing parameters and neutrino masses used in this work can be found in appendix~\ref{app2}.} $U_{\mathrm{PMNS}}$ appears in the charged current interactions
\begin{equation}
\label{eq:LCCdefPMNS}
- \frac{g}{\sqrt{2}} \, \overline{\hat{\ell}}_{L} \, \slash\!\!\!\!W^{-} \, U_{\mathrm{PMNS}} \, \hat{\nu}_L \;\; \mbox{with} \;\; U_{\mathrm{PMNS}} = U_\ell^\dagger \, U_\nu \; .
\end{equation}
When it comes to the implementation of $G_f$ and CP, and of the residual symmetries $G_\ell$ and $G_\nu$, we first specify the assignment of LH lepton doublets $L_\alpha$ and RH charged leptons $\ell_{\alpha R}$. In order to
constrain as much as possible the resulting lepton mixing pattern, we assign $L_\alpha$ to an irreducible, faithful (complex)\footnote{Only for the choice $n=2$ of the index of $G_f$ this representation is real.} three-dimensional representation ${\bf 3}$ of $G_f$. This representation can be chosen without loss of generality (see \cite{choiceof3} for details) as the representation ${\bf 3_{(n-1, 1)}}$ and ${\bf 3_{1 \, (1)}}$ in the convention of~\cite{Delta3n2} and~\cite{Delta6n2}, respectively. Right-handed charged leptons $\ell_{\alpha R}$ transform as the trivial singlet ${\bf 1}$ of $G_f$.
In order to distinguish the different flavours, we employ the $Z_3$ symmetry $Z_3^{(\mathrm{aux})}$ and assign $\ell_{eR} \sim 1$, $\ell_{\mu
R} \sim \omega$ and $\ell_{\tau R} \sim \omega^2$ with $\omega= e^{\frac{2 \, \pi \, i}{3}}$. Left-handed lepton doublets $L_\alpha$
do not carry a non-trivial charge under $Z_3^{(\mathrm{aux})}$. 

The residual symmetry $G_\ell$ is fixed to the diagonal subgroup of the $Z_3$ group, arising from the generator $a$ of $G_f$, see eqs.~(\ref{eq:genrep3}, \ref{eq:genrep1}) in appendix~\ref{app1}, and $Z_3^{(\mathrm{aux})}$. Since $a ({\bf 3})$ is diagonal, see eq.~(\ref{eq:genrep3}), the mass matrix $m_\ell$ of charged leptons is diagonal. In our analysis, we assume that charged lepton masses are canonically ordered\footnote{For results arising in the case of non-canonically ordered charged lepton masses, see~\cite{HMM}.} so that the contribution to lepton mixing from the charged lepton sector is trivial, i.e.
\begin{equation}
\label{eq:Ul}
U_\ell = \left( \begin{array}{ccc}
1 & 0 & 0\\
0 & 1 & 0\\
0 & 0 & 1
\end{array}
\right) \; .
\end{equation}
The lepton mixing pattern depends on the choice of $G_f$, the CP symmetry and the residual $Z_2$ symmetry among the neutral states. In general, the light neutrino mass matrix $m_\nu$ is constrained by the conditions~\cite{S4CPgeneral} 
\begin{equation}
\label{eq:ZXmnu}
Z ({\bf 3})^T \, m_\nu \, Z ({\bf 3}) = m_\nu \;\; \mbox{and} \;\; X ({\bf 3}) \, m_\nu \, X ({\bf 3}) = m_\nu^\star \; .
\end{equation}
The CP transformation $X ({\bf 3})$ can be written as
\begin{equation}
\label{eq:XOmega3}
X ({\bf 3}) = \Omega ({\bf 3}) \, \Omega ({\bf 3})^T
\end{equation}
with $\Omega ({\bf 3})$ being unitary; furthermore $\Omega ({\bf 3})$ can be chosen such that
\begin{equation}
\label{eq:ZOmega3}
\Omega ({\bf 3})^\dagger \, Z ({\bf 3}) \, \Omega ({\bf 3}) \;\; \mbox{is diagonal.}
\end{equation}
In this basis, rotated by $\Omega ({\bf 3})$, the light neutrino mass matrix is block-diagonal and real.
Since $Z ({\bf 3})$ generates a $Z_2$ symmetry, two of its eigenvalues are equal. This explains why the resulting matrix is block-diagonal and why a rotation around a free angle $\theta$, encoded in the rotation matrix $R_{fh} (\theta)$
(with the indices $f$ and $h$ determined by the pair of degenerate eigenvalues of $Z ({\bf 3})$), is necessary in order
to arrive at a basis in which $m_\nu$ is diagonal. 
Furthermore, positive semi-definiteness of the light neutrino masses is ensured by a diagonal matrix $K_\nu$, with entries taking values $\pm 1$ and $\pm i$. 
Hence, $U_\nu$ is given by
\begin{equation}
\label{eq:Unumodind}
U_\nu = \Omega ({\bf 3}) \, R_{fh} (\theta) \, K_\nu \; .
\end{equation}
The explicit form of $\Omega ({\bf 3})$ and the value of the indices $f$ and $h$ in the different cases, Case 1) through Case 3 b.1), will be presented in section~\ref{sec3}.
Since the charged leptons' physical basis coincides with the interaction basis, see eq.~\eqref{eq:Ul}, we have $U_\ell = \mathbb{1}$ and thus $U_{\mathrm{PMNS}}=U_\nu$.
The angle $\theta$ can take values between $0$ and $\pi$ and is fixed by accommodating the measured lepton mixing angles as well as possible.

\mathversion{bold}
\subsection{$(3,3)$ ISS framework}
\mathversion{normal}
\label{sec22}
In the $(3,3)$ ISS framework six neutral states, singlets under the SM gauge group, are added to the SM field content. In the following, these are denoted by $N_i$ and $S_j$ with $i,j = 1, 2, 3$. 
The Lagrangian giving rise to masses for the neutral particles (i.e. light neutrinos and heavy sterile states) reads
\begin{equation}
\label{eq:LISS}
- (y_D)_{\alpha i} \, \overline{L}^c_\alpha \, H \, N^c_i - (M_{NS})_{ij} \, \overline{N}_i \, S_j - \frac 12 \,  (\mu_S)_{kl} \, \overline{S}^c_k \, S_l + \mathrm{h.c.}
\end{equation}
with $\alpha=e, \mu, \tau$ and $i,j,k,l=1,2,3$. 
In the basis $\left(\nu_{\alpha L}, N^c_i, S_j \right)$,\footnote{In the following, we neglect possible contributions to the masses of the neutral particles arising from radiative corrections.} the mass matrix is of the form
\begin{equation}
\label{eq:MMaj}
\mathcal{M}_{\mathrm{Maj}} = 
\left( \begin{array}{ccc}
 \mathbb{0} & m_D & \mathbb{0} \\
 m_D^T & \mathbb{0} & M_{NS} \\
 \mathbb{0} & M_{NS}^T & \mu_S
\end{array}
\right) \;\; \mbox{with} \;\; m_D = y_D \, \frac{v}{\sqrt{2}}\,\text.
\end{equation}

In the limit $|\mu_S| \ll |m_D| \ll |M_{NS}|$ the light neutrino mass matrix is given at leading order 
in $(|m_D|/|M_{NS}|)^2$ by
\begin{equation}
\label{eq:mnuLO}
m_\nu = m_D \, \Big( M_{NS}^{-1} \Big)^{T} \, \mu_S \, M_{NS}^{-1} \, m_D^T\,\text.
\end{equation}
The contribution at subleading order reads~\cite{ISSsubleading}\footnote{Note the different choice of basis in~\cite{ISSsubleading}.}
\begin{equation}
\label{eq:mnuNLO}
\!\!m_\nu^1= - \frac 12 \, m_D \, \Big( M_{NS}^{-1} \Big)^T \, \Big[ \mu_S \, M_{NS}^{-1} \, m_D^T \, m_D^\star \, \Big( M_{NS}^{-1} \Big)^\dagger + \Big( M_{NS}^{-1} \Big)^\star \, m_D^\dagger \, m_D \, \Big( M_{NS}^{-1} \Big)^T \, \mu_S \Big] \, M_{NS}^{-1} \, m_D^T \, .
\end{equation}
The source of LN breaking in the ISS framework is $\mu_S$ and light neutrino masses vanish in the limit $\mu_S \, \rightarrow \, 0$, upon which LN conservation is restored. 

The matrix $\mathcal{M}_{\mathrm{Maj}}$ is diagonalised as
\begin{equation}
\label{eq:MMajdiag}
\mathcal{U}^ T \, \mathcal{M}_{\mathrm{Maj}} \, \mathcal{U} = \mathcal{M}_{\mathrm{Maj}}^{\mathrm{diag}} 
\end{equation}
with
\begin{equation}
\label{eq:formBigU}
\mathcal{U} = \left( \begin{array}{cc}
\widetilde{U}_\nu & S \\ T & V
\end{array}
\right)
\end{equation}
in which $\widetilde{U}_\nu$ is a three-by-three, $S$ a three-by-six, $T$ a six-by-three and $V$ a six-by-six matrix. 
The mass spectrum contains the three light (mostly active) neutrinos and six heavy (mostly sterile) states; their masses are denoted by $m_i$, with $i = 1, 2, 3$ corresponding to the light neutrinos, and $i = 4, ..., 9$ regarding the heavy neutral mass eigenstates.
For $|\mu_S| \ll |M_{NS}|$, the heavy masses are given to good approximation by $M_{NS}$, with $\mu_S$ determining the
mass splitting between the states forming pseudo-Dirac pairs.

We note that at leading order $\widetilde{U}_\nu$ approximately diagonalises the light neutrino mass matrix (c.f. eq.~\eqref{eq:mnuLO}) as
\begin{equation}
    \label{eq:Unutildemu}
    \widetilde{U}_\nu^T m_\nu\widetilde{U}_\nu\approx \:\mathrm{diag} (m_1, m_2, m_3)\,.
\end{equation}
While $\mathcal{U}$ is unitary, $\mathcal{U} \, \mathcal{U}^\dagger=\mathcal{U}^\dagger \, \mathcal{U} = \mathbb{1}$, none of the matrices $\widetilde{U}_\nu$, $S$, $T$ and $V$ has a priori this property. We can define the (in general non-unitary) PMNS mixing matrix as
\begin{equation}
\label{eq:defUPMNStilde}
\widetilde{U}_{\mathrm{PMNS}} = U_\ell^\dagger \, \widetilde{U}_\nu  \; .
\end{equation}
The non-unitarity of $\widetilde{U}_{\mathrm{PMNS}}$, induced by the mixing of the active neutrinos with the (heavy) sterile states, can be conveniently captured in the matrix $\eta$,
with flavour indices $\alpha, \beta=e, \mu, \tau$. It is defined as\footnote{Note the difference in sign with respect to the definition given in~\cite{ISSsubleading}.}
\begin{equation}
\label{eq:UPMNStildeeta}
\widetilde{U}_{\mathrm{PMNS}} = \Big( \mathbb{1} - \eta \Big) \, U_0
\end{equation}
with $\eta$ hermitian and $U_0$ unitary. 
Note that
\begin{equation}
\label{eq:UPMNStildeeta2}
\widetilde{U}_{\mathrm{PMNS}} \, \widetilde{U}_{\mathrm{PMNS}}^\dagger \approx \mathbb{1} - 2 \, \eta \; .
\end{equation}
For $U_\ell=\mathbb{1}$, which is always the case in our analysis, the following equality also holds 
\begin{equation}
\label{eq:UnuUl1}
\widetilde{U}_{\nu} = \Big( \mathbb{1} - \eta \Big) \, U_0 \; .
\end{equation}
The size of $\eta$ and its form in flavour space are given at leading order by
\begin{equation}
\label{eq:eta}
\eta = \frac 12 \, m_D^\star \, \Big( M_{NS}^{-1} \Big)^\dagger \, M_{NS}^{-1} \, m_D^T\,.
\end{equation}
We can estimate the form of the matrix $T$ as 
\begin{equation}
\label{eq:matT}
T = \left( \begin{array}{c}
\mathbb{0} \\ - M_{NS}^{-1} \, m_D^T \, \widetilde{U}_\nu
\end{array}
\right) \approx \left( \begin{array}{c}
\mathbb{0} \\ - M_{NS}^{-1} \, m_D^T \, U_0
\end{array}
\right) \; ,
\end{equation}
while for $S$ one has  
\begin{equation}
\label{eq:matS}
S = \left(\mathbb{0} \; , \;\; m_D^\star \, \Big( M_{NS}^{-1} \Big)^\dagger  
\right) \, V \; ,
\end{equation}
and $V$ approximately diagonalises the lower six-by-six matrix of $\mathcal{M}_{\mathrm{Maj}}$, i.e.~
\begin{equation}
\label{eq:defV}
V^T \, \left( \begin{array}{cc}
\mathbb{0} & M_{NS} \\
M_{NS}^T & \mu_S
\end{array}
\right) \, V \approx \:\mathrm{diag}\left(m_4, ..., m_9\right)\,.
\end{equation}
The matrix $\mu_S$, a complex symmetric matrix, is itself diagonalised by
\begin{equation}
 \label{eq:muS}
U_S^T \, \mu_S \, U_S = \left(
\begin{array}{ccc}
\mu_1 & 0 & 0 \\
0 & \mu_2 & 0 \\
0 & 0 & \mu_3
\end{array}
\right)\,,
\end{equation}
with $\mu_i$ real and positive semi-definite, and $U_S$ unitary.

\vspace{0.1in}
\noindent Like in the model-independent scenario, the charged lepton sector leaves the residual symmetry $G_\ell$ invariant. For this reason, we assign the three generations of LH lepton doublets $L_\alpha$ and of 
RH charged leptons $\ell_{\alpha R}$ to the same representations under $G_f$, the $Z_3$ group $Z_3^{(\mathrm{aux})}$ and the CP symmetry as in the model-independent scenario. As a consequence, also in the $(3,3)$ ISS framework
the charged lepton mass matrix $m_\ell$ is diagonal and the contribution to the lepton mixing matrix is $U_\ell=\mathbb{1}$.  The group $G_\nu=Z_2 \times CP$ is the residual symmetry among the neutral states. In the $(3,3)$ 
ISS framework, we also have to assign the heavy sterile states, $N_i$ and $S_j$ with $i,j= 1,2,3$, to representations of $G_f$, $Z_3^{(\mathrm{aux})}$ and CP.
In the following, we identify three minimal options to choose these representations.

\subsubsection*{Option 1}

For option 1, we assume that $N_i$ and $S_j$ each transform like the LH lepton doublets $L_\alpha$, namely as the triplet ${\bf 3}$ under $G_f$.  
Furthermore, the heavy sterile states are neutral under $Z_3^{(\mathrm{aux})}$. As a consequence of this assignment, the Dirac neutrino Yukawa matrix $y_D$, and 
consequently the mass matrix $m_D$ as well as the matrix $M_{NS}$, are non-vanishing in the limit of unbroken $G_f$, $Z_3^{(\mathrm{aux})}$ and CP. They take a particularly simple form 
 \begin{equation} 
 \label{eq:mDopt1}
 m_D = y_0 \, \left(
 \begin{array}{ccc}
 1 & 0 & 0\\
 0 & 1 & 0\\
 0 & 0 & 1
 \end{array}
 \right) \, \frac{v}{\sqrt{2}} \;\; \mbox{with} \;\; y_0 > 0
 \end{equation}
 and
 \begin{equation}
 \label{eq:MNSopt1}
 M_{NS} = M_0 \, \left(
 \begin{array}{ccc}
  1 & 0 & 0\\
 0 & 1 & 0\\
 0 & 0 & 1
 \end{array}
 \right) \;\; \mbox{with} \;\; M_0 > 0 \, .
 \end{equation}
Thus, the only source of $G_f$ and CP breaking in the sector of the neutral states is the matrix $\mu_S$. In order to preserve the residual symmetry $G_\nu$, the matrix $\mu_S$ is constrained by the following equations
\begin{equation}
\label{eq:muScond}
Z ({\bf 3})^T \, \mu_S \, Z ({\bf 3}) = \mu_S \;\; \mbox{and} \;\; X ({\bf 3}) \, \mu_S \, X ({\bf 3}) = \mu_S^\star \; ,
\end{equation}
implying that $\mu_S$ has to fulfil the same relations as $m_\nu$ (cf. eq.~(\ref{eq:ZXmnu})). Hence, the matrix $U_S$, which diagonalises $\mu_S$, is of the same form as $U_\nu$, see eq.~(\ref{eq:Unumodind}), 
\begin{equation}
 \label{eq:USopt1}
U_S = \Omega ({\bf 3}) \, R_{fh} (\theta_S) \; .
\end{equation}
Note that we do not mention explicitly a matrix equivalent to $K_\nu$ in eq.~(\ref{eq:Unumodind}), as we assume for concreteness in our analysis that it is the identity matrix. 

For option 1, $\mu_S$ is the unique source of LN violation and lepton flavour violation. Nevertheless, LN, $G_f$ and CP can be broken in different ways, explicitly or spontaneously, and at vastly
different scales in concrete models.

Plugging $m_D$, $M_{NS}$ and $\mu_S$
 from eqs.~(\ref{eq:mDopt1},\ref{eq:MNSopt1},\ref{eq:muS},\ref{eq:USopt1}) into the form of $m_\nu$ in eq.~(\ref{eq:mnuLO}), we find at leading order
 \begin{equation}
 \label{eq:mnuLOopt1}
m_\nu = \frac{y_0^2 \, v^2}{2 \, M_0^2} \, \mu_S = \frac{y_0^2 \, v^2}{2 \, M_0^2} \, U_S^\star \, \left( \begin{array}{ccc}
\mu_1 & 0 & 0 \\
0 & \mu_2 & 0 \\
0 & 0 & \mu_3
\end{array}
\right) \, U_S^\dagger \, .
\end{equation}
Consequently, the matrix $\widetilde{U}_\nu$, which diagonalises $m_\nu$ at leading order (neglecting the correction $\eta$ that encodes the deviation from unitarity of $\widetilde{U}_\nu$), is given by
\begin{equation}
\label{eq:UnuU0opt1}
\widetilde{U}_\nu \approx U_0 = U_S = \Omega ({\bf 3}) \, R_{fh} (\theta_S) \; , 
\end{equation}
and the light neutrino masses read
\begin{equation}
\label{eq:m123LOopt1}
m_i = \frac{y_0^2 \, v^2}{2 \, M_0^2} \, \mu_i \;\; \mbox{for} \;\; i=1,2,3 \; .
\end{equation}
Assuming $y_0 \sim 1$ and $M_0 \sim 1000 \, \mbox{GeV}$, we can estimate the size of $\mu_i$ to be of the order of $\mathrm{eV}$. The ratio between $m_D$ and $M_{NS}$, evaluating the impact 
of the heavy sterile states, is then $\frac{y_0 \, v}{\sqrt{2} \, M_0} \sim 0.17$.
Since the mass squared differences of neutrinos have been determined from neutrino oscillation data and the sum of neutrino masses is constrained by cosmological measurements, see appendix~\ref{app3}, 
the values of $\mu_i$ are further restricted. 
Since $U_\ell=\mathbb{1}$ and $\widetilde{U}_\nu$ is at leading order of the form given in eq.~(\ref{eq:UnuU0opt1}), we have for the PMNS mixing matrix
\begin{equation}
\label{eq:UPMNSLOopt1}
\widetilde{U}_{\mathrm{PMNS}}\approx\Omega ({\bf 3}) \, R_{fh} (\theta_S)
\end{equation}
with $\theta_S$ being constrained by the measured values of the lepton mixing angles, like $\theta$ in eq.~(\ref{eq:Unumodind}). We note that we consider the free angle $\theta_S$ to vary in the range $0$ and $\pi$.
The results in the $(3,3)$~ISS framework (at leading order) are thus identical to those obtained in the model-independent scenario. However, they can be altered by two effects: the inclusion of the subleading contribution $m_\nu^1$ to the light
neutrino mass matrix in eq.~(\ref{eq:mnuNLO}) and effects of non-unitarity of $\widetilde{U}_\nu$, see eqs.~(\ref{eq:UPMNStildeeta},\ref{eq:eta}). This is studied in detail analytically in section~\ref{sec4}
and numerically in section~\ref{sec5} (the experimental constraints on $\eta$ are discussed in section~\ref{sec50}). 

We briefly discuss the form of the matrices $S$, $T$ and $V$, as well as the mass spectrum of the heavy sterile states analytically. With eqs.~(\ref{eq:matT},\ref{eq:mDopt1},\ref{eq:MNSopt1},\ref{eq:UnuU0opt1}) the matrix $T$ reads at leading order
\begin{equation}
\label{eq:matTopt1}
T  = \left( \begin{array}{c}
\mathbb{0} \\ - \frac{y_0 \, v}{\sqrt{2} \, M_0} \, U_S
\end{array}
\right) \; .
\end{equation}
From the definition of $V$ in eq.~(\ref{eq:defV}) and with the form of $M_{NS}$ in eq.~(\ref{eq:MNSopt1}) and $\mu_S$ in eqs.~(\ref{eq:muS},\ref{eq:USopt1}), we find at leading order for $V$
 \begin{equation}
\label{eq:matVopt1}
V = \frac{1}{\sqrt{2}} \, \left(
\begin{array}{cc}
i \, U_S^\star & U_S^\star\\
-i \, U_S & U_S
\end{array}
\right) \; ,
\end{equation}
while the matrix $S$ in eq.~(\ref{eq:matS}) reads 
\begin{equation}
\label{eq:matSopt1}
S = \frac{y_0 \, v}{2 \, M_0} \, \left(- i \, U_S \; , \;\; U_S \right) \, . 
\end{equation}
We note that the approximate analytical results for $\widetilde{U}_\nu$, $S$, $T$ and $V$ have been compared to the numerical ones for one choice of parameters for Case 1) and we find good agreement in form and magnitude of their entries.
The mass spectrum of the heavy sterile states (arising from the diagonalisation through $V$ in eq.~(\ref{eq:matVopt1})) is at leading order
\begin{equation}
\label{eq:heavyMopt1}
m_{3 + i} = M_0 - \frac{\mu_i}{2} \;\; \mbox{and} \;\; m_{6+i}= M_0 + \frac{\mu_i}{2} \;\; \mbox{with} \;\; i=1,2,3 \,.
\end{equation}
All heavy sterile states are thus degenerate in mass to a very high degree for typical choices of $M_0$ and $\mu_S$, e.g.~$M_0 \sim 1000 \, \mathrm{GeV}$ and $\mu_S \lesssim 1 \, \mathrm{keV}$.

\vspace{0.2in}
Beyond option 1, there are two further minimal options, option 2 and option 3, in which only one of the mass matrices $m_D$, $M_{NS}$ and $\mu_S$ carries non-trivial flavour information.
These options share a common feature: in both the matrix $\mu_S$ has a trivial flavour structure.
Thus, for these options the sources of lepton flavour and LN violation are decoupled. 
For option 2,
$m_D$ contains all flavour information, while $M_{NS}$ is flavour-diagonal and flavour-universal, 
so that the mass spectrum of the heavy sterile
states will be degenerate to a high degree, like for option 1. 
Instead, for option 3 the entire flavour structure is encoded in the matrix $M_{NS}$, while 
$m_D$ is flavour-diagonal
and flavour-universal. In this way, the heavy sterile states have in general different masses. We note that the realisation of option 2 and option 3 requires in general that (at least) the assignment of the three sterile
states $S_i$, $i=1,2,3$, under the flavour symmetry $G_f$ be altered compared to option 1, in order to ensure that the matrix $\mu_S$ is non-vanishing in the limit of unbroken $G_f$, $Z_3^{(\mathrm{aux})}$ and CP.
However, this can always be achieved by an appropriate choice of $G_f$. Obviously, one can also consider less minimal options in which two of the three mass matrices $m_D$, $M_{NS}$ and $\mu_S$,
if not all three of them, have a non-trivial flavour structure.

\section{Lepton mixing in the model-independent scenario}
\label{sec3}

In this section, we revisit the four different types of lepton mixing patterns, Case~1) through Case 3~b.1), that have been identified in the study of~\cite{HMM}.
We mention for each case the generator $Z$, the CP transformation $X$ and the expressions for $\sin^2\theta_{ij}$, $J_{\mathrm{CP}}$, $I_1$ and $I_2$
and, where available, (approximate) formulae for the sines of the CP phases as well as (approximate) sum rules among the lepton mixing parameters.
We remind that the residual symmetry in the charged lepton sector, $G_\ell$,  is always chosen as the $Z_3$ group which corresponds to the diagonal 
subgroup of the $Z_3$ symmetry, contained in $G_f$ and arising from the generator $a$, and the $Z_3$ symmetry $Z_3^{(\mathrm{aux})}$. As discussed,
this leads to a diagonal charged lepton mass matrix and, consequently, to no contribution to lepton mixing from the charged lepton sector, see eq.~(\ref{eq:Ul}).

\subsection{Case 1)}
\label{sec31}

For Case 1), the generator $Z$ of the residual $Z_2$ symmetry and the CP transformation $X$ are given by
\begin{equation}
\label{eq:ZXCase1}
Z=c^{n/2} \;\; \mbox{and} \;\; X=a\, b \, c^s \, d^{2 \, s} \, X_0 \;\; \mbox{with} \;\; 0 \leq s \leq n-1 \; .
\end{equation}
Note that the index $n$ has to be even. The explicit form of the generators and of $X_0$ can be found in appendix~\ref{app1}.
The matrix $\Omega ({\bf 3})$ and the indices $f$ and $h$ of the rotation $R_{fh} (\theta)$, appearing in eq.~(\ref{eq:Unumodind}),
are
\begin{equation}
\label{eq:OmegaR13Case1}
\Omega ({\bf 3}) = e^{i \, \phi_s} \, U_{\mathrm{TB}} \, \left(
\begin{array}{ccc}
1 & 0 & 0\\
0 & e^{-3 \, i \, \phi_s} & 0\\
0 & 0 & -1
\end{array}
\right)
\;\; \mbox{and} \;\; 
R_{13} (\theta) = \left(
\begin{array}{ccc}
\cos \theta & 0 & \sin\theta\\
0 & 1 & 0\\
-\sin\theta & 0 & \cos\theta
\end{array}
\right)
\end{equation}
with
\begin{equation}
\label{eq:UTB}
U_{\mathrm{TB}} = \left(
\begin{array}{ccc}
\sqrt{\frac 23} & \frac{1}{\sqrt{3}} & 0\\
-\frac{1}{\sqrt{6}} & \frac{1}{\sqrt{3}} & \frac{1}{\sqrt{2}}\\
-\frac{1}{\sqrt{6}} & \frac{1}{\sqrt{3}} & -\frac{1}{\sqrt{2}}
\end{array}
\right) 
\end{equation}
and
\begin{equation}
\label{eq:phisdef}
\phi_s=\frac{\pi \, s}{n} \; .
\end{equation}
The matrix $K_\nu$, present in eq.~(\ref{eq:Unumodind}), is set to the identity matrix for concreteness.

The main results of Case 1) are the following:\\ 
$a)$ the solar mixing angle is constrained by 
\begin{equation}
\label{eq:sin2th12limitCase1}
\sin^2\theta_{12} \gtrsim \frac 13 \; ,
\end{equation}
$b)$ none of the mixing angles depends on the parameters $n$ and $s$
\begin{equation}
\label{eq:sin2thetaijCase1}
\sin^2\theta_{13}= \frac 23 \, \sin^2\theta \; , \;\; \sin^2\theta_{12} = \frac{1}{2+\cos 2 \theta} \; , \;\; \sin^2 \theta_{23} = \frac 12 \, \Big( 1+ \frac{\sqrt{3} \, \sin 2 \theta}{2+\cos 2 \theta} \Big) \; ,
\end{equation}
$c)$ the size of the free angle $\theta$ is (mainly) fixed by the measured value of the reactor mixing angle $(\theta_{13})$ and $\theta$ takes two different values in the interval between $0$ and $\pi$,
\begin{equation}
\label{eq:thetaCase1}
\theta \approx 0.18 \;\; \mbox{and} \;\; \theta \approx 2.96 \; ,
\end{equation}
$d)$ two approximate sum rules among the mixing angles can be established
\begin{equation}
\label{eq:sumrulesCase1}
\sin^2 \theta_{12} \approx \frac 13 \, \left( 1 + \sin^2\theta_{13} \right) \;\; \mbox{and} \;\; \sin^2\theta_{23} \approx \frac 12 \, \left( 1 \pm \sqrt{2} \, \sin \theta_{13} \right) 
\end{equation}
with $\pm$ depending on $\theta \lessgtr \pi/2$,\\
$e)$ the Dirac phase $\delta$ and the Majorana phase $\beta$ are both trivial, $\sin\delta=0$ and $\sin\beta=0$,\\
$f)$ the Majorana phase $\alpha$ only depends on the parameter $s$ (the ratio $s/n$) and its sine reads
\begin{equation}
\label{eq:sinalphaCase1}
\sin \alpha = - \sin 6 \, \phi_s \; ,
\end{equation}
$g)$ for $s=0$ and $s=\frac n2$, CP is not violated.

\subsection{Case 2)}
\label{sec32}

The residual $Z_2$ symmetry in the sector of the neutral states is the same as in Case 1), while the CP transformation $X$ depends on two different parameters
\begin{equation}
\label{eq:ZXCase2}
Z=c^{n/2} \;\; \mbox{and} \;\; X=c^s \, d^t \, X_0\,, \;\; \mbox{with} \;\; 0 \leq s, t \leq n-1 \; .
\end{equation}
Like for Case 1), the index $n$ of $G_f$ has to be even.
A more convenient choice of parameters than $s$ and $t$ are $u$ and $v$, which are related to the former by
\begin{equation}
\label{eq:uvdeffromst}
u=2 \, s-t \;\; \mbox{and} \;\; v=3 \, t \; .
\end{equation}
The matrix $\Omega ({\bf 3})$ and the indices $f$ and $h$ of the rotation matrix $R_{fh} (\theta)$ in eq.~(\ref{eq:Unumodind}) read
\begin{equation}
\label{eq:OmegaR13Case2}
\Omega ({\bf 3}) = e^{i \, \phi_v/6} \, U_{\mathrm{TB}} \, R_{13} \, \left( -\frac{\phi_u}{2} \right) \, \left(
\begin{array}{ccc}
1 & 0 & 0\\
0 & e^{-i \, \phi_v/2} & 0\\
0 & 0 & -i
\end{array}
\right)
\;\; \mbox{and} \;\; 
R_{13} (\theta) = \left(
\begin{array}{ccc}
\cos \theta & 0 & \sin\theta\\
0 & 1 & 0\\
-\sin\theta & 0 & \cos\theta
\end{array}
\right) 
\end{equation}
with
\begin{equation}
\label{eq:phiuphivdef}
\phi_u=\frac{\pi \, u}{n} \;\; \mbox{and} \;\; \phi_v=\frac{\pi \, v}{n} \; .
\end{equation}
For the definition of $U_{\mathrm{TB}}$ see eq.~(\ref{eq:UTB}). 
Like for Case 1) we set $K_\nu$ to the identity matrix.

The main features of the mixing pattern of Case 2) are:\\
$a)$ the solar mixing angle has a lower limit identical to the one of Case 1), see eq.~(\ref{eq:sin2th12limitCase1}),\\
$b)$ the lepton mixing angles depend on the parameters $u$ and $n$ as well as the free angle $\theta$
\begin{equation}
\label{eq:sin2thetaijCase2}
\!\!\!\sin^2 \theta_{13} = \frac 13 \, (1-\cos \phi_u \, \cos 2 \theta) \; , \;\; \sin^2 \theta_{12} = \frac{1}{2+\cos \phi_u \cos 2 \theta} \; , \;\;
\sin^2 \theta_{23} = \frac 12 \, \left( 1 + \frac{\sqrt{3} \, \sin \phi_u \cos 2\theta}{2+\cos\phi_u \cos 2 \theta} \right) \; ,
\end{equation}
$c)$ the size of $\cos \phi_u \, \cos 2 \theta$ (and thus of $u/n$ and $\theta$) is constrained by the measured value of the reactor mixing angle.
Taking into account symmetries of the formulae in $(u,\theta)$, discussed in~\cite{HMM}, it is sufficient to consider small values of $u/n$ and $\cos 2\theta \approx 1$.
The choice $u=0$ is associated with distinctive features (see point $g)$ below).\\
$d)$ the mixing angles fulfil two (approximate) sum rules: the one already found for Case 1), see first approximate equality in eq.~(\ref{eq:sumrulesCase1}), and 
\begin{equation}
\label{eq:sumruleCase2}
6 \, \sin^2 \theta_{23} \, (1-\sin^2 \theta_{13}) = 3+\sqrt{3} \, \tan\phi_u - 3 \, \left( 1+\sqrt{3} \, \tan\phi_u \right) \, \sin^2 \theta_{13} \; ,
\end{equation}
$e)$ the Dirac phase $\delta$ and the Majorana phase $\beta$ depend on the parameters $u$ and $n$ as well as on the free angle $\theta$. Information on them
is most conveniently given in terms of the CP invariants $J_{\mathrm{CP}}$ and $I_2$ (see appendix~\ref{app2})
\begin{equation}
\label{eq:JCPI2Case2}
J_{\mathrm{CP}} = - \frac{\sin 2\theta}{6 \, \sqrt{3}} \;\; \mbox{and} \;\; I_2= \frac 19 \, \sin 2 \,\phi_u \, \sin 2 \theta \; ,
\end{equation}
$f)$ the Majorana phase $\alpha$ depends, to very good accuracy, only on the parameters $v$ and $n$ (through the ratio $v/n$)
\begin{equation}
\label{eq:sinalphaCase2}
\sin\alpha \approx -\sin\phi_v \; ,
\end{equation}
$g)$ for the choice $u=0$, the atmospheric mixing angle and the Dirac phase are both maximal, $\sin^2 \theta_{23}=1/2$ and $|\sin\delta|=1$, while
the Majorana phase $\beta$ is trivial, $\sin\beta=0$, and the Majorana phase $\alpha$ exactly fulfils the approximate equality in eq.~(\ref{eq:sinalphaCase2}).\\
$h)$ if $v=0$ is permitted, this leads to a trivial Majorana phase $\alpha$, $\sin\alpha=0$,\\
$i)$ furthermore, three symmetry transformations of the formulae of the lepton mixing parameters (in the parameters $u$ and $\theta$) have been found in~\cite{HMM}. Two of them
are independent, e.g.~
\begin{equation}
\label{eq:symmtrafosCase2}
\begin{array}{lll}
u \; \rightarrow \; u+n\, ,& \theta \; \rightarrow \; \frac{\pi}{2} - \theta \, :& \sin^2 \theta_{ij}, \, J_{\mathrm{CP}}, \, I_2 \; \mbox{are invariant and} \;\; I_1 \; \mbox{changes sign;}
\\ 
u \; \rightarrow \; 2\, n-u\, ,& \theta \; \rightarrow \; \pi - \theta \, :& \sin^2 \theta_{13}, \, \sin^2 \theta_{12}, \, I_1, \, I_2 \; \mbox{are invariant,} \;\; J_{\mathrm{CP}} \; \mbox{changes sign} 
\\
&&\mathrm{and} \; \sin^2\theta_{23} \; \rightarrow \;\; 1-\sin^2\theta_{23} \; .
\end{array}
\end{equation}

\subsection{Case 3 a) and Case 3 b.1)}
\label{sec33}

\noindent Case 3 a) and Case 3 b.1) are based on a different residual $Z_2$ symmetry in the sector of the neutral states than that of Case 1) and Case 2). This $Z_2$ symmetry depends on the parameter $m$.
Similarly, the CP transformation $X$ depends on the parameter $s$.
The explicit form of the generator $Z$ and of $X$ is
\begin{equation}
\label{eq:ZXCase3}
Z=b \, c^m \, d^m \;\; \mbox{and} \;\; X=b \, c^s \, d^{n-s} \, X_0 \;\; \mbox{and} \;\; 0 \leq m, \, s \leq n-1 \; .
\end{equation}
Since $Z$ contains the generator $b$, Case 3 a) and Case 3 b.1) can only be realised with the flavour symmetry $G_f=\Delta (6 \, n^2)$. 
The value of the parameter $m$ and, consequently, the choice of the residual $Z_2$ symmetry are strongly constrained by the measured values of the lepton mixing angles.

A possible form of $\Omega ({\bf 3})$ and the matrix $R_{fh} (\theta)$ are given by 
\begin{equation}
\label{eq:OmegaR12Case3}
\!\!\!\Omega ({\bf 3}) = e^{i \, \phi_s}\, \left( \begin{array}{ccc}
1 & 0 & 0\\
0 & \omega & 0\\
0 & 0 & \omega^2 
\end{array}
\right) \, U_{\mathrm{TB}} \, \left(
\begin{array}{ccc}
1 & 0 & 0\\
0 & e^{-3 \, i \, \phi_s} & 0\\
0 & 0 & -1
\end{array}
\right) \, R_{13} (\phi_m) \;\; \mbox{and} \;\; R_{12} (\theta) = \left(
\begin{array}{ccc}
\cos\theta & \sin\theta & 0\\
-\sin\theta & \cos\theta & 0\\
0 & 0 & 1
\end{array}
\right) 
\end{equation}
with
\begin{equation}
\label{eq:phimdef}
\phi_m=\frac{\pi \, m}{n} \; . 
\end{equation}
Again, the matrix $K_\nu$ is set to the identity matrix.

Two viable types of mixing patterns are found~\cite{HMM}: in Case 3 a) the parameter $m$ fixes the values of the atmospheric and reactor mixing 
angles, while in Case 3 b.1) the parameter $m$ is around $n/2$ in order to successfully accommodate the solar mixing angle. We first recapitulate the results for Case 3 a)
and then those for Case 3 b.1).

\subsubsection{Case 3 a)}
\label{sec331}

The relevant properties of the mixing pattern of Case 3 a) are:\\
$a)$ the value of $m/n$ is strongly constrained by the measured value of the reactor mixing angle. This value has to be either close to $0$ or to $1$.
Not only $\sin^2\theta_{13}$ is fixed by $m/n$, but also the value of the atmospheric mixing angle
\begin{equation}
\label{eq:sin2theta1323Case3a}
\sin^2\theta_{13} = \frac 23 \, \sin^2 \phi_m \;\; \mbox{and} \;\; \sin^2 \theta_{23} = \frac 12 \, \left( 1+ \frac{\sqrt{3} \, \sin 2 \, \phi_m}{2+\cos 2 \, \phi_m} \right) \; ,
\end{equation}
$b)$ due to this strong correlation a sum rule can be derived for $\sin^2 \theta_{13}$ and $\sin^2 \theta_{23}$
\begin{equation}
\label{eq:sumruleCase3a}
\sin^2\theta_{23} \approx \frac 12 \, \left( 1 \pm \sqrt{2} \, \sin\theta_{13} \right) 
\end{equation}
with $\pm$ for $m/n$ close to $0$ or $1$, respectively,\\
$c)$ the solar mixing angle depends on the parameter $s$ and on the free angle $\theta$ as well
\begin{equation}
\label{eq:sin2theta12Case3a}
\sin^2\theta_{12} = \frac{1+\cos 2\,\phi_m\,\sin^2\theta+\sqrt{2}\,\cos\phi_m\,\cos 3\,\phi_s \, \sin 2\theta}{2+\cos 2 \, \phi_m}.
\end{equation}
Note that the solar mixing angle can be accommodated to its measured best-fit value for most of the choices of the parameter $s$. In particular, $\sin^2\theta_{12}$
is no longer constrained to be larger than $1/3$, as for Case 1) and Case 2). For most choices of $s$ two values of the free angle $\theta$, one close to $0$ or $\pi$ and another depending on the parameter $s$, permit an acceptable fit
to the measured value of $\sin^2\theta_{12}$.\\
$d)$ the CP invariants $J_{\mathrm{CP}}$, $I_1$ and $I_2$ depend in general on all parameters, $n$, $m$, $s$ and $\theta$,
\begin{eqnarray}
\label{eq:JCPI1I2Case3a}
&&J_{\mathrm{CP}} = - \frac{1}{6 \, \sqrt{6}} \, \sin 3 \, \phi_m \, \sin 3 \, \phi_s \, \sin 2 \theta \; ,
\\ \nonumber
&&I_1=-\frac 19 \, \cos \phi_m \, \sin 3 \, \phi_s \, \left( 4 \, \cos \phi_m \, \cos 3 \, \phi_s \, \cos 2 \theta + \sqrt{2} \, \cos 2 \, \phi_m \, \sin 2 \theta \right) \; , 
\\ \nonumber
&&I_2=\frac 49 \, \sin^2\phi_m \, \sin 3 \, \phi_s \, \sin\theta \, \left( \cos 3\, \phi_s \, \sin\theta -\sqrt{2} \, \cos\phi_m \, \cos\theta \right) \; , 
\end{eqnarray}
$e)$ approximate values can be found for the sines of the CP phases when the constraints on $m/n$ and $\theta$, arising from accommodating the lepton mixing angles, are used. These 
are
\begin{equation}
\label{eq:sinalphaCase3a}
|\sin\alpha| \approx |\sin 6 \, \phi_s| \; ,
\end{equation}
and
\begin{eqnarray}
\label{eq:sinbetasindeltaCase3a}
\mbox{for} \; \theta \approx 0, \, \pi&&\sin\delta \approx 0 \;\; \mbox{and} \;\; \sin\beta \approx 0 \; ,
\\ 
\nonumber
\mbox{for} \; \theta \not\approx 0, \, \pi&&|\sin\delta| \approx \left| \frac{3 \, \sin 6 \,\phi_s}{5+ 4 \, \cos 6 \, \phi_s}\right| \;\; \mbox{and} \;\; |\sin\beta| \approx 2 \, |\sin 6 \, \phi_s| \, \left| \frac{2+\cos 6 \, \phi_s}{5+4 \, \cos 6 \, \phi_s} \right| \; .
\end{eqnarray}
Note that the magnitude of $\sin\beta$ has an upper limit, $|\sin\beta| \lesssim 0.87$.\\
$f)$ if two values of the free angle $\theta$ permit an acceptable fit to the measured lepton mixing angles for a certain choice of $s$, the sine of the Majorana phase $\alpha$ for the two different values of $\theta$ has the same magnitude,
but opposite sign. If only one value of $\theta$ leads to a good fit to the experimental data, the Majorana phase $\alpha$ is trivial, $\sin\alpha=0$,\\
$g)$ for $s=0$, all CP phases are trivial, i.e.~$\sin\alpha=\sin\beta=\sin\delta=0$.\\
$h)$ for the choice $s=\frac n2$, the free angle $\theta$, that leads to the best accommodation of the measured values of the lepton mixing angles, is $\theta=0$. Consequently, 
the solar mixing angle is bounded from below, i.e.~$\sin^2\theta_{12} \gtrsim 1/3$, and all CP phases are trivial,\\ 
$i)$ like for Case 2), three symmetry transformations of the formulae of the lepton mixing parameters in the parameters $m$, $s$ and the free angle $\theta$ have been found in~\cite{HMM}. Two of them
are independent, e.g.~
\begin{equation}
\label{eq:symmtrafosCase3}
\begin{array}{lll}
s\; \rightarrow \; n-s \; ,&\theta \; \rightarrow \; \pi-\theta \; :&\sin^2\theta_{ij} \; \mbox{are invariant and} \;  J_{\mathrm{CP}}, \, I_1, \, I_2 \; \mbox{change sign;} 
\\ 
m \; \rightarrow \; n-m \; ,&\theta \; \rightarrow \; \pi-\theta \; :&\sin^2\theta_{13}, \, \sin^2\theta_{12}, \, I_1, \, I_2 \; \mbox{are invariant,} \; J_{\mathrm{CP}} \; \mbox{changes sign} 
\\ 
&&\mbox{and} \; \sin^2\theta_{23} \; \rightarrow 1-\sin^2\theta_{23} \; .
\end{array}
\end{equation}

\subsubsection{Case 3 b.1)}
\label{sec34}

The lepton mixing pattern of Case 3 b.1) arises from the matrices $\Omega ({\bf 3})$ and $R_{12} (\theta)$ in eq.~(\ref{eq:OmegaR12Case3}), if these are multiplied from the 
right with the cyclic permutation matrix $P_{\mathrm{cyc}}$
\begin{equation}
\label{eq:PcycCase3b1}
P_{\mathrm{cyc}} = \left( \begin{array}{ccc}
0 & 1 & 0\\
0 & 0 & 1\\
1 & 0 & 0
\end{array}
\right) \; \mbox{, i.e.~} \;\; U_{\mathrm{PMNS}}=U_\nu=\Omega ({\bf 3}) \, R_{12} (\theta) \, P_{\mathrm{cyc}} \; .
\end{equation}
This cyclic permutation corresponds to a re-ordering of the columns of the PMNS mixing matrix. 
The properties of the lepton mixing pattern of Case 3 b.1) can be summarised as follows:\\
$a)$ all lepton mixing parameters depend on $n$, $m$, $s$ and the free angle $\theta$,
\begin{eqnarray}
\label{eq:sin2thetaijCase3b1}
\sin^2 \theta_{13} &=& \frac 13 \, \left( 1+ \cos 2 \, \phi_m \, \sin^2\theta +\sqrt{2} \, \cos\phi_m \, \cos 3 \, \phi_s \, \sin 2 \theta \right) \; ,
\\ \nonumber
\sin^2 \theta_{23} &=& \frac 12 \, \left( 1+ \frac{2 \, \sqrt{3}\, \sin\phi_m\,\sin\theta \, (\sqrt{2}\,\cos 3 \, \phi_s \, \cos\theta-\cos\phi_m \, \sin\theta)}{2-\cos 2 \, \phi_m \,\sin^2\theta-\sqrt{2}\,\cos\phi_m\,\cos3 \, \phi_s \, \sin2 \theta} \right) \; ,
\\ \nonumber
\sin^2\theta_{12} &=& 1 - \frac{2 \, \sin^2 \phi_m}{2-\cos 2 \, \phi_m \,\sin^2\theta-\sqrt{2}\,\cos\phi_m \, \cos 3 \, \phi_s \, \sin 2\theta}
\end{eqnarray}
and
\begin{eqnarray}
\label{eq:JCPI1I2Case3b1}
J_{\mathrm{CP}} &=& -\frac{1}{6 \, \sqrt{6}} \, \sin 3 \, \phi_m \, \sin 3 \, \phi_s \, \sin 2 \theta \; ,
\\ \nonumber
I_1 &=& -\frac 49 \, \sin^2 \phi_m \, \sin 3 \, \phi_s \, \sin\theta \, \left( \cos 3 \, \phi_s \, \sin\theta-\sqrt{2} \, \cos\phi_m \, \cos\theta \right) \; ,
\\ \nonumber
I_2 &=& -\frac 49 \, \sin^2 \phi_m \, \sin 3 \, \phi_s \, \cos\theta \, \left( \cos 3 \, \phi_s \, \cos\theta+\sqrt{2} \, \cos \phi_m \, \sin\theta \right) \; ,
\end{eqnarray}
$b)$ the parameter $m$ is strongly constrained by the measured value of $\sin^2\theta_{12}$, i.e.~$m\approx \frac n2$,\\
$c)$ for $m=\frac n2$, two approximate sum rules among the lepton mixing angles are found
\begin{equation}
\label{eq:sumrulesCase3b1}
\sin^2\theta_{12} \approx \frac 13 \, \left( 1-2 \, \sin^2\theta_{13} \right) \;\; \mbox{and} \;\; \sin^2\theta_{23} \approx \frac 12 \left( 1+\sqrt{\frac23} \, \frac{\cos 3 \, \phi_s \, \sin2 \theta_0}{1-\sin^2 \theta_{13}} \right)
\end{equation}
with $\theta_0 \approx 1.31$ or $\theta_0 \approx 1.83$, constrained by the measured value of the reactor mixing angle,\\
$d)$ for $m= \frac n2$ and $s= \frac n2$, the atmospheric mixing angle is maximal, $\sin^2\theta_{23}= \frac 12$,\\
$e)$ for $m= \frac n2$, the Majorana phases only depend on the parameter $s$ (the ratio $s/n$), and have the same magnitude, 
\begin{equation}
\label{eq:sinaplhabetaCase3b1}
\sin\alpha=\sin\beta=-\sin 6 \, \phi_s
\end{equation}
and the Dirac phase fulfils the approximate relation 
\begin{equation}
\label{eq:sindeltaCase3b1}
\sin\delta \approx \pm \sin 3 \,\phi_s \;\; \mbox{with} \; \pm \; \mbox{referring to} \;\; \theta \lessgtr \pi/2\; .
\end{equation}
Taking into account the constraints on the free angle $\theta$ and the parameter $s$, arising from the experimental data on lepton mixing angles, the magnitude of the sine of the Dirac phase is bounded from below, $|\sin\delta| \gtrsim 0.71$,\\ 
$f)$ for $m= \frac n2$ and $s= \frac n2$, the Dirac phase is maximal, $|\sin\delta|=1$, while both Majorana phases are trivial, $\sin\alpha=0$ and $\sin\beta=0$,\\ 
$g)$ for $s=0$, CP is not violated,\\ 
$h)$ for Case 3 b.1) the same symmetry transformations hold as for Case 3 a), see point $i)$ in section~\ref{sec331}, eq.~(\ref{eq:symmtrafosCase3}).

\mathversion{bold}
\section{Impact of heavy sterile states of the $(3,3)$ ISS on lepton mixing}
\mathversion{normal}
\label{sec4}

As already mentioned in section~\ref{sec22}, there are two possible effects that can have an impact on lepton mixing: the inclusion of the subleading
contribution $m_\nu^1$ to the light neutrino mass matrix $m_\nu$ and effects of non-unitarity of $\widetilde{U}_\nu$, which are encoded in  $\eta_{\alpha\beta}$.
A numerical analysis of examples for each case, Case 1) through Case 3 b.1), can be found in
section~\ref{sec5} and confirms the analytical results, which we proceed to discuss.

\subsection{Subleading contribution to the light neutrino mass matrix}
\label{sec41}

When plugging in the form of the matrices
$m_D$, $M_{NS}$ and $\mu_S$ for option 1, see eqs.~(\ref{eq:mDopt1},\ref{eq:MNSopt1},\ref{eq:muS},\ref{eq:USopt1}), the subleading contribution to the light neutrino mass matrix, shown in eq.~(\ref{eq:mnuNLO}), takes a simple form:
\begin{equation}
\label{eq:mnuNLOopt1}
m_\nu^1 = - \frac{y_0^4 \, v^4}{4 \, M_0^4}  \, \mu_S = - \frac{y_0^4 \, v^4}{4 \, M_0^4}  \,  \, U_S^\star \, \left( \begin{array}{ccc}
\mu_1 & 0 & 0 \\
0 & \mu_2 & 0 \\
0 & 0 & \mu_3
\end{array}
\right) \, U_S^\dagger \; .
\end{equation}
Comparing with the leading order contribution $m_\nu$, found in eq.~(\ref{eq:mnuLOopt1}), we see that $m_\nu^1$ has exactly the same form in flavour space and is suppressed by
a factor $\frac{y_0^2 \, v^2}{2 \, M_0^2}$. Thus, this subleading contribution does not introduce any change in the lepton mixing parameters and only slightly corrects the values of the 
light neutrino masses, e.g.~for $y_0 \sim 1$ and $M_0 \sim 1000 \, \mathrm{GeV}$ the correction is around $0.03$ with respect to the leading order result, see eq.~(\ref{eq:m123LOopt1}).
Such a correction can be compensated by re-adjusting the values of the parameters $\mu_i$.

\mathversion{bold}
\subsection{Effects of non-unitarity of $\widetilde{U}_\nu$}
\label{sec42}
\mathversion{normal}
The deviation from unitarity of $\widetilde{U}_\nu$ is encoded in $\eta$, see eqs.~(\ref{eq:UPMNStildeeta},\ref{eq:eta}). For option 1, the form of $\eta$ turns out to be flavour-diagonal and
flavour-universal, since both $m_D$ and $M_{NS}$ have this property, see eqs.~(\ref{eq:mDopt1},\ref{eq:MNSopt1})
\begin{equation}
\label{eq:etaopt1}
\eta = \frac{y_0^2 \, v^2}{4 \, M_0^2} \, \mathbb{1} \equiv \eta_0 \, \mathbb{1} \; .
\end{equation}
Furthermore, it is independent of the particular case, Case 1) through Case 3 b.1), which we confirm numerically.

For $y_0 \sim 1$ and $M_0 \sim 1000 \, \mathrm{GeV}$ we have $\eta_0 \sim 0.015$, while for $y_0 \sim 0.1$ it is suppressed by further two orders of magnitude (at constant $M_0$).
The features of being flavour-diagonal and flavour-universal are numerically confirmed. 
The size of $\eta$ and its dependence on $y_0^2$ as well as $\frac{1}{M_0^2}$ are also very well fulfilled.
For further details and a comparison with experimental bounds on $\eta_{\alpha\beta}$
see section~\ref{sec50}.

Since $\eta$ is flavour-diagonal as well as flavour-universal and $\eta_0$ is positive, the presence of $\eta$ effectively leads to a suppression of all elements of $U_0 = U_S$, see eqs.~(\ref{eq:UPMNStildeeta},\ref{eq:UnuU0opt1}).
We can thus easily estimate the deviations expected in the results for the lepton mixing parameters (mixing angles and CP invariants/CP phases) extracting them in the same
way as for the unitary case, i.e.~the $(3,3)$ ISS framework at leading order and the model-independent scenario (MIS).\footnote{We extract the lepton mixing parameters using eqs.~(\ref{eq:sin2thij},\ref{eq:JCP},\ref{eq:I1I2}) in appendix~\ref{app2} (with $U_\text{PMNS}$ replaced by $\widetilde{U}_\nu$).} We consider relative deviations between the non-unitary results, $(\sin^2\theta_{ij})_{\mathrm{ISS}}$, $(J_{\mathrm{CP}})_{\mathrm{ISS}}$ and 
$(I_i)_{\mathrm{ISS}}$, and the unitary ones, $(\sin^2\theta_{ij})_{\mathrm{MIS}}$, $(J_{\mathrm{CP}})_{\mathrm{MIS}}$ and $(I_i)_{\mathrm{MIS}}$,\footnote{When considering these relative deviations, we always assume that
$(\sin^2\theta_{ij})_{\mathrm{MIS}}$, $(J_{\mathrm{CP}})_{\mathrm{MIS}}$ and $(I_i)_{\mathrm{MIS}}$ do not vanish.}
\begin{equation}
\Delta \sin^2\theta_{ij}=\frac{(\sin^2\theta_{ij})_{\mathrm{ISS}} - (\sin^2\theta_{ij})_{\mathrm{MIS}}}{(\sin^2\theta_{ij})_{\mathrm{MIS}}} \; , \;\;
\Delta J_{\mathrm{CP}}=\frac{(J_{\mathrm{CP}})_{\mathrm{ISS}}-(J_{\mathrm{CP}})_{\mathrm{MIS}}}{(J_{\mathrm{CP}})_{\mathrm{MIS}}} \;\; \mbox{and} \;\;
\Delta I_i=\frac{(I_i)_{\mathrm{ISS}}-(I_i)_{\mathrm{MIS}}}{(I_i)_{\mathrm{MIS}}} 
\end{equation}
and alike for the sines of the CP phases $\delta$, $\alpha$ and $\beta$. In doing so, we can find formulae for the relative deviations that are valid for all cases, Case 1) through Case 3 b.1).
The exact numerical values of these deviations can in general (slightly) depend on the chosen case and other parameters, such as the index of $G_f$, the choice of the residual $Z_2$ symmetry in the sector of the neutral states, and the value of the free angle 
$\theta_S$. We comment on this in the numerical analysis to be carried in section~\ref{sec5}.

For $\Delta \sin^2 \theta_{ij}$ we have
\begin{equation}
\label{eq:Deltasin2thetaij_gen}
\!\!\!\!\!\!\Delta \sin^2 \theta_{13} \approx -2 \, \eta_0 \; , \;\; \Delta \sin^2 \theta_{12} \approx - \frac{2 \, \eta_0}{1-|U_{e3}|^2} \approx -2.04 \, \eta_0 \; , \;\; \Delta \sin^2 \theta_{23} \approx - \frac{2 \, \eta_0}{1-|U_{e3}|^2} \approx -2.04 \, \eta_0
\end{equation}
for $|U_{e3}|^2 \approx 0.022$~\cite{NuFIT50}. For the CP invariants $J_{\mathrm{CP}}$, $I_1$ and $I_2$ we find
\begin{equation}
\label{eq:DeltaJCPI1I2_gen}
\Delta J_{\mathrm{CP}} \approx - 4 \, \eta_0 \; , \;\; \Delta I_1 \approx  - 4 \, \eta_0 \; , \;\; \Delta I_2 \approx  - 4 \, \eta_0 \; .
\end{equation}
With this information we can also extract $\Delta \sin\delta$, $\Delta \sin\alpha$ and $\Delta \sin\beta$ and arrive at
\begin{equation}
\label{eq:DeltasinCPphases_gen}
\Delta\sin\delta \approx -2.82 \, \eta_0 \; , \;\; \Delta\sin\alpha \approx -2.95 \, \eta_0 \; , \;\; \Delta\sin\beta \approx -2.95 \, \eta_0
\end{equation}
for $|U_{e2}|^2 \approx 0.30$, $|U_{e3}|^2 \approx 0.022$ and $|U_{\mu3}|^2 \approx 0.56$~\cite{NuFIT50}.
For $y_0 \sim 1$ and $M_0 \sim 1000 \, \mathrm{GeV}$ we expect 
\begin{equation}
\label{eq:estimateDsy01}
\Delta \sin^2 \theta_{ij} \approx -0.03 \; , \;\; \Delta J_{\mathrm{CP}} \approx \Delta I_i \approx -0.06 \; , \;\; \Delta\sin\delta\approx -0.042, \; \Delta\sin\alpha\approx \Delta\sin\beta \approx -0.044  \; .
\end{equation}
Due to the suppression of all elements of $U_0 = U_S$, all relative deviations are expected to be negative. Furthermore, their size slightly depends on 
the considered quantity and is generally not expected to exceed values of a few percent.
These estimates are confirmed numerically, as we discuss in section~\ref{sec5}.
It is important to note that certain features, like the vanishing of the sine
and the periodicity of some of the CP phases in terms of the group theory parameters, remain preserved exactly, since the flavour structure of 
the light neutrino mass matrix is not changed and the deviation from unitarity only amounts to a common rescaling of all elements of the PMNS mixing matrix.

\vspace{0.11in}
\noindent Furthermore, we can estimate the deviations in the (approximate) sum rules induced by effects of non-unitarity of the lepton mixing matrix, such as the ones in eq.~(\ref{eq:sumrulesCase1}). 
These are discussed in turn for each of the cases, Case~1) through Case 3 b.1).

\subsection*{Case 1)}

Two approximate sum rules have been found for Case 1), see eq.~(\ref{eq:sumrulesCase1}). The effects of non-unitarity of the lepton mixing matrix on these are expected to be as follows: for the first sum rule,
relating the solar and the reactor mixing angle, using the best-fit value $|U_{e 3}|^2\approx 0.022$~\cite{NuFIT50}, we have
\begin{equation}
\label{eq:sumrule1Case1nonuni}
\Delta \Sigma_1 \approx -2 \, \eta_0 \, \left( \frac{1+|U_{e3}|^4}{1-|U_{e3}|^4} \right) \approx -2 \, \eta_0
\end{equation}
with $\Delta\Sigma_1$ corresponding to the relative deviation of the non-unitary result from the unitary one and defined as

\begin{equation}
\Delta\Sigma_1 = \frac{\left( \frac{3 \, \sin^2 \theta_{12}}{1 + \sin^2\theta_{13}} \right)_{\mathrm{ISS}} - \left( \frac{3 \, \sin^2 \theta_{12}}{1 + \sin^2\theta_{13}} \right)_{\mathrm{MIS}}}{\left( \frac{3 \, \sin^2 \theta_{12}}{1 + \sin^2\theta_{13}} \right)_{\mathrm{MIS}}} \;\; \mbox{with} \;\; \left( \frac{3 \, \sin^2 \theta_{12}}{1 + \sin^2\theta_{13}} \right)_{\mathrm{MIS}} \approx 1 \;\; \mbox{from eq.~(\ref{eq:sumrulesCase1}),}
\end{equation}
while the deviation for the second sum rule, the one involving the atmospheric and the reactor mixing angle, is of the form
\begin{equation}
\label{eq:sumrule2Case1nonuni}
\Delta\Sigma_2 \approx - \sqrt{2} \, \eta_0 \, \left( \frac{\sqrt{2} \pm |U_{e3}|\, (1+|U_{e3}|^2)}{(1\pm \sqrt{2} \, |U_{e3}|)\, (1-|U_{e3}|^2)} \right) \approx -1.87 \, (-2.31) \, \eta_0
\end{equation}
for $+ (-)$.
$\Delta\Sigma_2$ is defined analogously to $\Delta\Sigma_1$ with the help of the second approximate sum rule in eq.~(\ref{eq:sumrulesCase1}).
The different signs refer to the different signs in the sum rule.
We note that none of the relative deviations, $\Delta\Sigma_1$ and $\Delta\Sigma_2$, depends on the parameters $n$, $s$ 
or on the precise value of the free angle $\theta_S$, up to the sign in $\Delta\Sigma_2$. 
For $y_0 \sim 1$ and $M_0 \sim 1000$~GeV we expect these to be 
\begin{equation}
\label{eq:estimateDSigma12y01}
\Delta\Sigma_1 \approx -0.03 \;\; \mbox{and} \;\; \Delta\Sigma_2 \approx -0.028 \, (-0.035) 
\end{equation}
for $+ (-)$ from the expression for $\Delta\Sigma_2$ in eq.~(\ref{eq:sumrule2Case1nonuni}).

\subsection*{Case 2)}

For Case 2) we also have two approximate sum rules: one which coincides with the first sum rule of Case 1) and another one, relating the atmospheric and the reactor mixing angles, shown in eq.~(\ref{eq:sumruleCase2}).
The effects of non-unitarity (of the PMNS mixing matrix) on the latter one are estimated to be of the order of 
\begin{equation}
\label{eq:sumruleCase2nonuni}
\Delta\Sigma_3 \approx - 2 \, \eta_0 \, \left( \frac{\sqrt{3}+\tan\phi_u}{\sqrt{3}\, (1-|U_{e3}|^2) + (1-3\, |U_{e3}|^2) \, \tan\phi_u} \right) \; ,
\end{equation}
where $\Delta\Sigma_3$ is defined in the analogous way as $\Delta\Sigma_1$. The form of $\Delta\Sigma_3$ can be simplified by remembering that $u/n$ is required to be small and thus we expand in $\phi_u=\frac{\pi \, u}{n}$ up to the linear order.
At the same time, we use the best-fit value for $|U_{e3}|^2 \approx 0.022$~\cite{NuFIT50} so that we have
\begin{equation}
\label{eq:sumruleCase2nonunisimp}
\Delta\Sigma_3 \approx  -2.05 \, \eta_0 \, (1+ 0.026 \, \phi_u) \; .
\end{equation}
This shows that there is only a very mild dependence of $\Delta\Sigma_3$ on $\phi_u = \frac{\pi\, u}{n}$. 
Furthermore, there is no explicit dependence of $\Delta\Sigma_3$ on the parameter $v$ and the free angle $\theta_S$.
Numerically we find for $y_0 \sim 1$ and $M_0 \sim 1000$~GeV that 
\begin{equation}
\label{eq:estimateDSigma3y01}
\Delta\Sigma_3 \approx -0.031 \; ,
\end{equation}
which is of a size very similar to the other relative deviations.

\subsection*{Case 3 a) and Case 3 b.1)}

For Case 3 a) the approximate sum rule, found in eq.~(\ref{eq:sumruleCase3a}), is actually identical to the second sum rule for Case 1), see eq.~(\ref{eq:sumrulesCase1}), taking into account the different signs in both of them.
We thus expect very similar results also for Case 3 a).

For Case 3 b.1) two approximate sum rules are derived for $m=\frac n2$, see eq.~(\ref{eq:sumrulesCase3b1}). For the first of these two, we find as relative deviation of the non-unitary result from the unitary one
\begin{equation}
\label{eq:sumrule1Case3b1nonuni}
\Delta\Sigma_4 \approx -2 \, \eta_0 \, \left( \frac{1-2 \, |U_{e3}|^4}{1-3 \, |U_{e3}|^2+2 \, |U_{e3}|^4} \right) \approx -2.14 \, \eta_0 \; ,
\end{equation}
while for the second one we have
\begin{equation}
\label{eq:sumrule2Case3b1nonuni}
\Delta\Sigma_5 \approx - 2 \, \eta_0 \, \left( \frac{\sqrt{3}+ \sqrt{2} \, \cos 3 \, \phi_s \, \sin 2 \, \theta_0}{\sqrt{3} \, (1-|U_{e3}|^2) + \sqrt{2} \, \cos 3 \, \phi_s \, \sin 2 \, \theta_0} \right) \approx -2.05 \, \eta_0 \mp 0.019 \, \eta_0 \, \cos 3 \, \phi_s \; ,
\end{equation}
where we have again used $|U_{e3}|^2 \approx 0.022$~\cite{NuFIT50} and $\theta_0 \approx \frac{\pi}{2} \pm \epsilon$ with $\epsilon \approx 0.26$, cf. text below eq.~(\ref{eq:sumrulesCase3b1}). We thus see that the relative deviation $\Delta\Sigma_5$ only weakly depends on the value
of the parameter $s$, related to the chosen CP transformation $X$. 
Furthermore, we infer that neither $\Delta\Sigma_4$ nor $\Delta\Sigma_5$ depends strongly on the parameter $n$ or the free angle $\theta_S$.
Using $y_0 \sim 1$ and $M_0 \sim 1000$~GeV, we have for the two relative deviations 
\begin{equation}
\label{eq:estimateDSigma45y01}
\Delta\Sigma_4 \approx -0.032 \;\; \mbox{and} \;\; \Delta\Sigma_5 \approx -0.031 \; .
\end{equation}

\section{Numerical analysis}
\label{sec5}

In this section, we study numerically the impact of the heavy sterile states of the $(3,3)$ ISS framework on the results for the lepton mixing parameters, and if available, on the approximate sum rules among these.
We do so for each of the different cases, Case 1) through Case 3 b.1), for some viable choices of the group theory parameters, e.g.~the index $n$ of the flavour symmetry $G_f$. 
We also compare these findings to the analytical estimates, presented in section~\ref{sec4}.

Before detailing results for the different cases in sections~\ref{sec51}-\ref{sec54}, we present constraints on the Dirac neutrino Yukawa coupling $y_0$, the mass scale $M_0$ of the heavy sterile states, as well as on the parameters $\mu_i$, emphasising the role of the bounds on the unitarity of the PMNS mixing matrix.

\subsection{Symmetry endowed (3,3) ISS: setup and unitarity constraints on \mathversion{bold}{$\widetilde{U}_\nu$}}
\label{sec50}

We begin by briefly discussing the constraints arising from the
violation of unitarity of the PMNS mixing matrix $\widetilde{U}_\nu$, 
as encoded in the  
matrix $\eta$.  
As can be seen from eq.~(\ref{eq:etaopt1}), $\eta$ is determined by 
the chosen regimes for $y_0$ and $M_0$, which 
characterise the impact of the heavy sterile states on the lepton
mixing parameters. Thus, the experimental limits on the quantities
$\eta_{\alpha\beta}$, $\alpha,\beta=e,\mu,\tau$, are at the source of
the most important constraints on the present (3,3) ISS framework.

Before discussing how the limits on $\eta_{\alpha\beta}$ crucially 
constrain $y_0$ and hence the combination of $y_0$ and $M_0$, let us
first emphasise two points: 
we have checked numerically that the form of the 
quantities $\eta_{\alpha\beta}$ does not depend on the specific case, 
Case 1) through Case 3 b.1), as expected from the analytical estimate
in eq.~(\ref{eq:etaopt1}); 
furthermore, we also confirm numerically that the matrix 
$\eta$ is flavour-diagonal and flavour-universal, and that
$\eta_0$ is proportional to $y_0^2$ (and inversely proportional to
$M_0^2$), as can be also seen from eq.~(\ref{eq:etaopt1}). 
The (indirect) experimental constraints on $\eta_{\alpha\beta}$ are taken
from~\cite{etaconstraints} and  
are given by\footnote{
We use the bounds obtained in~\cite{etaconstraints}, although
the form of $\eta$ is flavour-diagonal and flavour-universal in the case at hand.}
\begin{equation}
\label{eq:etaexp}
\left|\eta_{\alpha\beta}\right| \leq \left(
\begin{array}{ccc}
1.3 \times 10^{-3} & 1.2 \times 10^{-5} & 1.4 \times 10^{-3}\\
1.2 \times 10^{-5} & 2.2 \times 10^{-4} & 6.0 \times 10^{-4}\\
1.4 \times 10^{-3} & 6.0 \times 10^{-4} & 2.8 \times 10^{-3}
\end{array}
\right) \;\; \mbox{at the $1\,\sigma$ level.}
\end{equation}
As can be verified, the diagonal element subject to the
strongest experimental bounds is $\eta_{\mu\mu}$, 
$\left|\eta_{\mu\mu}\right| \leq 2.2 \, (4.4) \, [6.6] \times 10^{-4}$ at
the $1 (2) [3]\, \sigma$ level. 
We thus use this limit in the subsequent analysis.

\begin{figure}[t!]
\begin{center}
\parbox{3in}{\hspace{-0.3in}\includegraphics*[scale=0.55]{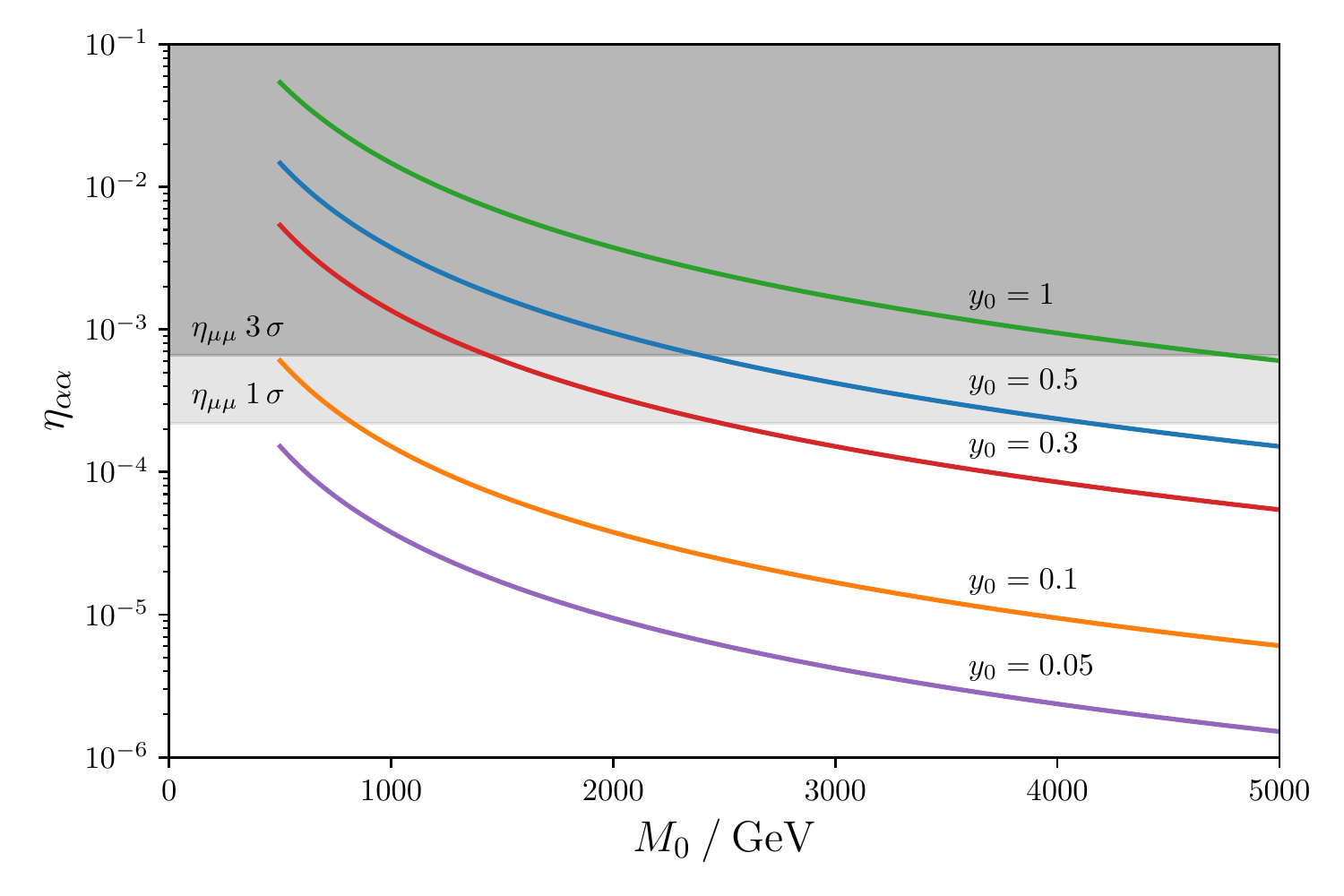}} 
\hspace{-0.1in}
\parbox{3in}{\includegraphics*[scale=0.55]{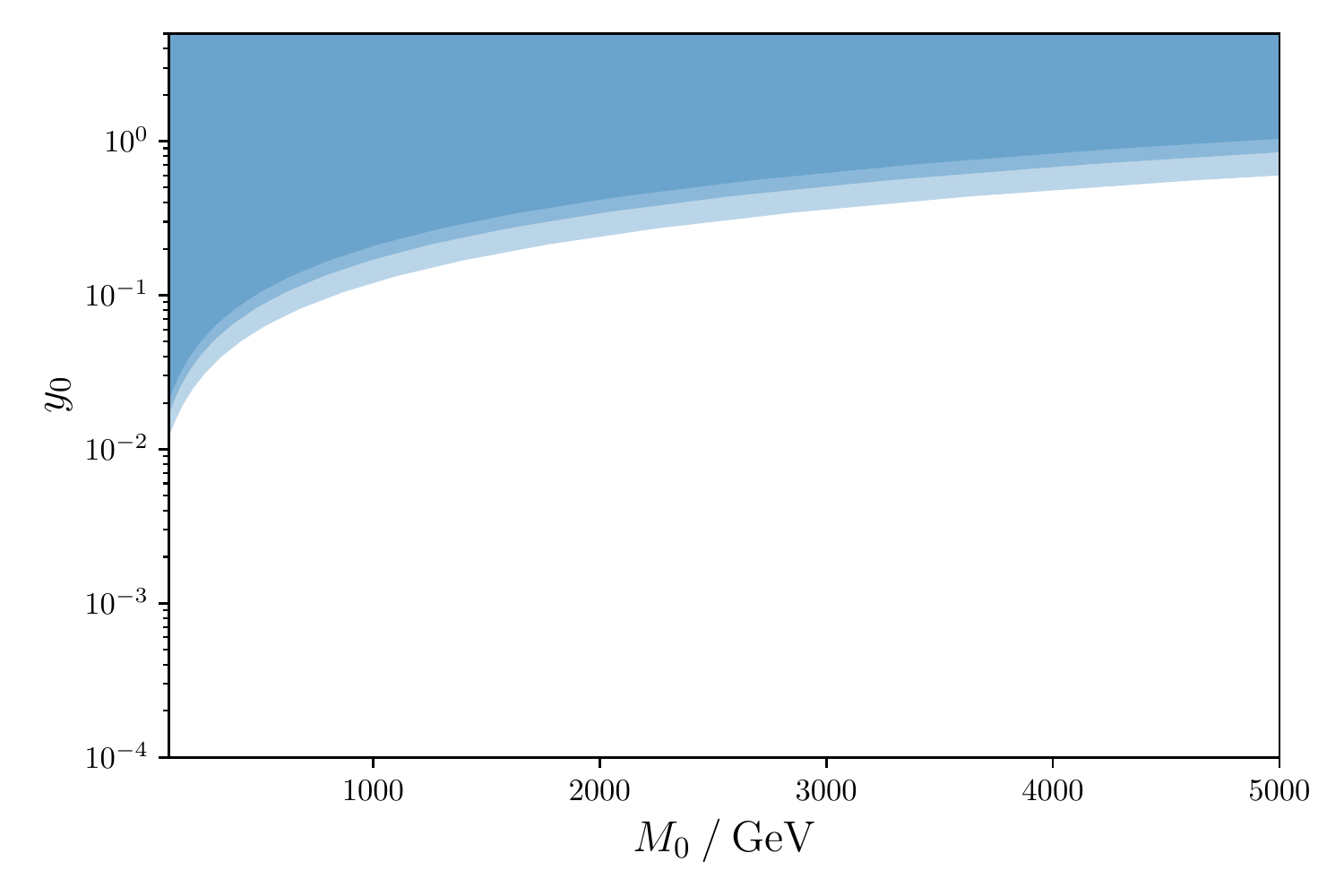}} 
\end{center}
\caption{{\small {\bf Constraints from unitarity of 
\mathversion{bold}{$\widetilde{U}_\nu$}}. {\bf Left plot}:
    $\eta_{\alpha\alpha}$  with respect to the mass scale $M_0$ (in GeV) 
for different values of the Yukawa coupling: $y_0=0.05$ (purple line), $y_0=0.1$ (orange line), $y_0=0.3$ (red line), $y_0=0.5$ (blue line)  
and $y_0=1$ (green line). The grey-shaded regions denote the
areas excluded by the strongest constraint on the flavour-diagonal
entries of $\eta$ (arising from $\eta_{\mu\mu}$)
at $1 \, \sigma$ level (light grey)~\cite{etaconstraints} 
and $3 \, \sigma$ level (dark grey). 
{\bf Right plot}: Disfavoured regions of the ($M_0-y_0$) plane, with $M_0$
given in GeV, due to 
conflict with experimental bounds on $\eta_{\alpha\alpha}$, at $1 \, \sigma$, $2 \, \sigma$ and 
$3 \, \sigma$ (respectively denoted by light, medium and dark blue). 
\label{fig:Constraints_eta}}}
\end{figure}

The maximal size of the Yukawa coupling $y_0$, compatible with the
experimental constraints on $\eta_{\alpha\beta}$, can be read from
the left plot in fig.~\ref{fig:Constraints_eta}: for
$y_0$ as small as $y_0=0.1$, the unitarity constraints can be evaded 
for values of $M_0$ as low as $M_0 \gtrsim 500$~GeV (at the $3 \,
\sigma$ level); larger values,  
$y_0=0.5$, already require $M_0 \gtrsim 2400$~GeV, and for 
$y_0=1$ one must have $M_0 \gtrsim 4800$~GeV in order to be in
agreement with the bounds of eq.~(\ref{eq:etaexp}) at the $3 \, \sigma$ level, i.e. $|\eta_{\mu\mu}|\lesssim 6.6\times10^{-4}$. 
This is illustrated in the right panel of
fig.~\ref{fig:Constraints_eta} by an exclusion plot in the
($M_0-y_0$) plane. The subsequent numerical analyses will rely on 
regimes of $y_0$ and $M_0$ compatible with experimental data at the $3\, \sigma$ 
level,\footnote{As will be discussed in detail 
in section~\ref{sec:clfv}, the predictions for
cLFV observables will not lead to any additional
constraints on the parameter space of the $(3,3)$ ISS framework 
with $G_f$ and CP in the case of option 1.
Thus, the only relevant constraints are those arising from the effects
of non-unitarity of $\widetilde{U}_\nu$.} and regimes in conflict with experimental bounds on $\eta_{\alpha\beta}$ will be clearly indicated in the discussion.
  
In view of the above, 
we will in general
assume that the mass scale $M_0$ varies in the range 
\begin{equation}
\label{eq:M0range}
500 \, \mathrm{GeV} \lesssim M_0 \lesssim 5000 \, \mathrm{GeV} \; .
\end{equation}
Although mostly lying beyond future collider reach~\cite{Antusch:2016ejd}, 
the chosen range for $M_0$ (and thus for $M_{NS}$ and the heavy mass 
spectrum) is motivated by its 
phenomenological interest, as it is in general associated 
with extensive observational imprints, being thus indirectly
accessible in numerous dedicated facilities~\cite{Abada:2014vea,Arganda:2014dta,Abada:2014kba,Abada:2014cca,Arganda:2015naa,Abada:2015oba,DeRomeri:2016gum}.  

Concerning the Yukawa coupling $y_0$, and following the results
displayed in fig.~\ref{fig:Constraints_eta},
we will in general illustrate our results for 
two different values of the Yukawa coupling $y_0$,
\begin{equation} 
\label{eq:y0choice}
y_0= 0.5 \;\; \mbox{and} \;\; y_0=0.1 \; .
\end{equation}
Nevertheless, we will exceptionally consider larger values of 
the Yukawa coupling $y_0=1$, in order to better illustrate 
the effects of the deviations from unitarity of the PMNS mixing matrix. 
These cases will be clearly 
identified in the discussion; 
unless otherwise stated,
disfavoured regimes associated with bounds on $\eta_{\alpha\beta}$ will be indicated by a grey-shaded
area in the corresponding plots.

Finally, we consider the free parameters $\mu_i$.
As can be seen from eq.~(\ref{eq:m123LOopt1}), in the case of option 1,
$\mu_i$ are directly proportional to the light neutrino masses
$m_i$. Thus, they are experimentally constrained 
by the measured mass squared differences and by the bound on the sum
of the light neutrino masses coming from cosmology. The 
latest experimental data are collected in
appendix~\ref{app3}. 
We notice that in our numerical study, the two mass squared differences are always adjusted to their experimental best-fit value~\cite{NuFIT50}.

A few comments are still in order concerning the light neutrino mass 
spectrum - the value of the lightest neutrino mass $m_0$, and the ordering
(NO vs. IO).  
Regarding $m_0$,
we have verified that the results 
for the lepton mixing parameters are always independent of its
choice. 
Throughout this section, we have thus fixed its value to
\begin{equation}
\label{eq:m0fixed}
m_0 = 0.001 \, \mathrm{eV}.
\end{equation}
Furthermore, we note that we have performed the numerical analysis for
both NO and IO, 
and no (numerically significant) differences were found, neither for the relative deviations of the lepton mixing parameters, nor for the (approximate) sum rules.
Accordingly, all the
results of this section will be only illustrated for the case of a NO 
light neutrino mass spectrum. 
However, notice that upon discussion of the prospects of the current framework 
concerning $0\nu\beta\beta$ decay in section~\ref{sec6}, 
we will consider both orderings of the mass spectrum, and also vary $m_0$.

\bigskip
Leading to the fits presented in the following subsections, 
we only consider experimental constraints on the lepton mixing angles 
and the two mass squared differences, but not on the CP phase
$\delta$, since the latter 
is only very mildly experimentally constrained (a summary of the
relevant neutrino oscillation data is given in appendix~\ref{app3}). 
Additional information on the numerical fit procedure can be found in
appendix~\ref{APPFIT}.  

\subsection{Case 1)}
\label{sec51}

In order to scrutinise the effects of the $(3,3)$ ISS framework and its heavy states on the lepton mixing parameters,
we choose a value of the index $n$ that allows studying several different values of the parameter $s$ (and thus CP transformations $X$) for Case 1). In this way, 
the behaviour of the Majorana phase $\alpha$, see eq.~(\ref{eq:sinalphaCase1}), can be studied systematically. Concretely, in the following we use
\begin{equation}
\label{eq:numchoiceparaCase1}
n=26 \;\; \mbox{and} \;\; 0 \leq s \leq 25 \,.
\end{equation}

\begin{figure}[t!]
\begin{center}
\parbox{3in}{\includegraphics*[scale=0.5]{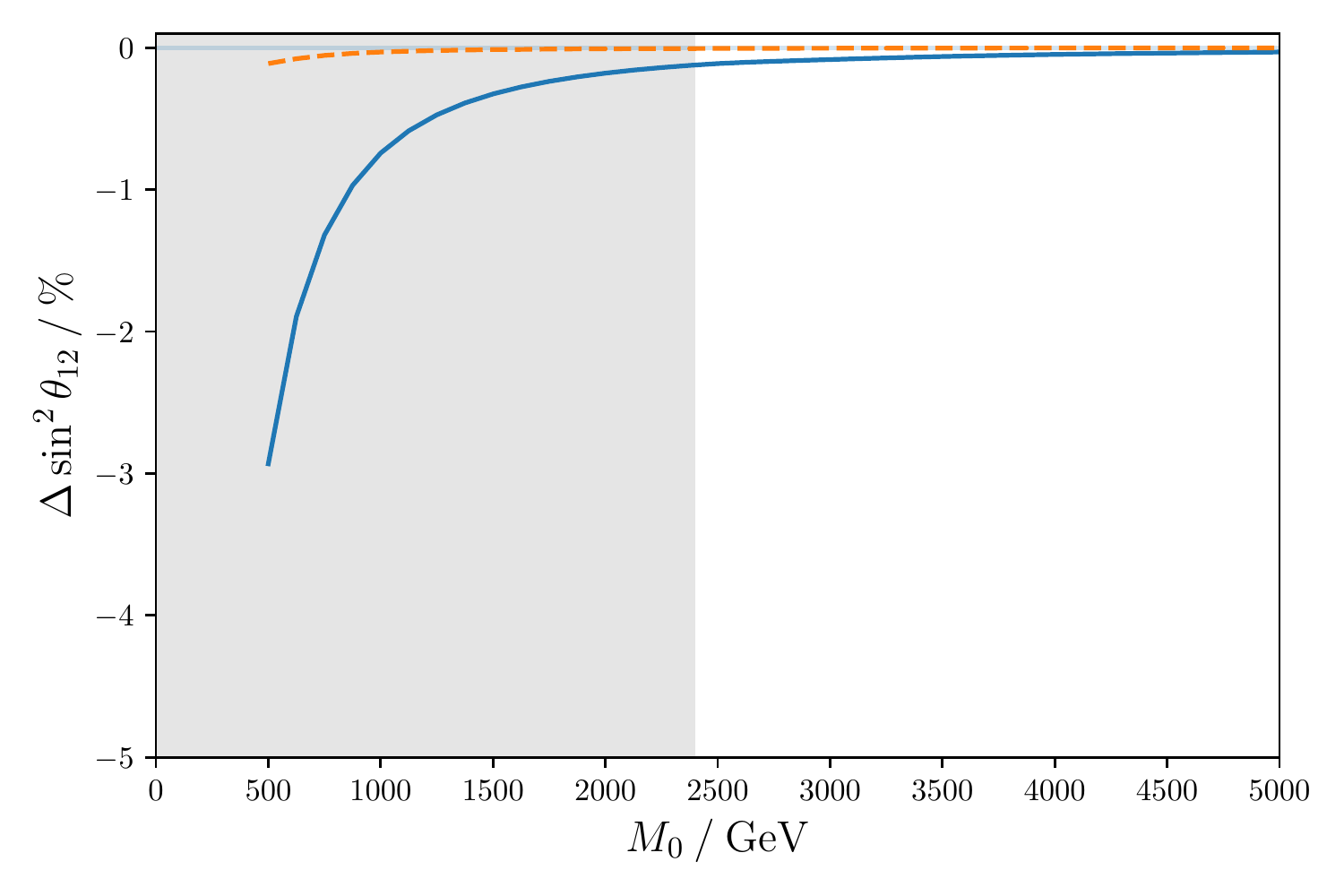}}
\hspace{0.2in}
\parbox{3in}{\includegraphics*[scale=0.5]{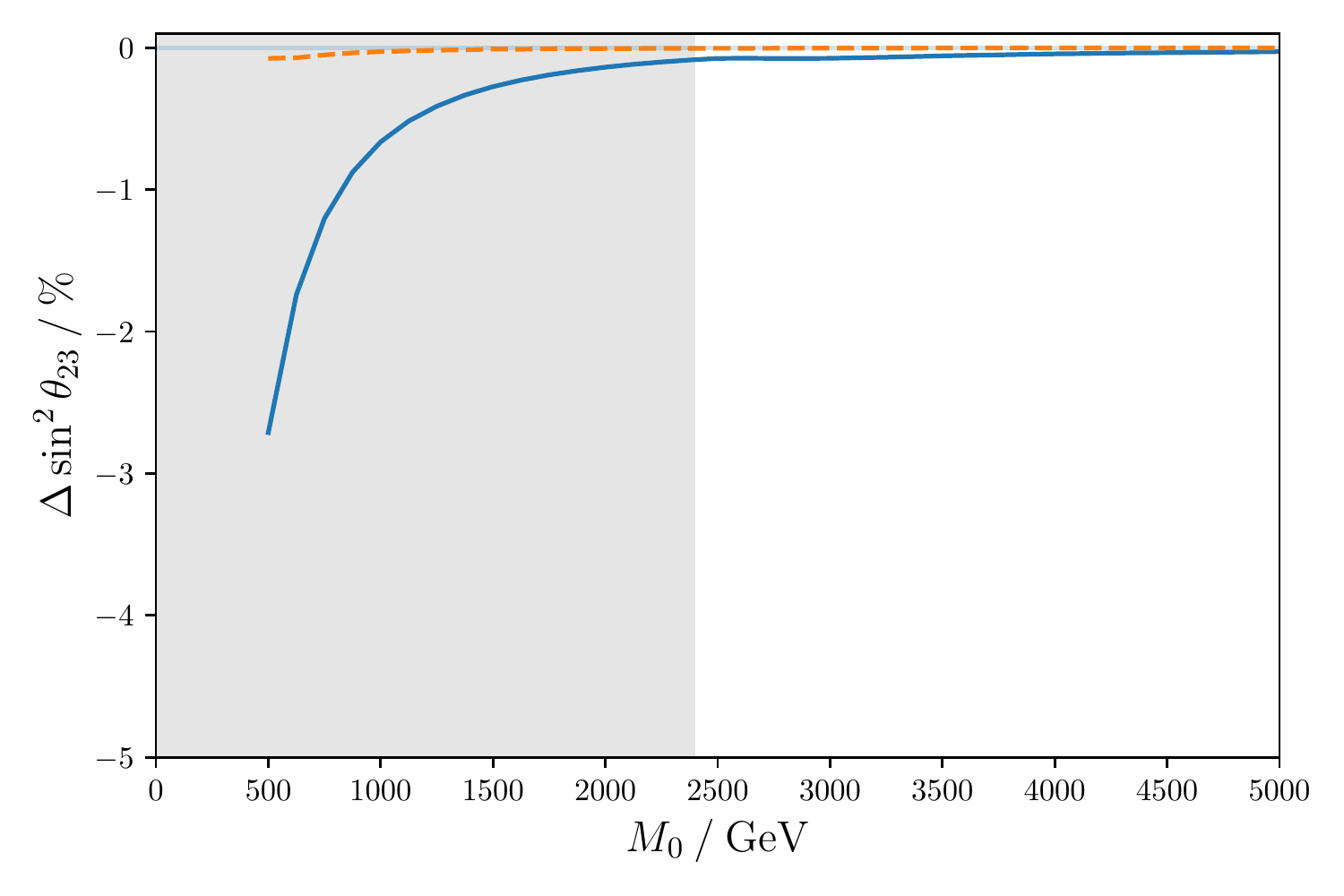}}
\end{center}
 \caption{{\small {\bf Case 1)} 
 Relative deviation of $\sin^2\theta_{12}$ (left) and
$\sin^2\theta_{23}$ (right) as obtained for option 1 of the $(3,3)$ ISS
from the corresponding values derived in the 
model-independent scenario, as a function of 
$M_0$ (in GeV).  
For concreteness, we have fixed $s=1$ and $n=26$.
The curves are associated with distinct values of the Yukawa coupling $y_0$:
the orange (dashed) curve corresponds to $y_0=0.1$ and the blue
(solid) one to $y_0=0.5$. A grey-shaded area denotes regimes
disfavoured due to conflict with experimental bounds (see detailed discussion in section~\ref{sec50}). 
\label{fig:Case1_s2thij}}}
\end{figure}

Based on the results obtained in the model-independent scenario (see section~\ref{sec31}), and the analytical estimates of the effects due to the heavy sterile states
of the $(3,3)$ ISS framework carried in section~\ref{sec4}, only the CP phase $\alpha$ is expected to show a dependence on the parameters $n$ and $s$ (through the ratio $s/n$). 
This is confirmed by our numerical analysis. 
Without loss of generality we thus set  $s=1$ to study the relative deviations $\Delta\sin^2\theta_{12}$ and $\Delta\sin^2\theta_{23}$. 
These are shown in fig.~\ref{fig:Case1_s2thij}, respectively in the left and right plots, as a function of $M_0$, which determines the scale of the heavy mass spectrum. 
We notice that their sign and size is consistent with the estimate found in 
eq.~(\ref{eq:estimateDsy01}).\footnote{Notice that following eq.~(\ref{eq:etaopt1}), $y_0 \sim 1$ and $M_0 \sim 1000$~GeV lead to the same result for the quantity $\eta_0$ as $y_0 \sim 0.5$ and $M_0 \sim 500$~GeV.}
The relative deviation of the reactor mixing angle, $\Delta\sin^2\theta_{13}$, is not shown and does not fulfil the expectations from the analytical estimate, since it turns out to be positive and always below
$0.5\%$ for values of $y_0 \lesssim 0.5$ and $M_0 \gtrsim 500$~GeV. 
This is a consequence of having $\theta_{13}$ driving the fit to determine $\theta_S$, due to its associated experimental precision, see appendix~\ref{app3}.
Consequently, we find for $\theta_S$ values around $0.19$, which are slightly larger than those obtained in the model-independent scenario, see eq.~(\ref{eq:thetaCase1}).
We note that in the plots shown here, we always have $\theta_S < \pi/2$, since this leads to a much better agreement with the experimentally preferred value of the atmospheric mixing angle: $\sin^2 \theta_{23} \approx 0.604$ to be compared to the experimental values 
$\sin^2 \theta_{23} = 0.570^{+0.018} _{-0.024}$ for light neutrinos with NO and $\sin^2 \theta_{23} = 0.575^{+0.017} _{-0.021}$ for light neutrinos with IO~\cite{NuFIT50}.
\begin{figure}[t!]
\begin{center}
\parbox{3in}{\includegraphics*[scale=0.5]{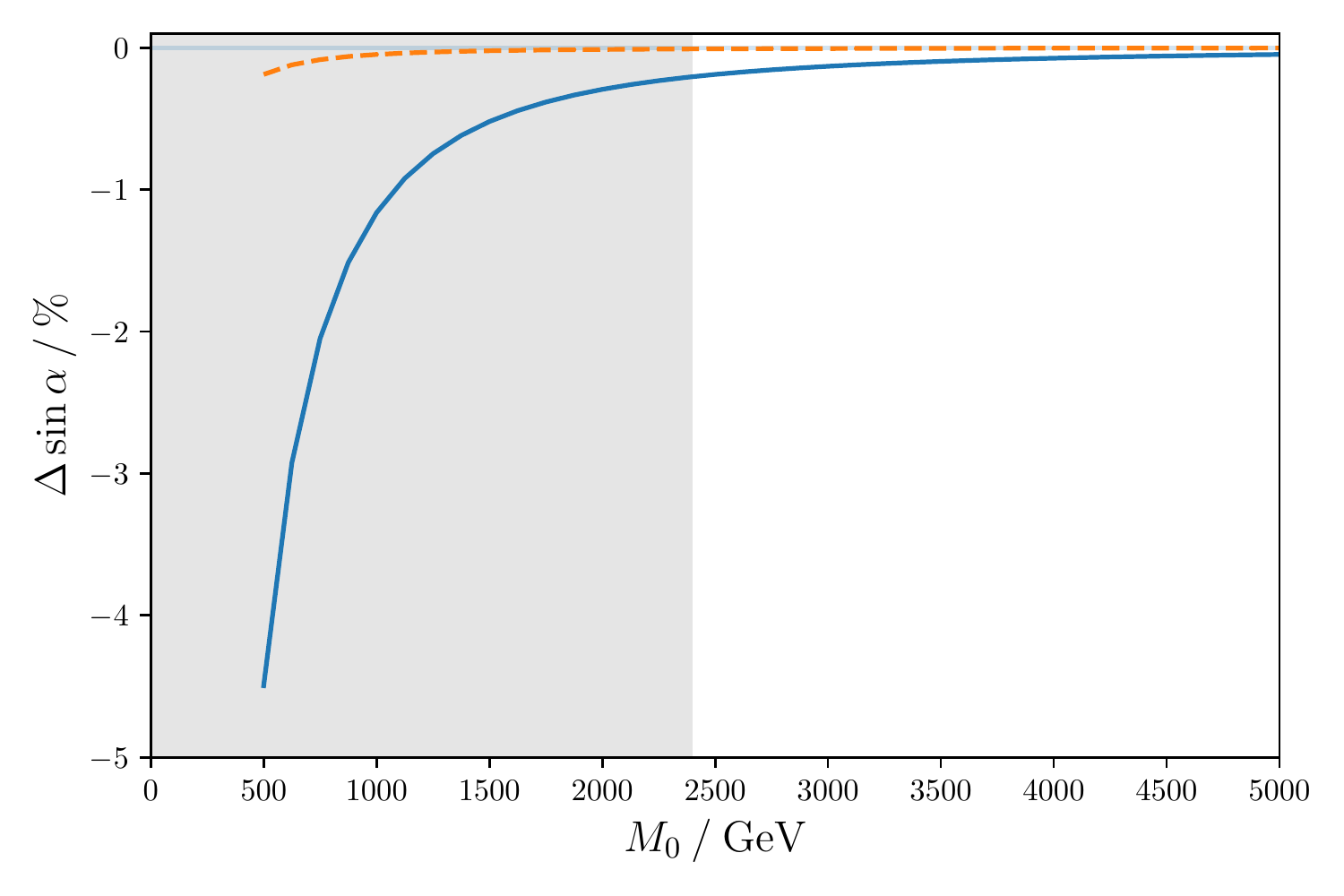}}
\hspace{0.2in}
\parbox{3in}{\includegraphics*[scale=0.5]{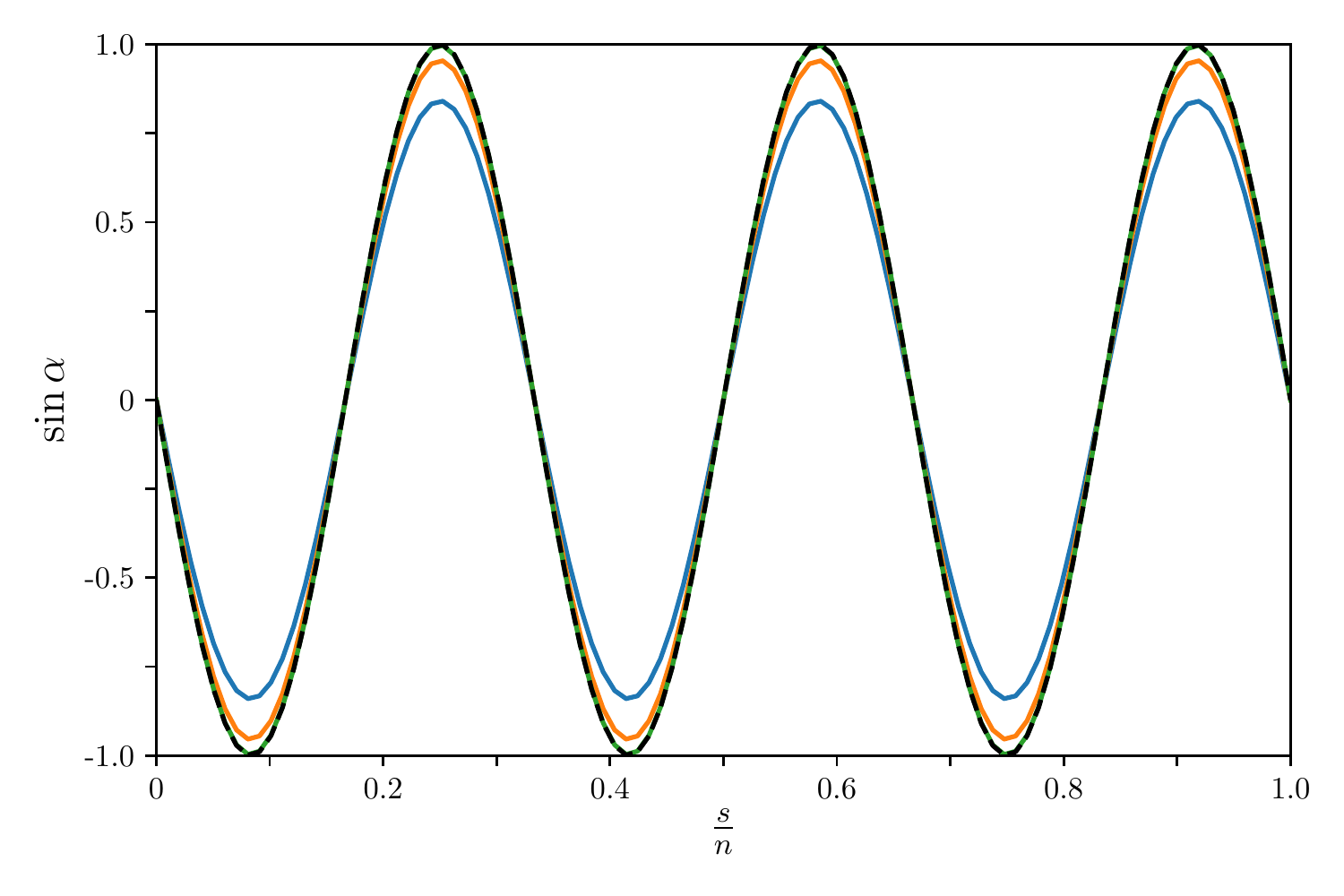}}
\end{center}
\caption{\small {{\bf Case 1)} {\bf Left plot}: 
Relative deviation of $\sin\alpha$ as obtained for option 1 of the $(3,3)$ ISS framework from the corresponding model-independent
prediction, with respect to $M_0$ (in GeV).
Line and colour code as in fig.~\ref{fig:Case1_s2thij}. 
{\bf   Right plot}:  
$\sin\alpha$ with respect to $s/n$ (fixing $n=26$ and continuously varying $0\leq  s < 26$). 
The black (dashed) curve
displays the result for $\sin\alpha$ obtained in the model-independent
scenario, see eq.~(\ref{eq:sinalphaCase1}).
The coloured (solid) curves refer to distinct values of $M_0$:
blue for $M_0=500$~GeV, orange for $M_0=1000$~GeV and green for $M_0=5000$~GeV.  
We have chosen $y_0=1$ in order to better display the
deviation from the model-independent scenario (notice that such a
value of $y_0$ requires 
$M_0 \gtrsim 4800$~GeV to comply with the experimental bounds on $\eta_{\alpha\alpha}$
at the $3\, \sigma$ level, cf. section~\ref{sec50}).
\label{fig:Case1_salpha}}}
\end{figure}

Moving on to the relative deviation of the Majorana phase $\alpha$, we note that also in this case
the size, sign and behaviour of the relative deviation $\Delta\sin\alpha$ (depending on $y_0$ and $M_0$) does not depend on the actual choice of the parameter $s$.
Thus, we have again taken $s=1$. 
In the left plot in fig.~\ref{fig:Case1_salpha}, we present the relative deviation of $\sin\alpha$ as obtained for option 1 of the $(3,3)$ ISS framework from the corresponding model-independent
prediction, with respect to $M_0$ (in GeV).
Comparing the maximal size of the relative deviation of $\sin\alpha$ ($\Delta\sin\alpha$) with the ones of the solar and the atmospheric mixing angles, $\Delta\sin^2\theta_{12}$
and $\Delta\sin^2\theta_{23}$, previously displayed in fig.~\ref{fig:Case1_s2thij}, we confirm that the latter are slightly smaller than the former, as expected from the analytical estimate in eq.~(\ref{eq:estimateDsy01}). 
The right plot in fig.~\ref{fig:Case1_salpha} illustrates the suppression of the value of $\sin\alpha$ depending on $s/n$
for three different values of $M_0$, 
$M_0=500$~GeV, $1000$~GeV and $5000$~GeV, and these are compared to the result expected in the model-independent scenario, see eq.~(\ref{eq:sinalphaCase1}).
We have chosen here $y_0=1$ in order to enhance the visibility 
of the deviations between the model-independent scenario and the (3,3) ISS presented in this plot, although such a large value of the Yukawa coupling requires $M_0$ to be at least $M_0 \gtrsim 4800$~GeV in order to comply with the experimental bounds on
the quantities $\eta_{\alpha\beta}$, see section~\ref{sec50}. Beyond this suppression of the value of $\sin\alpha$, we note that the periodicity in $s/n$ is still the same, independently of the effects of non-unitarity of $\widetilde{U}_\nu$, confirming the analytical estimates of section~\ref{sec42}. We have also numerically verified the analytical expectation that the Dirac phase $\delta$ as well as the Majorana phase $\beta$
remain trivial, i.e.~$\sin\delta=0$ and $\sin\beta=0$.

\begin{figure}[t!]
\begin{center}
\parbox{3in}{\includegraphics*[scale=0.5]{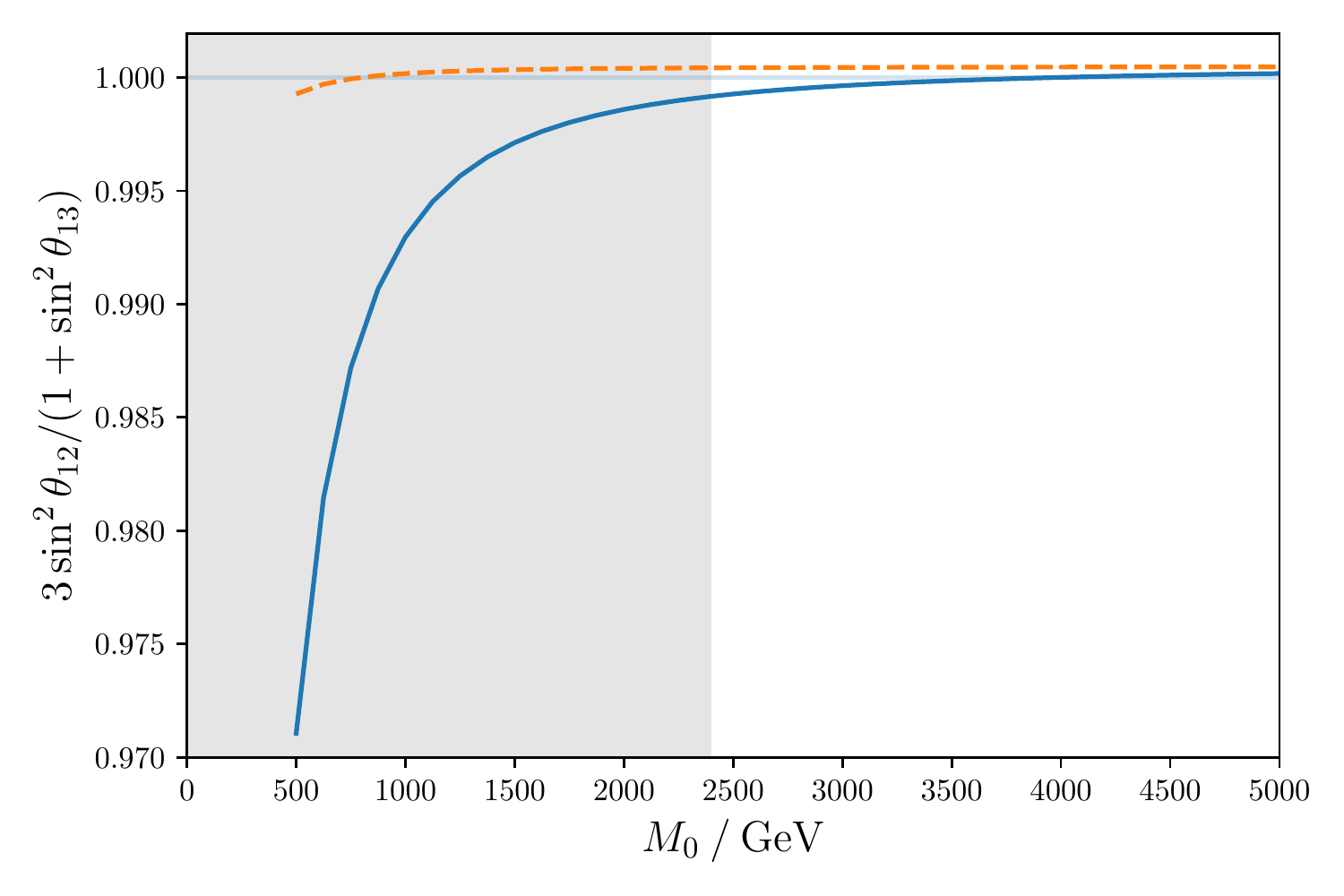}}
\hspace{0.2in}
\parbox{3in}{\includegraphics*[scale=0.5]{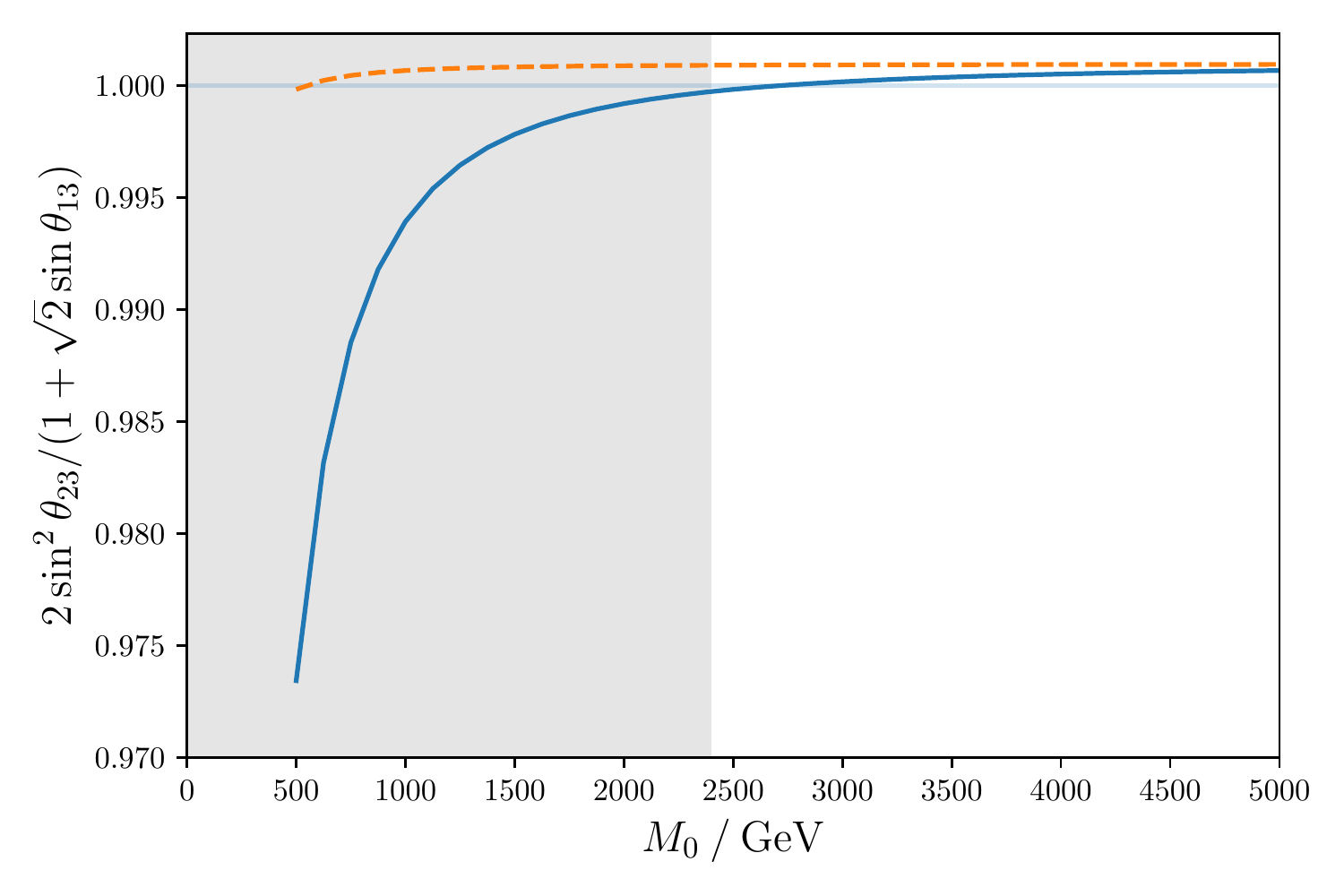}}
\end{center}
\caption{{\small {\bf Case 1)} 
Validity check of approximate sum rules
for option 1 of the $(3,3)$ ISS framework 
with respect to the mass $M_0$ (in GeV).
 Line and colour code as in fig.~\ref{fig:Case1_s2thij}.
 We note that for the second sum rule (right plot) we focus on the approximate sum rule with a plus sign, since we present results for $\theta_S < \pi/2$, see eq.~(\ref{eq:sumrulesCase1})
and below. 
\label{fig:Case1_Sigma12}}}
\end{figure}

Finally, we address the validity of the two approximate sum rules, see eq.~(\ref{eq:sumrulesCase1}). As can be seen from the plots in fig.~\ref{fig:Case1_Sigma12},
 deviations do not exceed the level of $-3\%$, in agreement with the analytical estimate. Furthermore, we numerically confirm that the maximally achieved relative deviation is slightly
larger for the first sum rule than for the second, for $\theta_S < \pi/2$. 
We also note that for large values of $M_0$, where effects of the non-unitarity of $\widetilde{U}_\nu$ should be suppressed, both ratios related to the two different sum rules become slightly larger than one. This is consistent with the fact that
these sum rules only hold approximately.

\subsection{Case 2)}
\label{sec52}

  %
 \begin{figure}[t!]
\begin{center}
\parbox{3in}{\includegraphics*[scale=0.5]{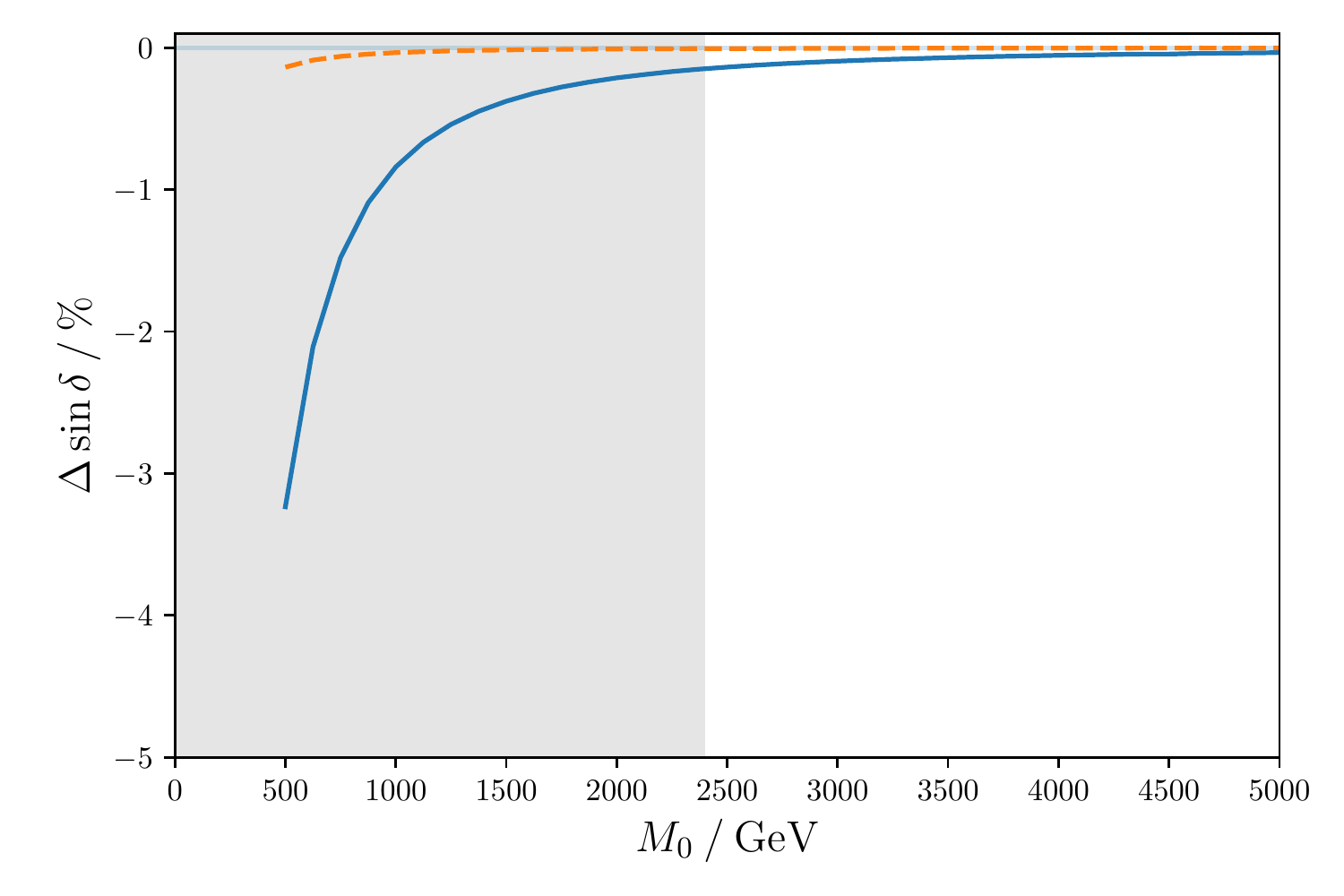}}
\hspace{0.2in}
\parbox{3in}{\includegraphics*[scale=0.5]{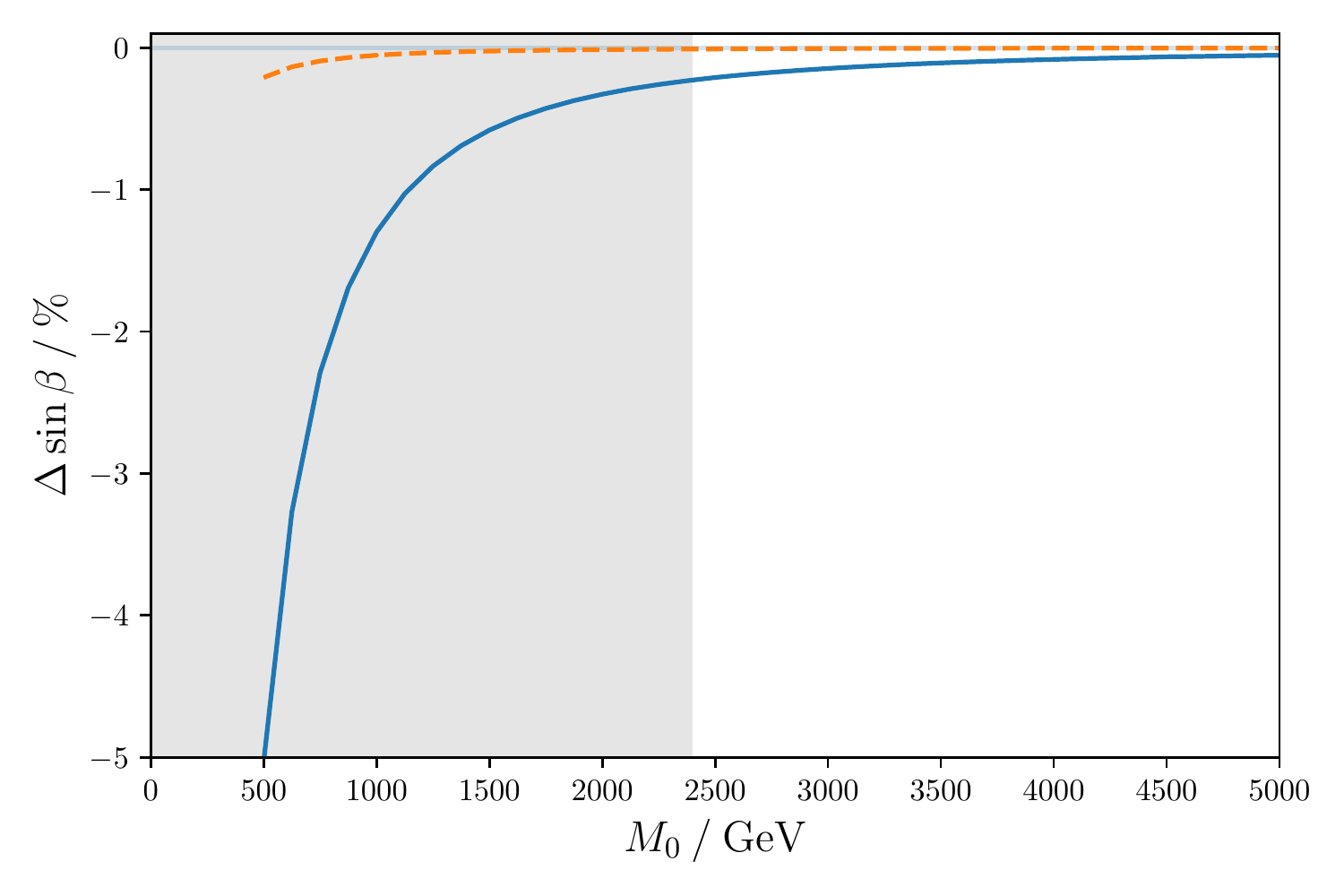}}
\end{center}
\caption{{\small {\bf Case 2)} Relative deviations $\Delta\sin\delta$ (left plot) and $\Delta\sin\beta$ (right plot), as obtained for option 1 of the $(3,3)$ ISS framework, from the values obtained in the model-independent scenario, 
 with respect to the mass $M_0$ (in GeV). 
 The concrete choice of $v$ is irrelevant and thus we 
 have set $v=3$. Line and colour code as in fig.~\ref{fig:Case1_s2thij}.
\label{fig:Case2_sdelta_sbeta}}}
\end{figure}
In our numerical study, we choose as representative values of the index $n$ and of the parameter $u$
\begin{equation}
\label{eq:numchoiceparaCase2}
n=14 \;\; \mbox{and} \;\; u=1,
\end{equation}
also commenting on results for the choices $u=-1$, $u=15$ as well as $u=0$ in order to comprehensively analyse the features of Case 2). 
For the parameter $v$, we consider all permitted values according to the relations in eqs.~(\ref{eq:ZXCase2},\ref{eq:uvdeffromst}) and the chosen value of $u$, e.g.~for $u=1$ we have
\begin{equation}
\label{eq:numchoiceparavCase2u1}
v= 3, \, 9, \, 15, \, 21, \, 27, \, 33, \, 39  \; .
\end{equation}
We start by discussing the relative deviations of $\sin^2\theta_{ij}$. The results for $\Delta\sin^2\theta_{12}$ and $\Delta\sin^2\theta_{23}$ are consistent with the analytical expectations, see eq.~(\ref{eq:Deltasin2thetaij_gen}).
 Indeed, the plots for $\Delta\sin^2\theta_{12}$ and $\Delta\sin^2\theta_{23}$ look very similar to those presented in fig.~\ref{fig:Case1_s2thij} for Case 1).
However, the relative deviation $\Delta\sin^2\theta_{13}$ does not agree with the analytical expectations and instead is always very small, showing that like in Case 1),  $\sin^2\theta_{13}$
is typically adjusted to its experimental best-fit value (since it also drives the fit for the present case).
 
We confirm numerically that the deviations of $\sin^2 \theta_{ij}$ 
do not depend on the choice of the parameter $v$ and we
have thus fixed  $v=3$. 
As regards the dependence of $\Delta\sin^2\theta_{ij}$ on the parameter $u$, we have also checked that the
 aforementioned different choices of $u$ all lead to the same result. 
 
\begin{figure}[t!]
\begin{center}
\parbox{3in}{\includegraphics*[scale=0.5]{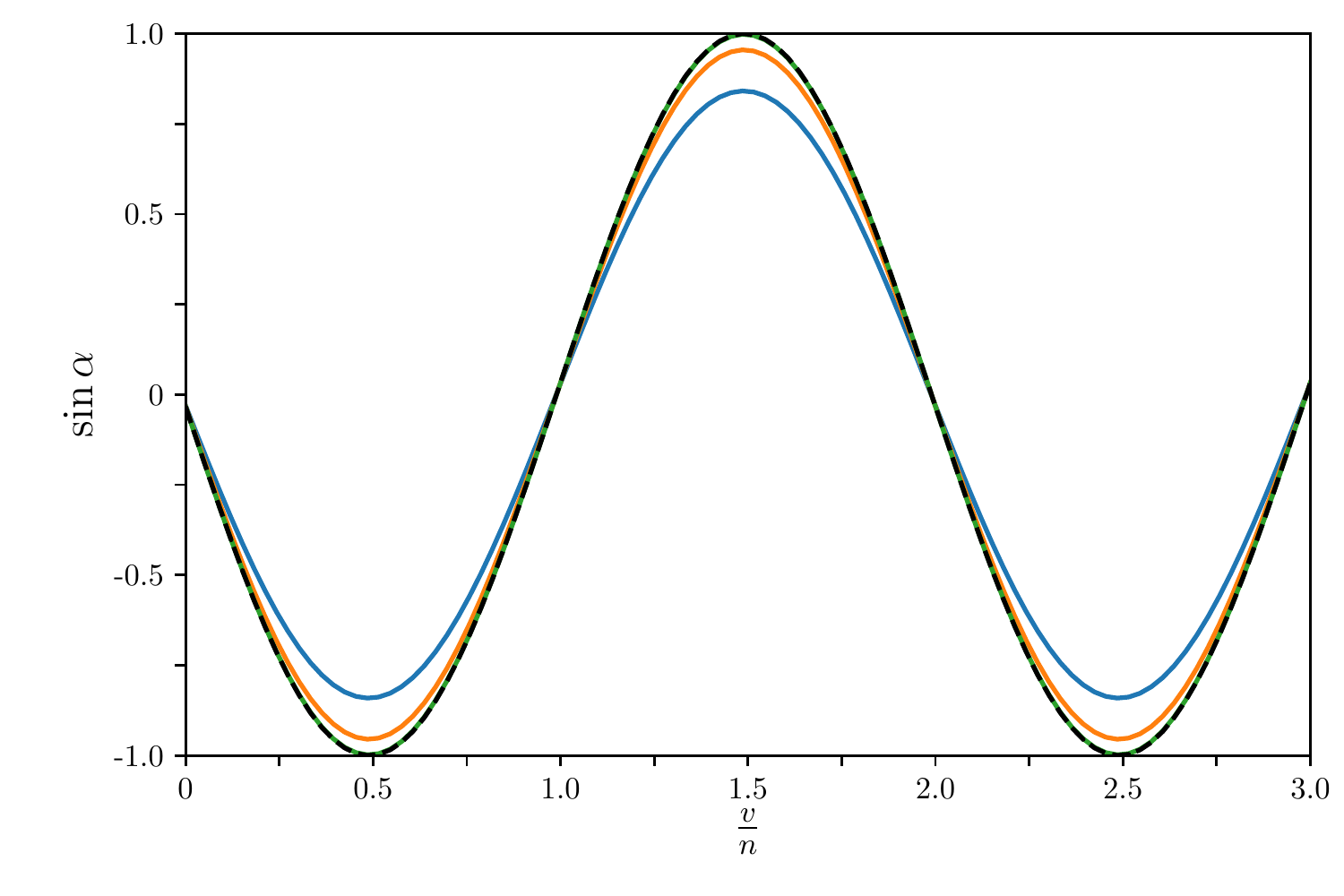}}
\hspace{0.2in}
\parbox{3in}{\includegraphics*[scale=0.5]{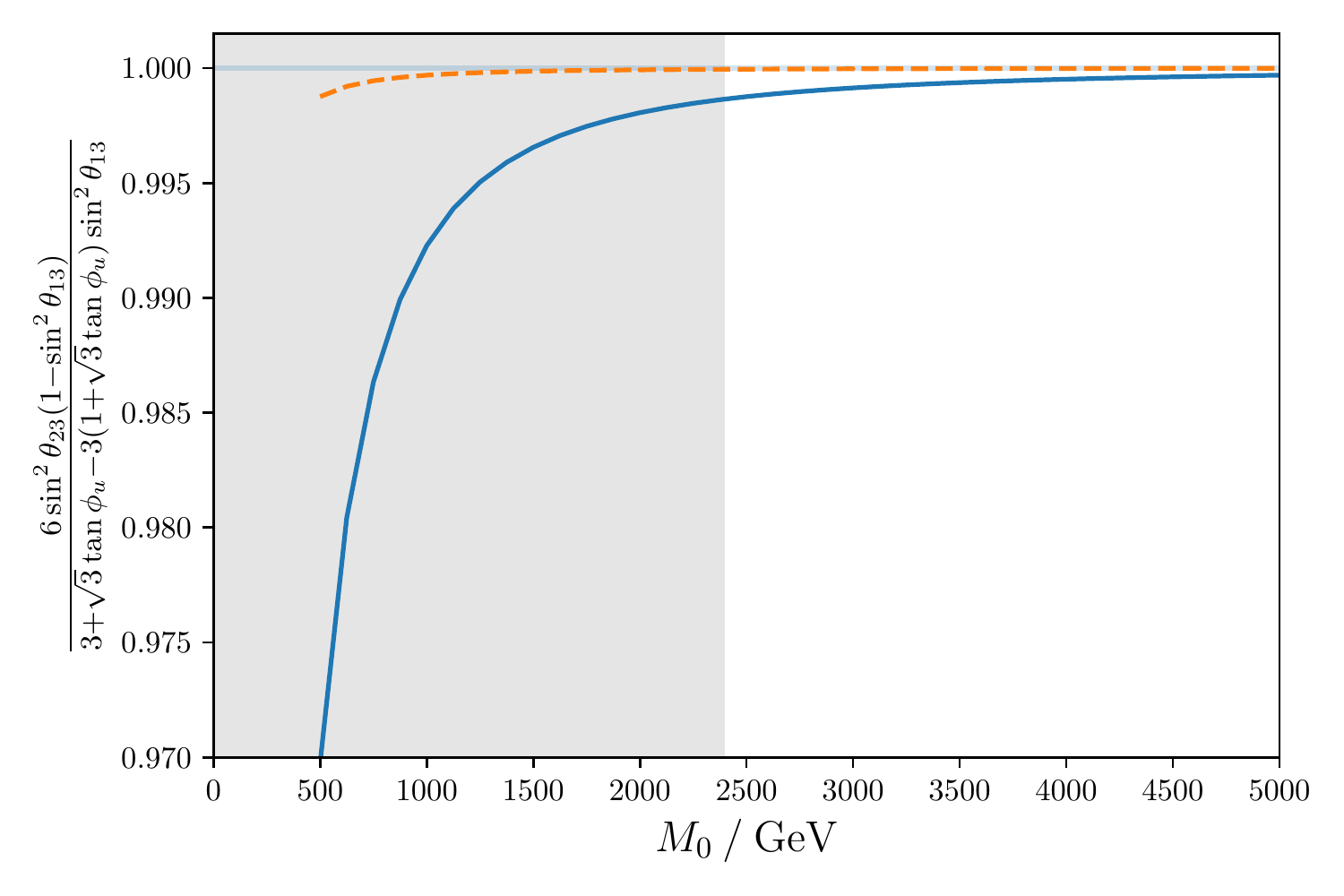}}
\end{center}
\caption{\small {{\bf Case 2)} {\bf Left plot}: $\sin\alpha$ with respect to $v/n$ (fixing $n=14$ and continuously varying $0 \leq v < 3n = 42$). 
The different (coloured) curves refer to three different masses $M_0$ like in fig.~\ref{fig:Case1_salpha}, also setting $y_0=1$.
 The black (dashed) curve displays the result for $\sin\alpha$, obtained in the model-independent scenario, see eq.~(\ref{eq:sinalphaCase2}).
 {\bf Right~plot}: Validity check of the exact sum rule in eq.~(\ref{eq:sumruleCase2}) for option 1 of the $(3,3)$ ISS framework
 with respect to the mass $M_0$ (in GeV). Line and colour code as in fig.~\ref{fig:Case2_sdelta_sbeta}.
 We have chosen $n=14$ and $u=1$ so that $\tan\phi_u \approx 0.23$. 
\label{fig:Case2_salpha_Sigma3}}}
\end{figure}

 For the relative deviations of the CP phases $\delta$ and $\beta$, $\Delta\sin\delta$ and $\Delta\sin\beta$, we present our findings in fig.~\ref{fig:Case2_sdelta_sbeta}.
  Since these deviations are also independent of the choice of $v$, we choose $v=3$ for concreteness. 
The plot for $\Delta\sin\alpha$ looks very similar to the corresponding one of Case 1), see left plot in fig.~\ref{fig:Case1_salpha}.
The sign and size of the deviations are in accordance with the analytical expectations, see 
 eqs.~(\ref{eq:DeltasinCPphases_gen},\ref{eq:estimateDsy01}). We note that both Majorana phases $\alpha$ and $\beta$ experience slightly larger effects from the non-unitarity of the lepton mixing matrix (i.e., the presence of the heavy sterile states) than the Dirac phase $\delta$.
 The effects of the non-unitarity of $\widetilde{U}_\nu$ on the behaviour of $\sin\alpha$ with respect to $v/n$, shown in the left plot of fig.~\ref{fig:Case2_salpha_Sigma3}, are very similar to those
 encountered when studying $\sin\alpha$ with respect to $s/n$ for Case 1), see the right plot in fig.~\ref{fig:Case1_salpha}.
 Again, we emphasise that the periodicity of $\sin\alpha$ in $v/n$ is not altered by the effects of the non-unitarity of the PMNS mixing matrix.

Next, we detail our numerical results for the relative deviations of the two (approximate) sum rules found for Case 2), see section~\ref{sec32}. We have checked that for the sum rule which is common for Case 1) and Case 2)
(see first approximate equality in eq.~(\ref{eq:sumrulesCase1})), the results do coincide with those shown in the left plot in fig.~\ref{fig:Case1_Sigma12}. 
Concerning the exact sum rule, shown in eq.~(\ref{eq:sumruleCase2}), the numerical results are given in the right plot in fig.~\ref{fig:Case2_salpha_Sigma3}. We see that the size and sign 
of the relative deviation agree with the analytical estimate shown in eq.~(\ref{eq:estimateDSigma3y01}). We have also checked numerically that the results do not depend on the choice of $u$ and $v$; while the plot presented relies on
$u=1$ and $v=3$, similar results have been found for the other mentioned choices of $u$ and the admitted values of $v$. 

We comment on the choice $u=0$ that predicts maximal atmospheric mixing and maximal Dirac phase $\delta$, $\sin\beta=0$ and the exact equality in eq.~(\ref{eq:sinalphaCase2}):
the relative deviations $\Delta\sin^2\theta_{23}$ and $\Delta\sin\delta$ are of the same sign and size,  
and exhibit the same dependence on the Yukawa coupling $y_0$ and on the mass scale $M_0$ as occurs for the choice $u=1$.
Furthermore, the fact that the Majorana phase $\beta$ is trivial is not altered by the effects of non-unitarity of $\widetilde{U}_\nu$, as expected from the analytical estimates, see section~\ref{sec4}.
 The results for the Majorana phase $\alpha$ look very similar
to those displayed in fig.~\ref{fig:Case1_salpha} (left plot) and fig.~\ref{fig:Case2_salpha_Sigma3} (left plot). 
Moreover, we confirm that whenever the choice $v=0$ is permitted,  the Majorana phase $\alpha$ vanishes 
independently of the deviations of $\widetilde{U}_\nu$
from unitarity. 

Finally, we notice that we have performed a numerical check to confirm that  the symmetry transformations in the parameters $u$ and $\theta$, see eq.~(\ref{eq:symmtrafosCase2}) under point $i)$ in section~\ref{sec32}, are still valid.

\subsection{Case 3 a)}
\label{sec53}

As representative values for $n$ and $m$, we take
\begin{equation}
\label{eq:numchoiceparaCase3a}
n=17 \;\; \mbox{and} \;\; m=1 \; ,
\end{equation}
since these can  satisfactorily accommodate the experimental data on the reactor and the atmospheric mixing angles, $ \sin^2 \theta_{13} = 0.02221^{+0.00068} _{-0.00062}$ and 
$\sin^2 \theta_{23} = 0.570^{+0.018} _{-0.024}$ for light neutrinos with NO and $ \sin^2 \theta_{13} = 0.02240^{+0.00062} _{-0.00062}$ and 
$\sin^2 \theta_{23} = 0.575^{+0.017} _{-0.021}$ for light neutrinos with IO~\cite{NuFIT50},
according to the expectations from the model-independent scenario, see section~\ref{sec331} and, especially, eq.~(\ref{eq:sin2theta1323Case3a}).
We consider all possible values of the parameter $s$. In addition to $m=1$, we also study the results on lepton mixing for the choice $m=16$. 
The rather large value of the index $n$ of the flavour symmetry is needed in order to achieve a sufficiently small value of $m/n$ (or $1-m/n$).

 \begin{figure}[t!]
\begin{center}
\parbox{3in}{\includegraphics*[scale=0.55]{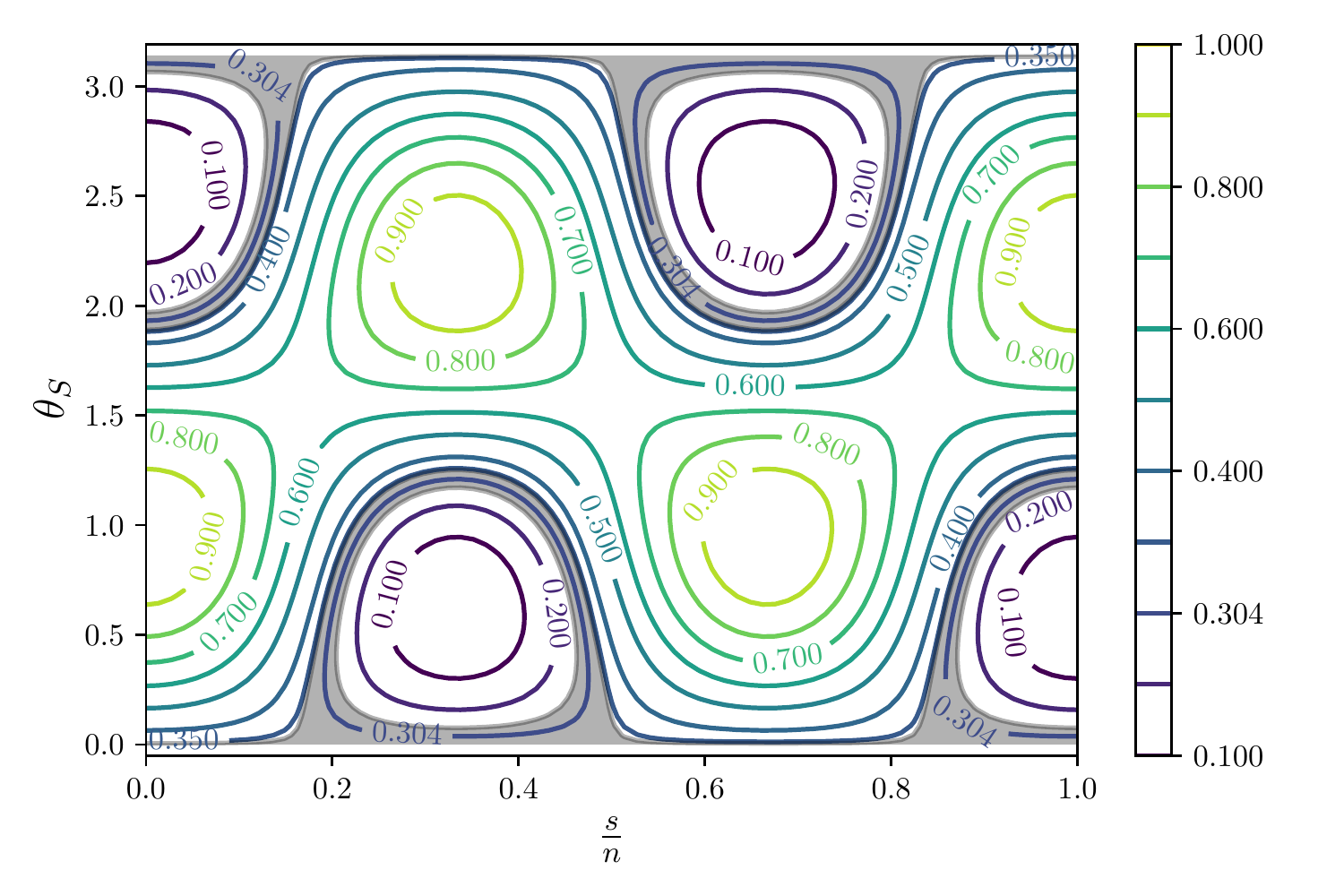}}
\hspace{0.2in}
\parbox{3in}{\includegraphics*[scale=0.55]{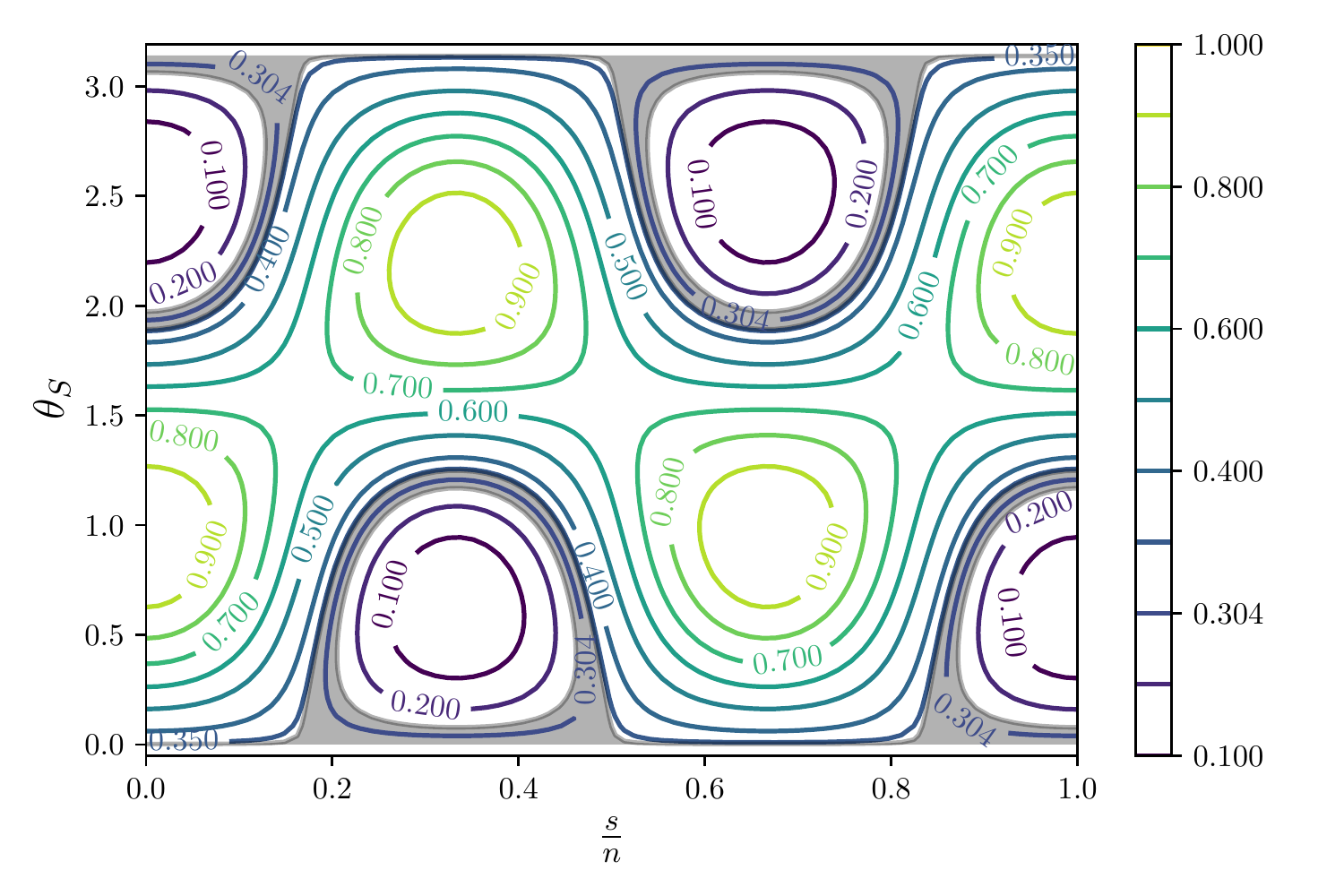}}
\end{center}
\caption{{\small {\bf Case 3 a)} Contour plots for $\sin^2\theta_{12}$ in the $(s/n-\theta_S)$ plane, obtained for option 1 of the $(3,3)$ ISS framework.
The left plot is for $M_0=1000$~GeV and the right one for $M_0=5000$~GeV. We fix $y_0=0.5$ in order to amplify the differences between the two plots.
Here the grey-shaded areas denote 
values of $\sin^2\theta_{12}$ which are 
experimentally favoured at the $3 \, \sigma$ level~\cite{NuFIT50}.
\label{fig:Case3a_s2th12_contour}}}
\end{figure}
Since fixing $n$ and $m$ determines completely the value of the reactor and the atmospheric mixing angles, we only consider, like for Case 1) and Case 2), 
the relative deviations $\Delta\sin^2\theta_{13}$ and $\Delta\sin^2\theta_{23}$. We note that their size and sign do agree with the analytical expectations, see eq.~(\ref{eq:estimateDsy01}).
 Furthermore, we confirm numerically that there is no dependence of these results on the parameter $s$ and the free angle $\theta_S$.
Since in Case 3 a) $\theta_{12}$ is the only lepton mixing angle that depends on the free angle $\theta_S$, $\sin^2\theta_{12}$ naturally drives the fit, and thus the relative deviation $\Delta\sin^2\theta_{12}$ is always very small.
 Given that $\sin^2\theta_{12}$ further depends on the parameter $s$, we present
 in fig.~\ref{fig:Case3a_s2th12_contour}
 plots for $\sin^2\theta_{12}$ in the $(s/n-\theta_S)$ plane
for two different values of the mass scale $M_0=1000$~GeV (left plot) and $M_0=5000$~GeV (right plot). 
We fix the Yukawa coupling to $y_0=0.5$ in order to better perceive the differences in the plots for the two different values of $M_0$,
although such a large value of $y_0$ does require $M_0 \gtrsim 2400$~GeV in order to comply with the experimental constraints on $\eta_{\alpha\beta}$, see section~\ref{sec50}. 
As one observes in fig.~\ref{fig:Case3a_s2th12_contour}, the visible differences are still very small. 
We stress that here the grey-shaded areas indicate the values of $\sin^2\theta_{12}$ that are experimentally favoured at the $3 \, \sigma$ level~\cite{NuFIT50}.
As can be clearly seen from fig.~\ref{fig:Case3a_s2th12_contour}, for most values of $s$
a successful accommodation of the experimental data can be obtained for 
two different values of the free angle $\theta_S$.
 One of these values is close to $\theta_S \approx 0$ or $\theta_S \approx \pi$.
These plots can be compared with a very similar one shown in the original analysis of the different mixing patterns,  see~\cite{HMM}.

 \begin{figure}[t!]
\begin{center}
\parbox{3in}{\includegraphics*[scale=0.55]{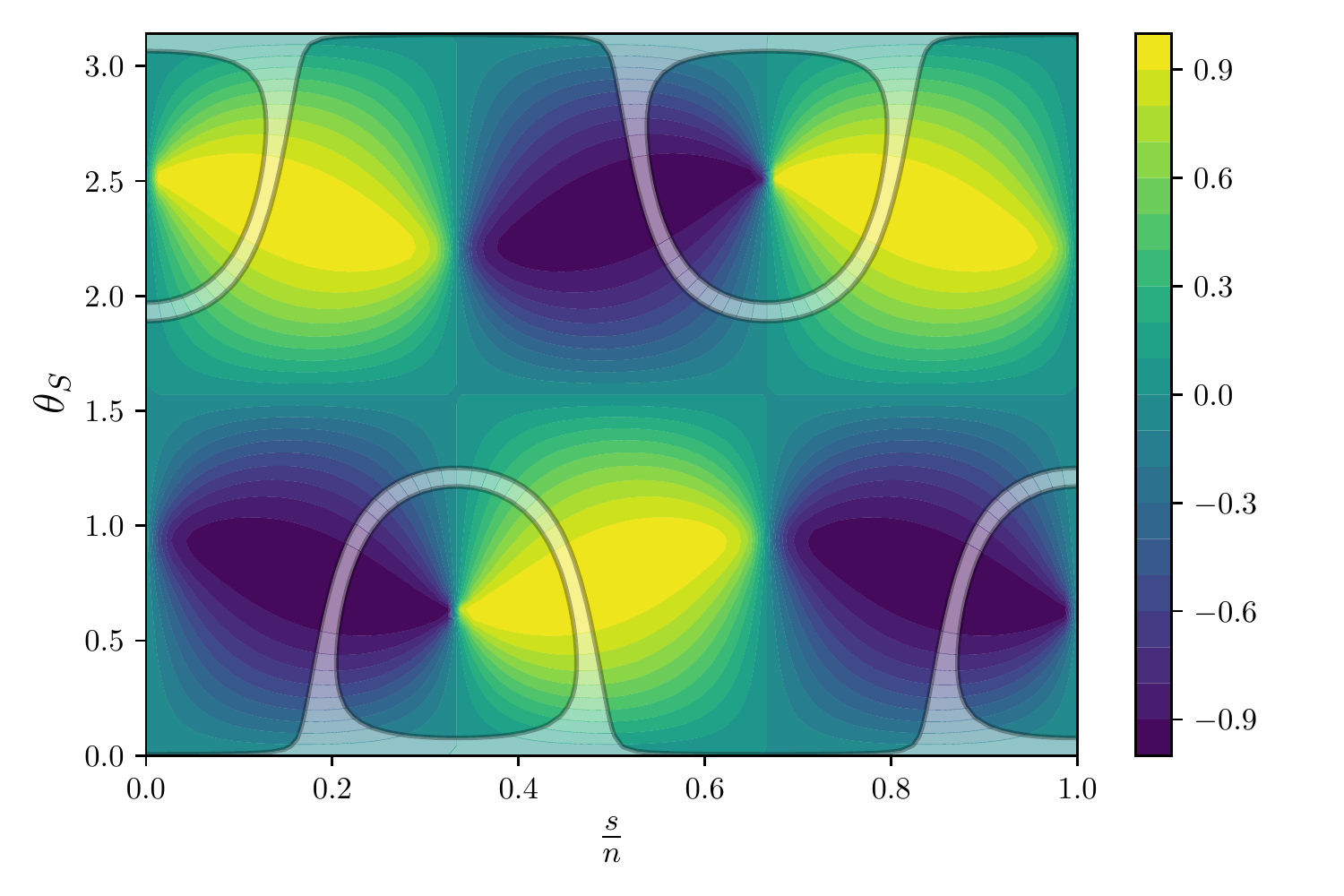}}
\hspace{0.2in}
\parbox{3in}{\includegraphics*[scale=0.55]{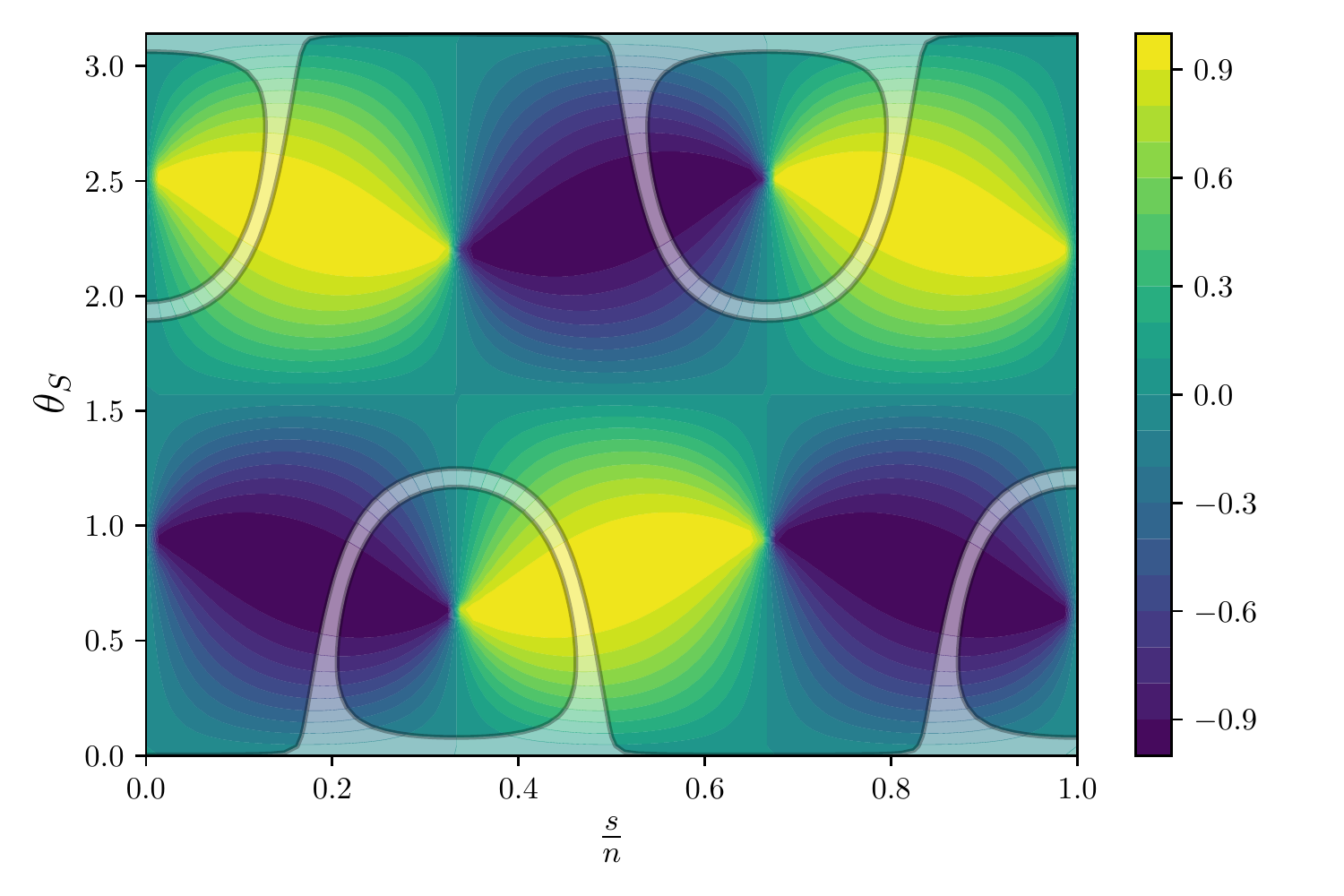}}
\vspace{0.1in}
\parbox{3in}{\includegraphics*[scale=0.55]{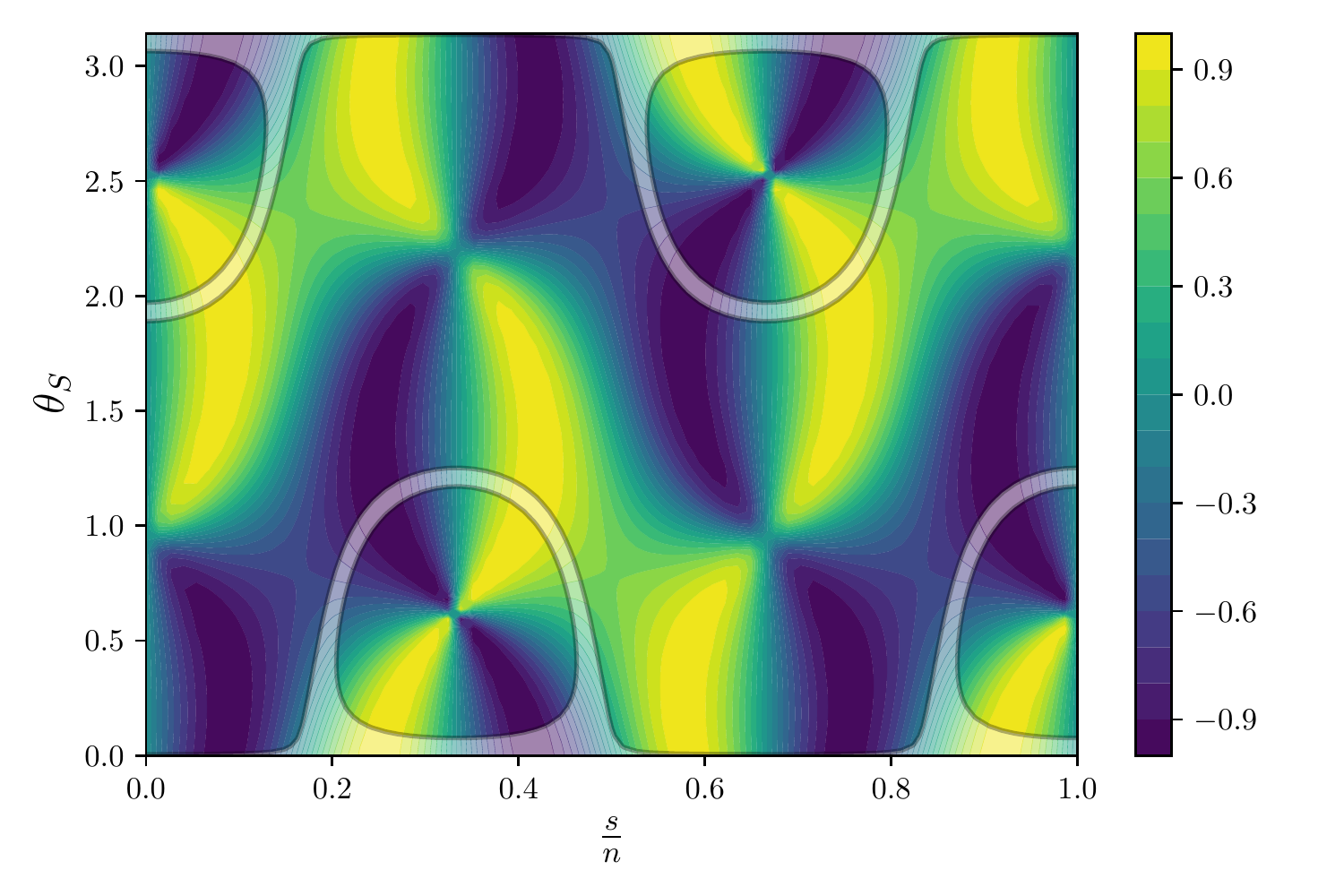}}
\hspace{0.2in}
\parbox{3in}{\includegraphics*[scale=0.55]{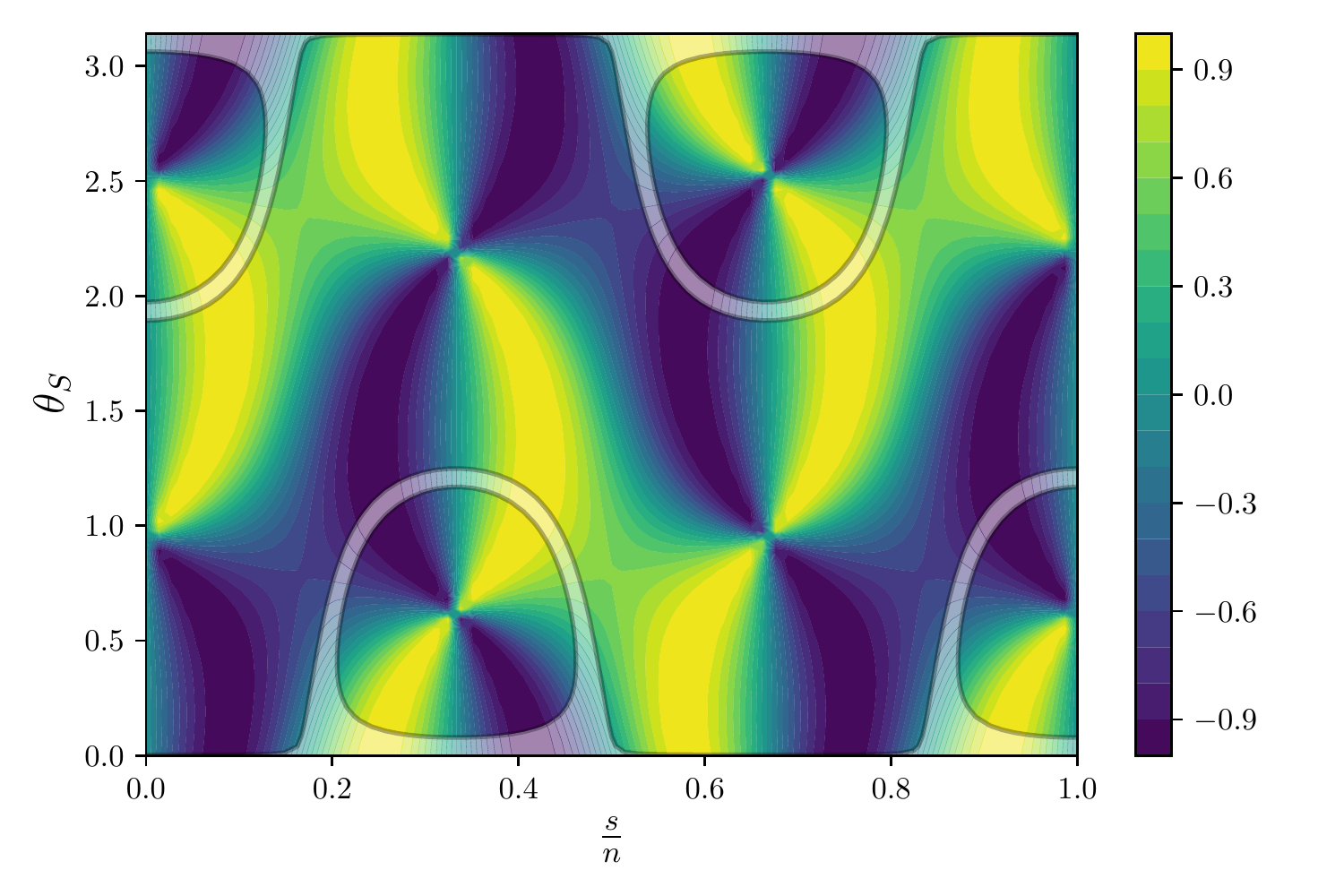}}
\vspace{0.1in}
\parbox{3in}{\includegraphics*[scale=0.55]{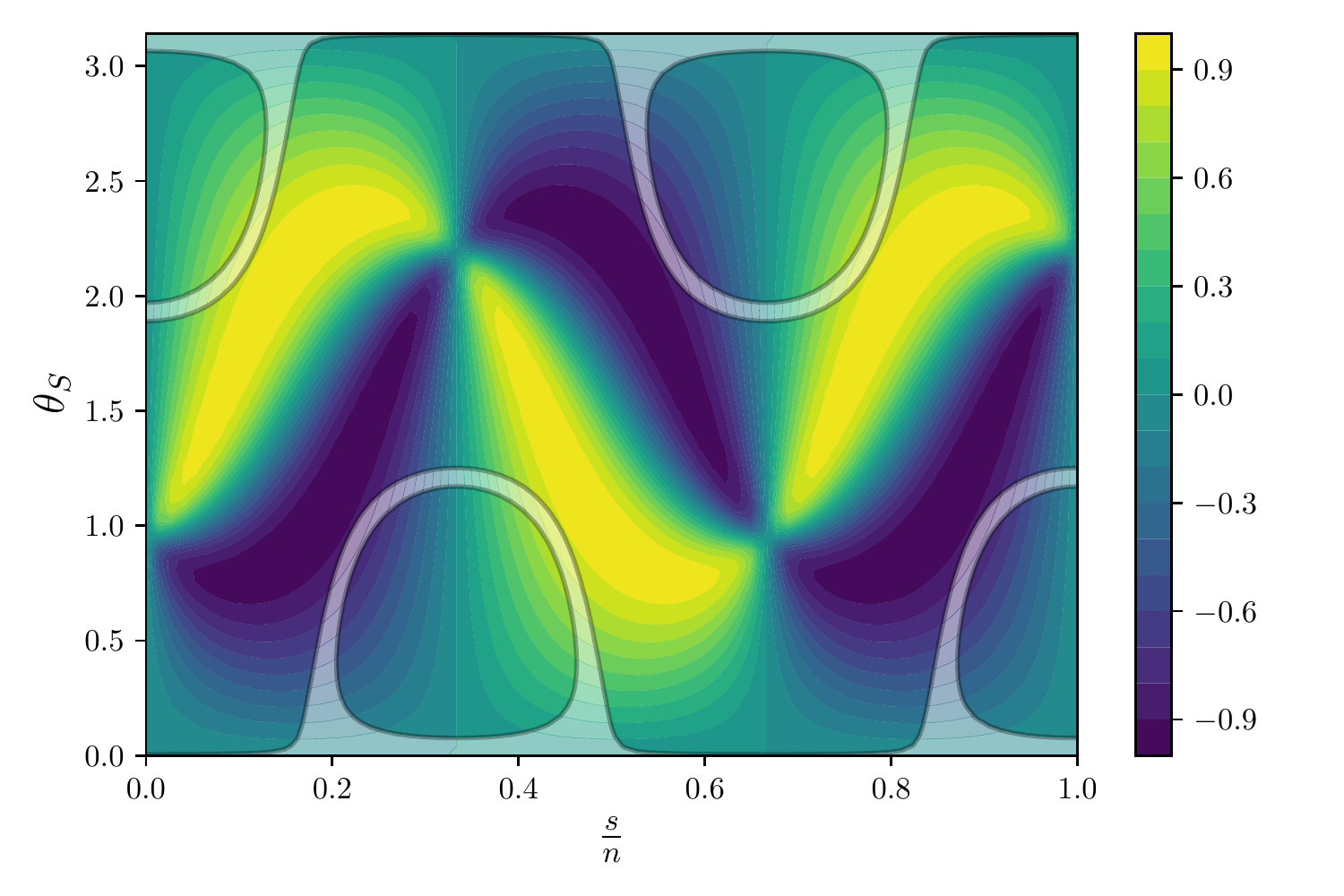}}
\hspace{0.2in}
\parbox{3in}{\includegraphics*[scale=0.55]{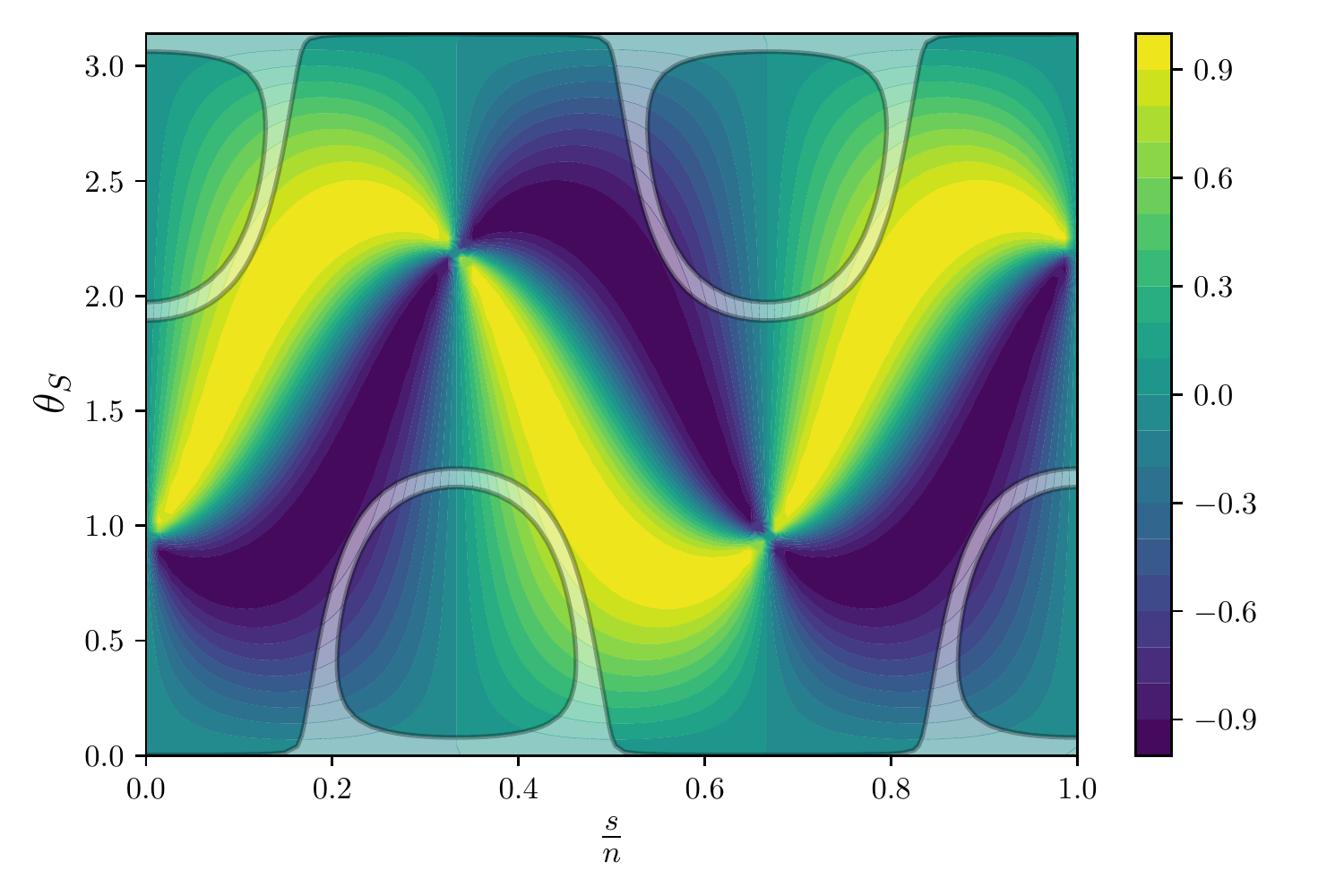}}
\end{center}
\caption{{\small {\bf Case 3 a)} Contour plots for the sines of the CP phases in the $(s/n-\theta_S)$ plane, 
obtained for option 1 of the $(3,3)$ ISS framework. 
From top to bottom (first to third row), $\sin\delta$, $\sin\alpha$ and $\sin\beta$. 
On the left column plots, $M_0=1000$~GeV while on the right $M_0=5000$~GeV. We again fix $y_0=0.5$ (see fig.~\ref{fig:Case3a_s2th12_contour}). 
The colour scheme denotes the values of the sines, from $-1$ (dark blue) to $+1$ (light yellow), as indicated by the colour bar on the right of each plot.
The white/grey-shaded areas correspond here to those of fig.~\ref{fig:Case3a_s2th12_contour}, and indicate the values of the solar mixing angle that are experimentally preferred at the $3 \, \sigma$ level.
\label{fig:Case3a_sdelta_salpha_sbeta_contour}}}
\end{figure}
The results for the relative deviations of the CP phases, $\Delta\sin\delta$, $\Delta\sin\alpha$ and $\Delta\sin\beta$, look similar to those obtained for the already presented cases, Case 1) and Case 2). 
For this reason, we prefer to show contour plots for the sines of all three CP phases in the $(s/n-\theta_S)$ plane.
These can be found in fig.~\ref{fig:Case3a_sdelta_salpha_sbeta_contour}, 
where we display $\sin\delta$, $\sin\alpha$ and $\sin\beta$, for two different values of $M_0$, $M_0=1000$~GeV (left plots) and $M_0=5000$~GeV (right plots). The colour scheme denotes the values of the sines (indicated by the colour bar on the right of each plot). We again take $y_0=0.5$ in order to enhance the visibility of differences in the plots. The white/grey-shaded areas indicate the values of the solar mixing angle that are experimentally preferred at the $3 \, \sigma$ level. It turns out that visible differences between the plots for $M_0=1000$~GeV and 
$M_0=5000$~GeV are (mainly) found in regions of the $(s/n-\theta_S)$ plane that
are not compatible with the experimental value of $\sin^2\theta_{12}$ at the $3 \, \sigma$ level. Nevertheless, the results presented in these plots are interesting, since the validity of the approximate formulae for the sines of the CP phases
(found in eqs.~(\ref{eq:sinalphaCase3a},\ref{eq:sinbetasindeltaCase3a}) under point $e)$ for the model-independent scenario) as well as the fact that the absolute value of $\sin\beta$ is bounded to be smaller than $\sim 0.9$, can be checked.
Furthermore, they can be directly compared with the results for the model-independent scenario presented in~\cite{HMM}.
Again, we confirm numerically that the effects of non-unitarity of the PMNS mixing matrix do not affect the 
vanishing of $\sin\delta$, $\sin\alpha$ and/or $\sin\beta$ (occurring for certain choices of group theory parameters).   
The approximate sum rule, quoted in eq.~(\ref{eq:sumruleCase3a}), is valid with a plus sign for the choice $n=17$ and $m=1$. Studying its behaviour depending on the Yukawa coupling $y_0$ and on the mass scale $M_0$ thus
leads to results 
very similar to those obtained for the second approximate sum rule (see second approximate equality in eq.~(\ref{eq:sumrulesCase1})), for values of the free angle $\theta_S <\pi/2$, as shown
in the right plot of fig.~\ref{fig:Case1_Sigma12}. 

In the end, we note that we have numerically confirmed that the symmetry transformations, given under point $i)$ in section~\ref{sec331}, are valid.

\subsection{Case 3 b.1)}
\label{sec54}

For the last case, we focus on
\begin{equation}
\label{eq:numchoiceparaCase3b1}
n=20 \;\; \mbox{and} \;\; m=11 \; .
\end{equation}
All viable values of the parameter $s$ are studied. We choose the index $n$ of the flavour symmetry to be rather large\footnote{As shown in~\cite{HMM}, values 
of $n$ as small as $n=2$ are sufficient in order to successfully accommodate the experimental data on lepton mixing angles.} in order to allow studying different values of $m$, while achieving  good agreement with
experimental data on the solar mixing angle. In addition to the value $m=11$ we also perform a numerical analysis for $m=9$
and $m=10$.

\begin{figure}[t!]
\begin{center}
\parbox{3in}{\hspace{-0.3in}\includegraphics*[scale=0.55]{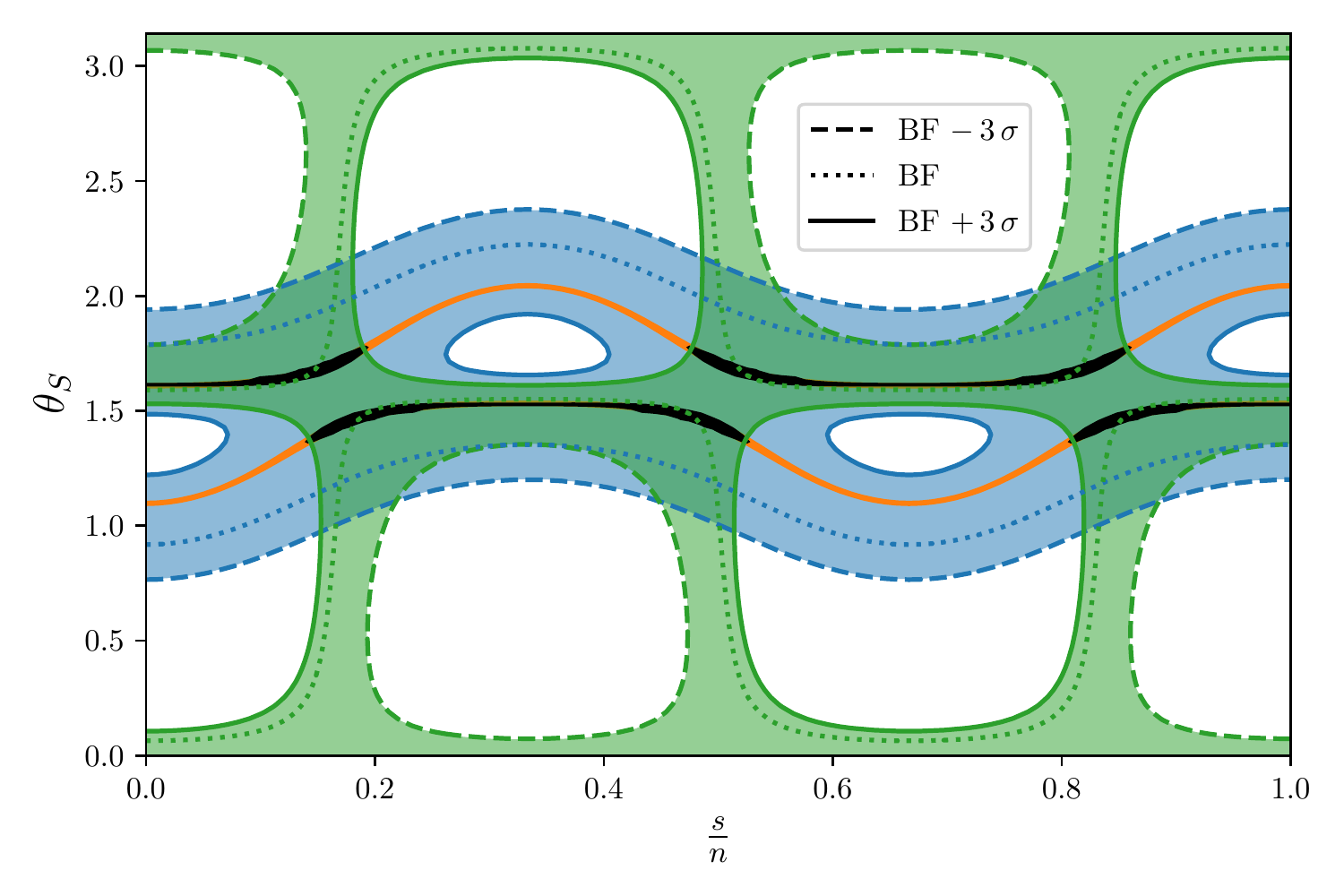}}
\hspace{-0.1in}
\parbox{3in}{\includegraphics*[scale=0.55]{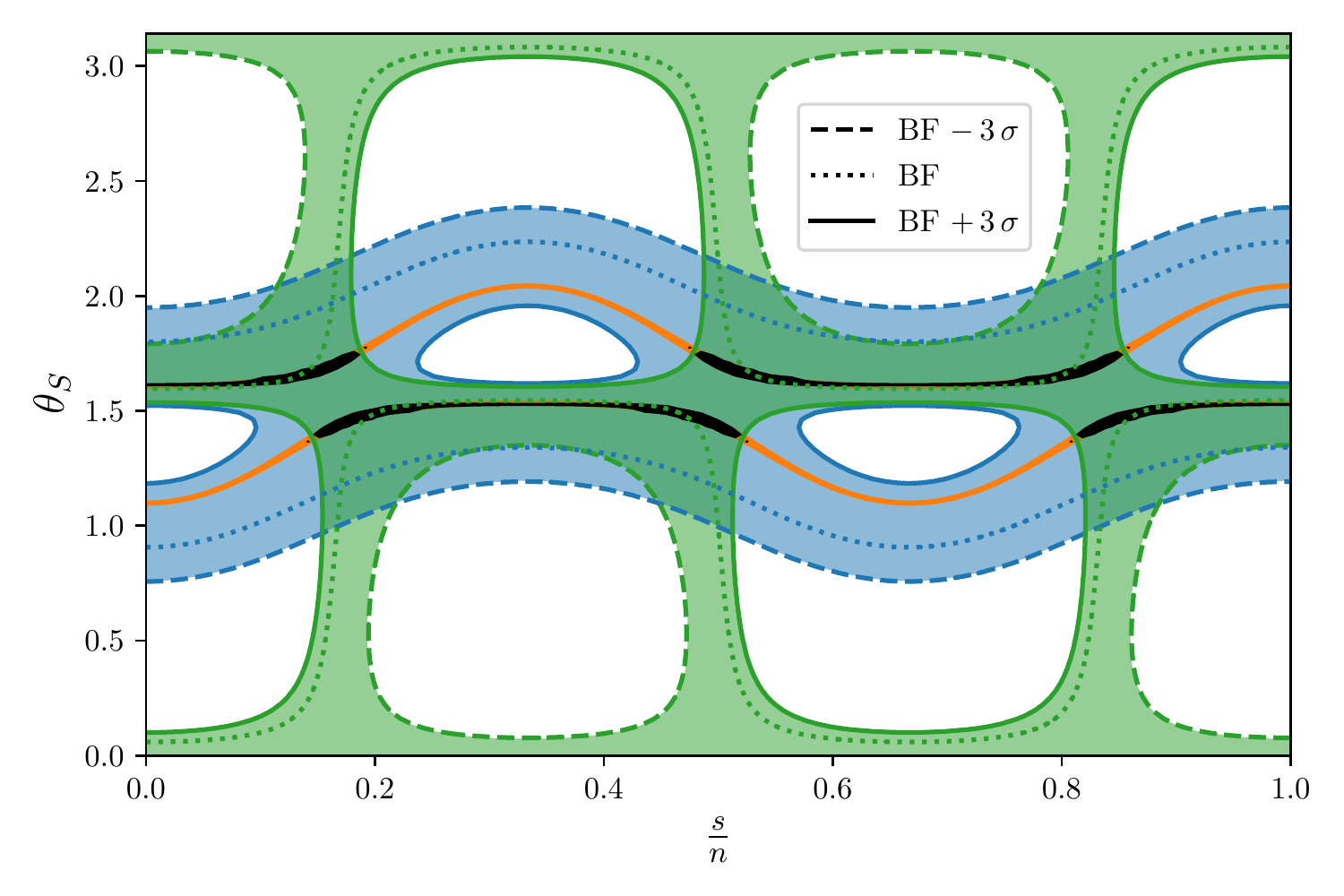}}
\end{center}
\caption{{\small {\bf Case 3 b.1)} Contour plots for $\sin^2\theta_{ij}$, obtained for option 1 of the $(3,3)$ ISS framework, in the $(s/n-\theta_S)$ plane. Blue, green and orange respectively correspond to $\sin^2\theta_{12}$, $\sin^2\theta_{23}$ and $\sin^2\theta_{13}$.
Dotted lines indicate the experimental best-fit (BF) value for each $\sin^2\theta_{ij}$, while the coloured surfaces correspond to 
a $3 \sigma$ interval: dashed (solid) lines respectively define the BF$\mp 3 \sigma$ boundaries.  
Their overlap is highlighted in black. On the left, $M_0=1000$~GeV, while on the right $M_0=5000$~GeV. We again fix $y_0=0.5$ in order to amplify the differences between the two plots.
\label{fig:Case3b1_s2thij_contour}}}
\end{figure}

For Case 3 b.1) all mixing angles turn out to depend on the parameter $s$ and the free angle $\theta_S$, in addition to the two parameters $n$ and $m$ which we have fixed, see  eq.~(\ref{eq:sin2thetaijCase3b1}).
In what follows we identify the areas in the $(s/n-\theta_S)$ plane in which the three mixing angles (individually and simultaneously) are in agreement with the experimental data at the $3 \, \sigma$ level~\cite{NuFIT50}.
This is shown in the contour plots in fig.~\ref{fig:Case3b1_s2thij_contour}, for $\sin^2\theta_{12}$ (blue), $\sin^2\theta_{23}$ (green) and $\sin^2\theta_{13}$ (orange) and their combination (black), for two different values of the mass 
scale $M_0$, $M_0=1000$~GeV (left plot) and $M_0=5000$~GeV (right plot).
We note that we have again chosen $y_0=0.5$ for better visibility of the differences in the plots, although in this case $M_0=1000$~GeV leads to conflict with the experimental constraints on the quantities $\eta_{\alpha\beta}$, see section~\ref{sec50}. We see that the areas of agreement with experimental data at the $3 \,\sigma$ level slightly differ between $M_0=1000$~GeV and $M_0=5000$~GeV. 
However, their overlap (shown in black in the two plots) is not visibly affected, and thus the parameter space in the $(s/n-\theta_S)$ plane compatible with the experimental data on lepton mixing angles hardly depends on 
the mass scale $M_0$. Indeed, comparing these two plots to a similar one, presented in the original analysis of the mixing pattern Case 3 b.1) for the model-independent scenario~\cite{HMM}, we confirm that all agree very well.
We note that the by far strongest constraint on the allowed parameter space in the $(s/n-\theta_S)$ plane is imposed by the reactor mixing angle $\sin^2\theta_{13}$.
The results shown in the plots in fig.~\ref{fig:Case3b1_s2thij_contour} also confirm that all values of the parameter $s$ lead to a successful accommodation of the experimental data of the mixing angles for $n=20$ and $m=11$. 
The values of the free angle $\theta_S$
are then close to $\pi/2$.
 Regarding the size and sign of the relative deviations $\Delta\sin^2\theta_{12}$ and $\Delta\sin^2\theta_{23}$, we note that these are consistent with the analytical estimates, see eq.~(\ref{eq:estimateDsy01}), whereas for $\Delta\sin^2\theta_{13}$ 
we always find it to be very small due to the pull in the fit that drives the adjustment of the free angle $\theta_S$ to match the best-fit value of the reactor mixing angle. This is analogous to what has been observed for Case 1) 
and Case 2).  
 \begin{figure}[t!]
\begin{center}
\parbox{3in}{\includegraphics*[scale=0.55]{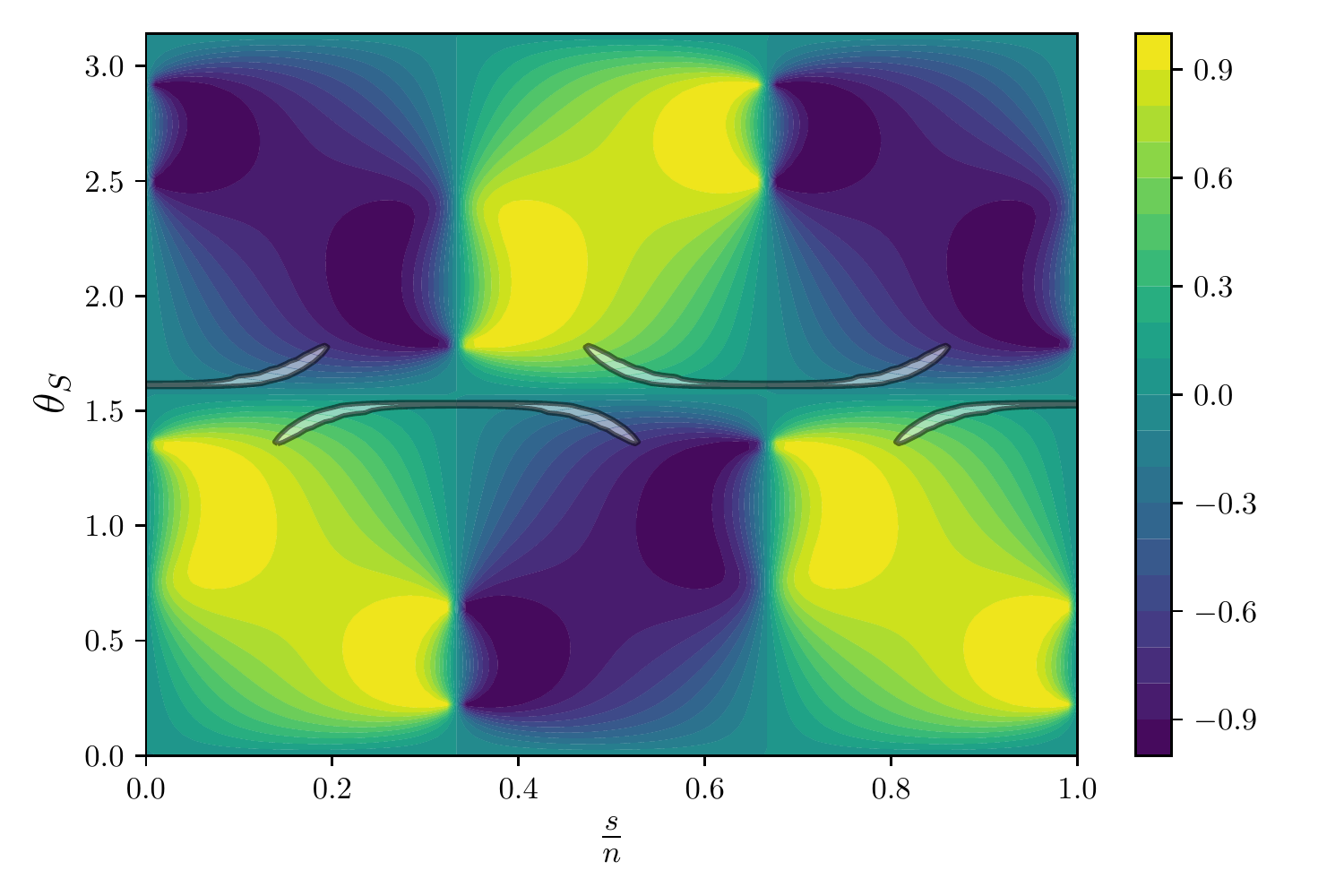}}
\hspace{0.2in}
\parbox{3in}{\includegraphics*[scale=0.55]{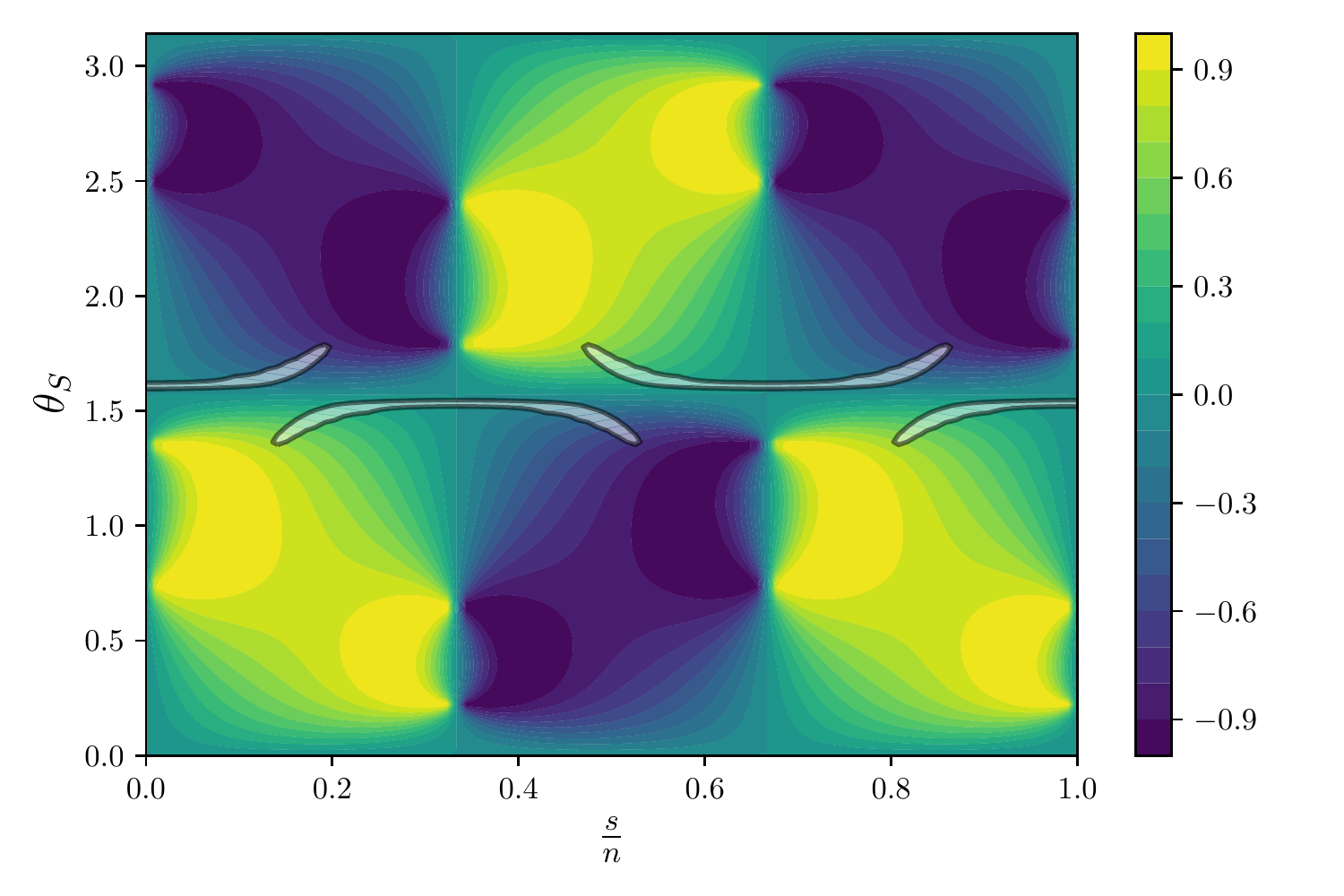}}
\vspace{0.1in}
\parbox{3in}{\includegraphics*[scale=0.55]{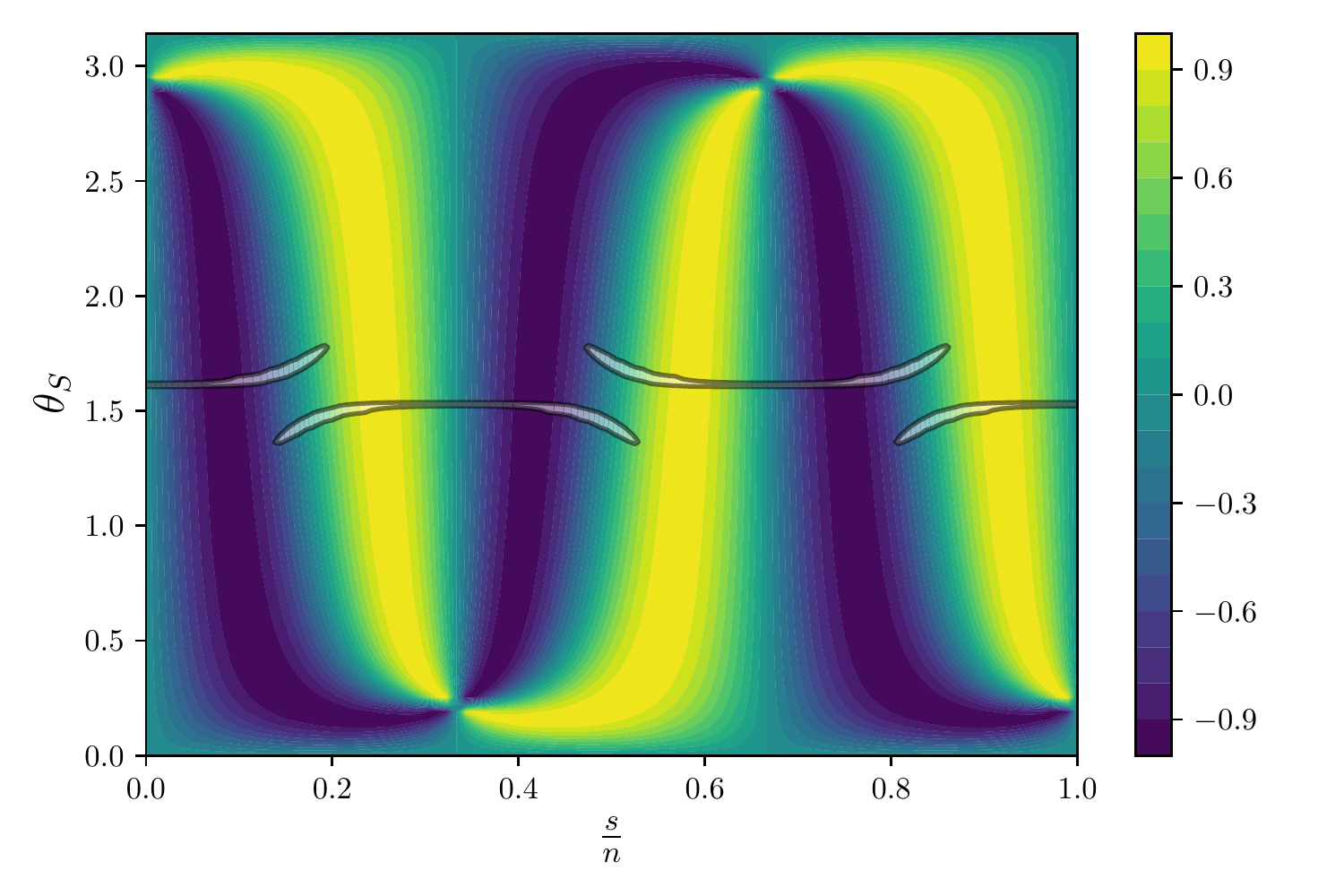}}
\hspace{0.2in}
\parbox{3in}{\includegraphics*[scale=0.55]{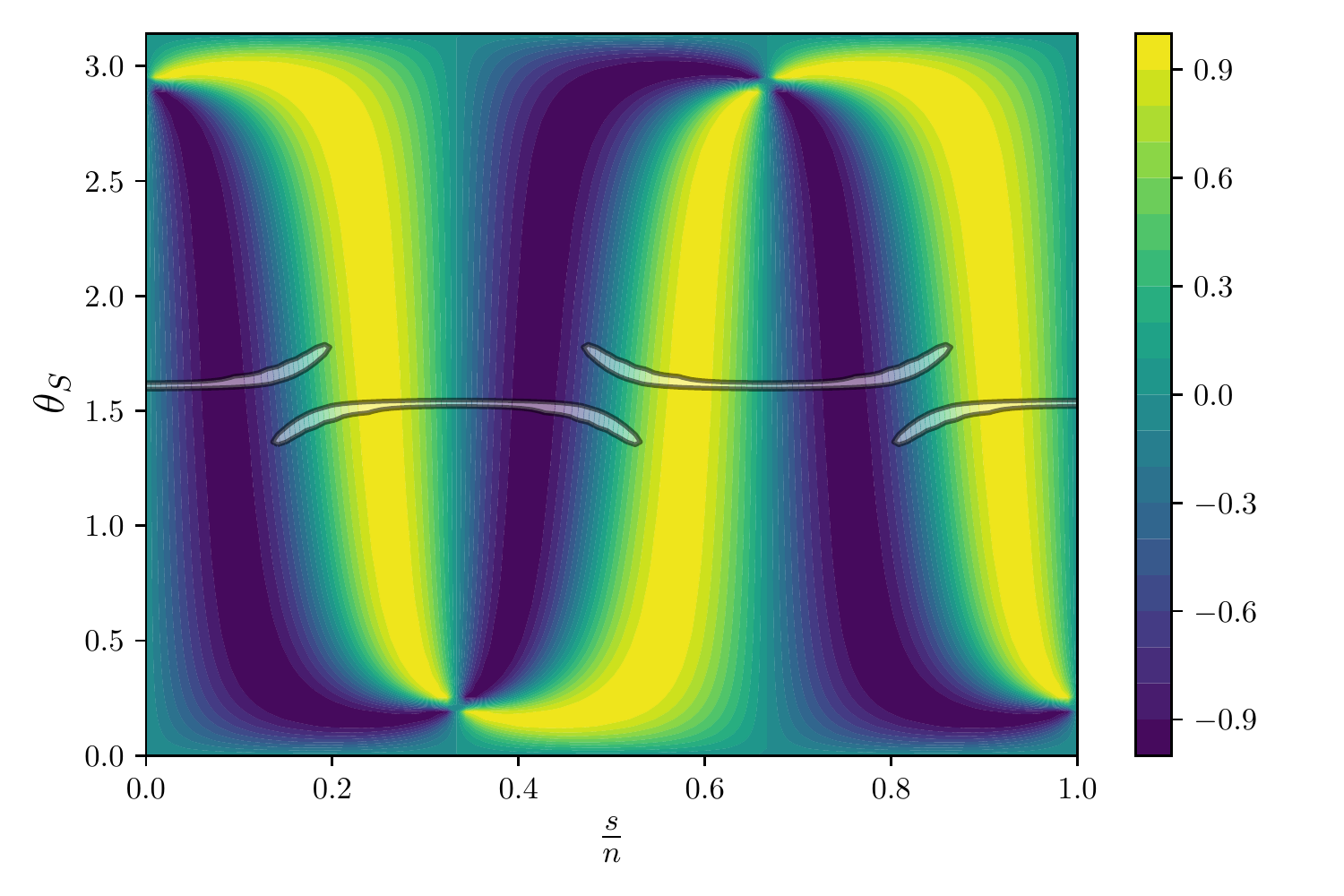}}
\vspace{0.1in}
\parbox{3in}{\includegraphics*[scale=0.55]{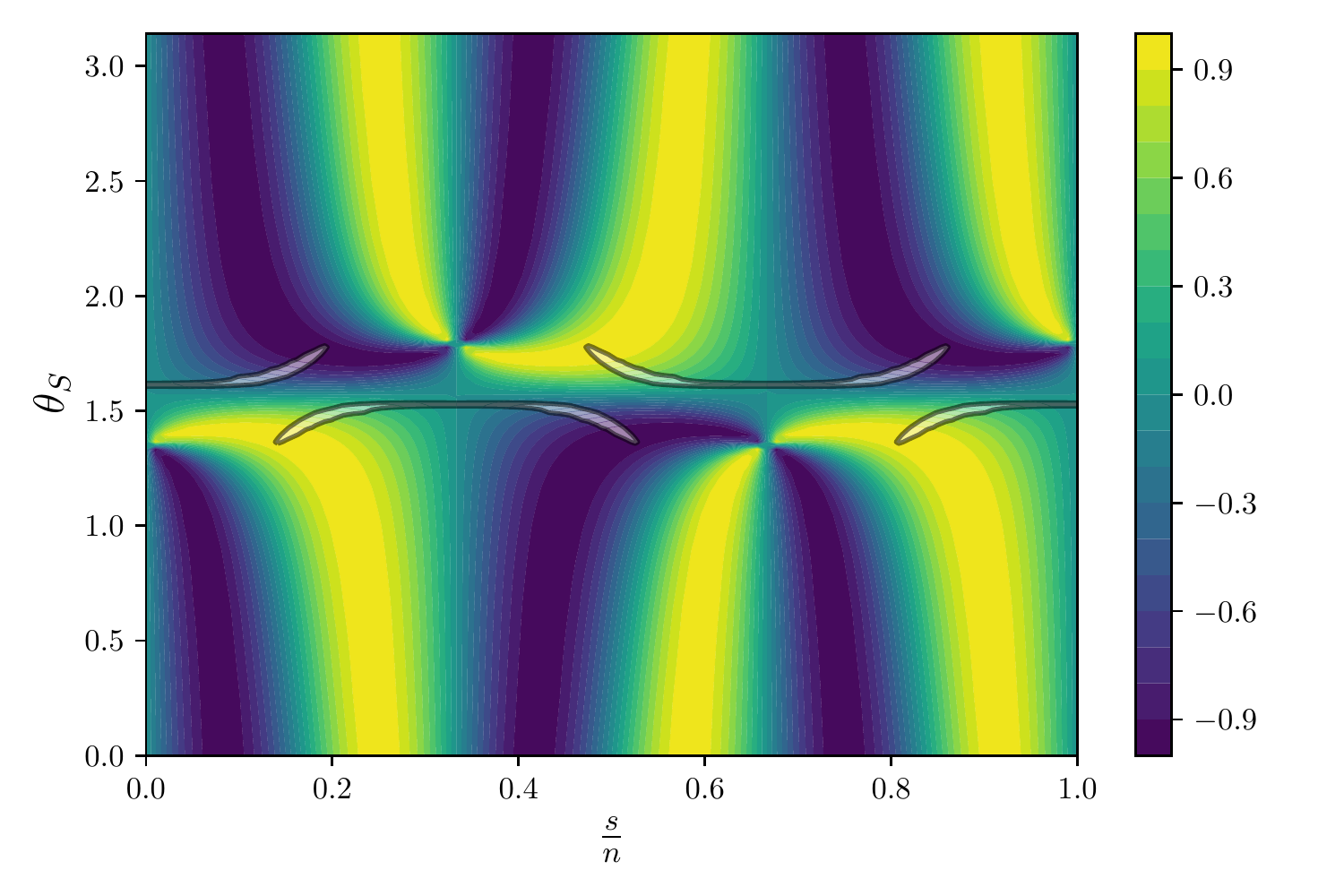}}
\hspace{0.2in}
\parbox{3in}{\includegraphics*[scale=0.55]{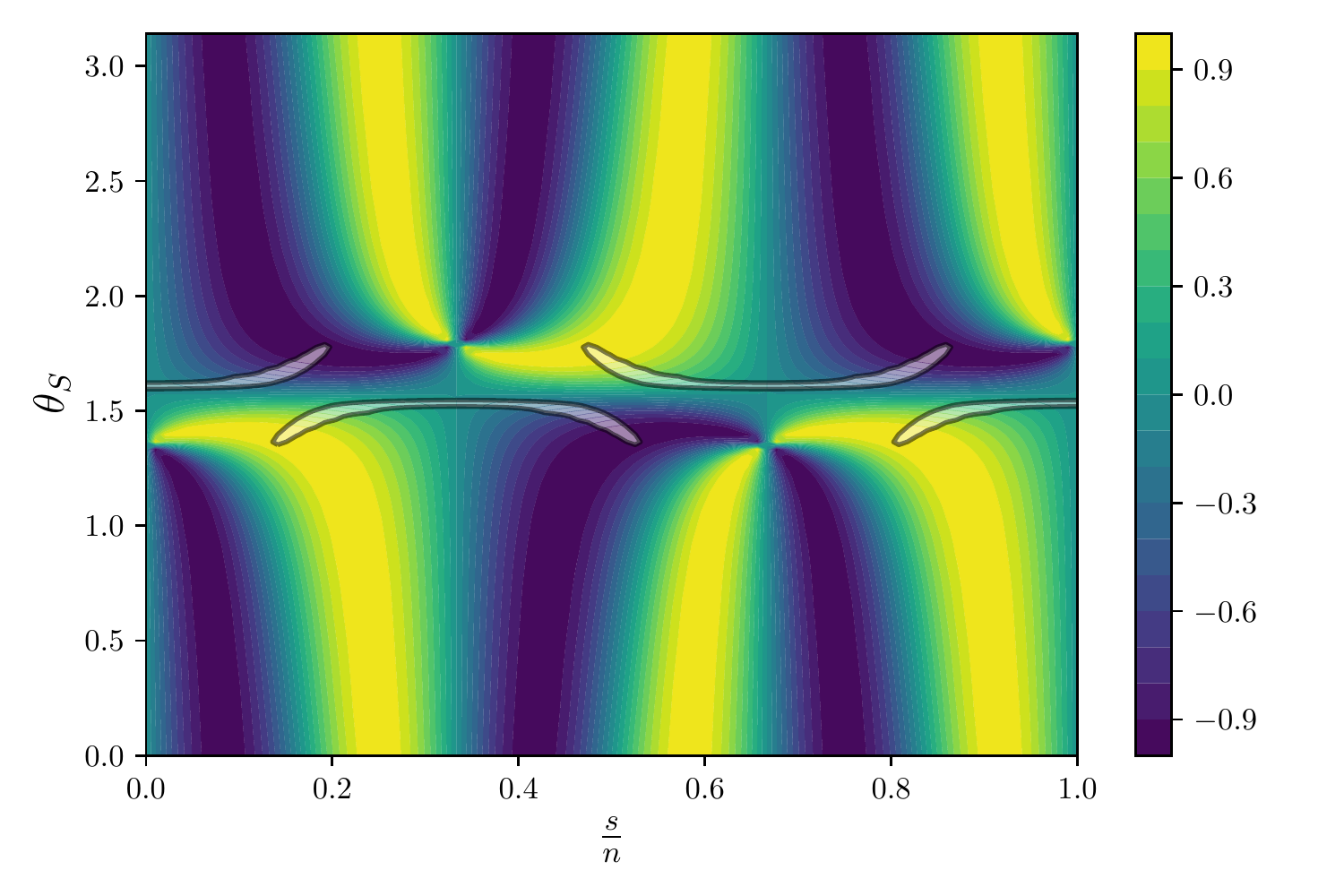}}
\end{center}
\caption{{\small {\bf Case 3 b.1)} Contour plots for the sines of the CP phases, obtained for option 1 of the $(3,3)$ ISS framework, in the $(s/n-\theta_S)$ plane. 
From top to bottom (first to third row), $\sin\delta$, $\sin\alpha$ and $\sin\beta$.
The white/grey-shaded areas correspond to the black regions in the plots in fig.~\ref{fig:Case3b1_s2thij_contour}, and indicate the regions in which all three mixing angles agree with experimental data at the $3 \, \sigma$ level.
Input parameters ($M_0$ and $y_0$) and colour coding as in fig.~\ref{fig:Case3a_sdelta_salpha_sbeta_contour}.
\label{fig:Case3b1_sdelta_salpha_sbeta_contour}}}
\end{figure}
In what concerns the CP phases, we proceed in the same way as for the three mixing angles, and show in fig.~\ref{fig:Case3b1_sdelta_salpha_sbeta_contour} several contour plots in the $(s/n-\theta_S)$ plane. 
We choose the same values of $M_0$ and $y_0$ as for the analogous study done for Case 3 a); conventions and colour-coding are identical to fig.~\ref{fig:Case3a_sdelta_salpha_sbeta_contour}.
Like in Case 3 a), the visible differences for the different values of $M_0$ are mostly found in regions of the $(s/n-\theta_S)$ plane that disagree with experimental data on the three mixing angles by more than $3\, \sigma$.
We can observe that the absolute value of $\sin\delta$ has an upper bound $\sim 0.8$ for the choice $n=20$ and $m=11$, whereas the sines of both Majorana phases 
are a priori not constrained.
 Comparing the relative deviations of the sines of the CP phases, $\Delta\sin\alpha$, $\Delta\sin\beta$ and $\Delta\sin\delta$, with the analytical estimates, see eq.~(\ref{eq:estimateDsy01}), we find agreement in the size; notice however that 
 the sign of the relative deviations $\Delta\sin\beta$ and $\Delta\sin\delta$ is positive. 

\begin{figure}[t!]
\begin{center}
\parbox{3in}{\includegraphics*[scale=0.5]{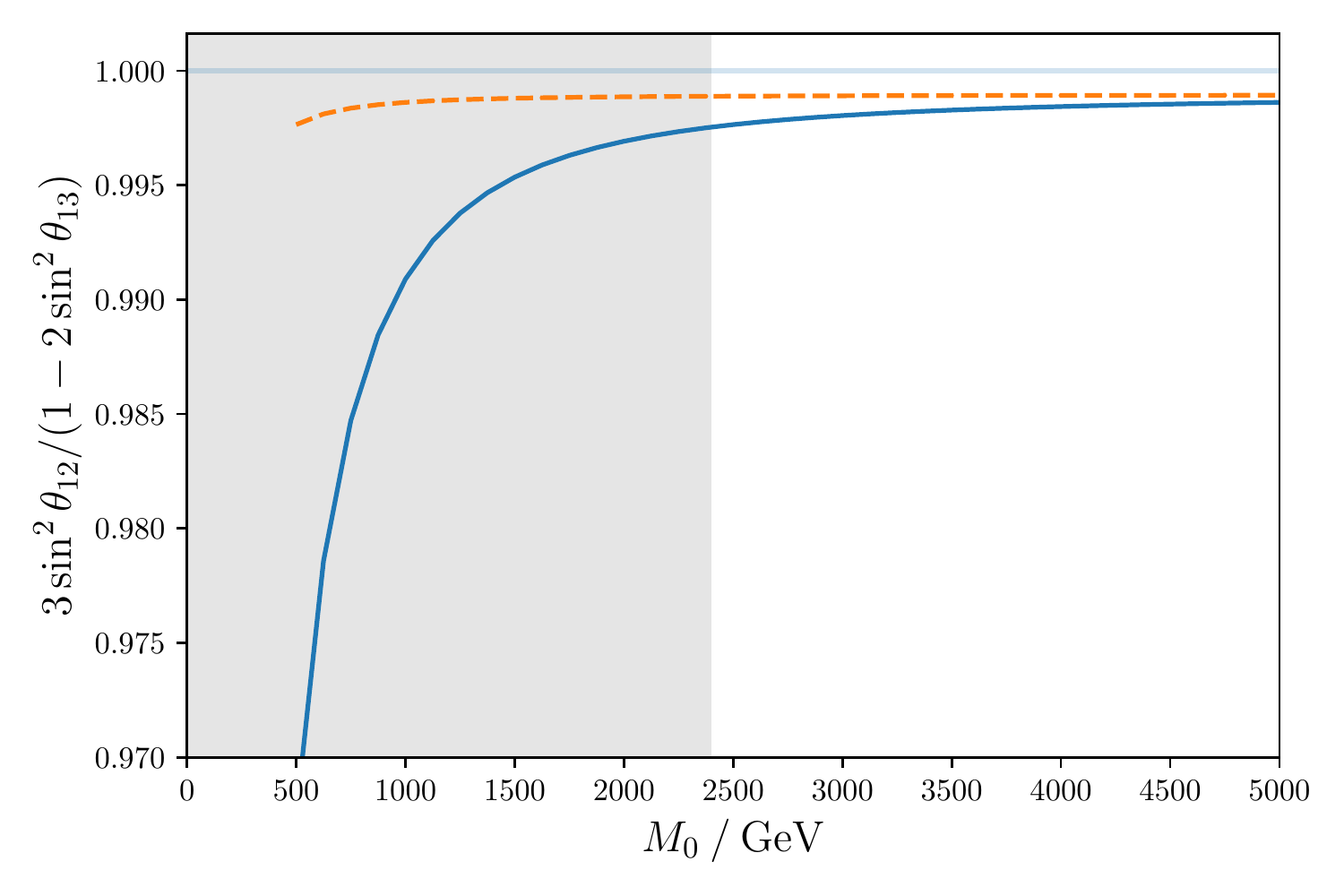}}
\hspace{0.2in}
\parbox{3in}{\includegraphics*[scale=0.5]{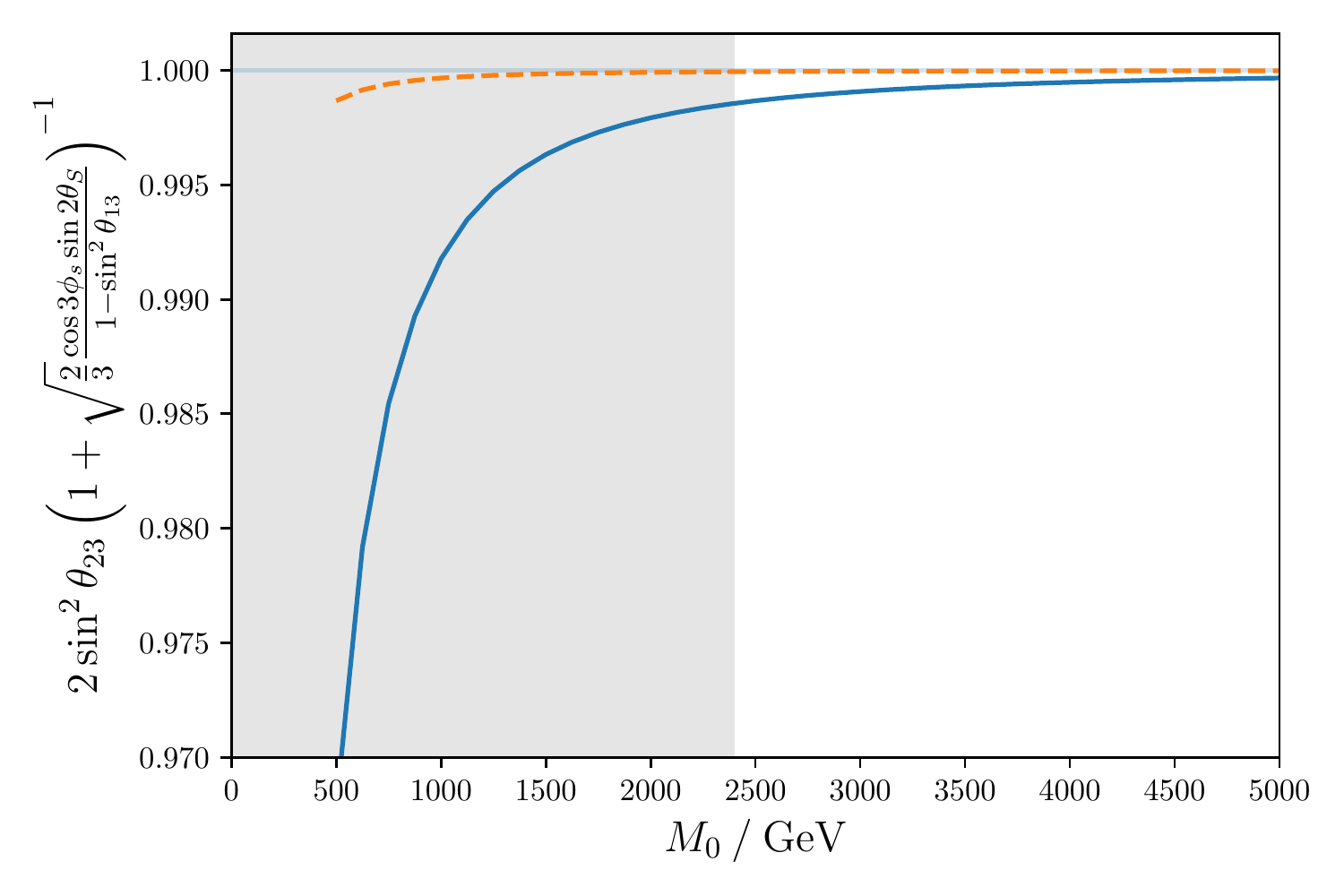}}
\end{center}
\caption{{\small {\bf Case 3 b.1)} 
 Validity check of approximate sum rules in eq.~(\ref{eq:sumrulesCase3b1}) for option 1 of the $(3,3)$ ISS framework 
 with respect to the mass $M_0$ (in GeV).
 In addition to $n=20$ and $m=10$, we fix $s=4$ ($\cos3 \, \phi_s \approx -0.31$) and $\theta_S \approx 1.83$ for the evaluation of both approximate sum rules.
Otherwise, same conventions and colour-coding as in fig.~\ref{fig:Case1_s2thij}.
\label{fig:Case3b1_Sigma45}}}
\end{figure}
As shown in the model-independent scenario, several simplifications of the formulae in eqs.~(\ref{eq:sin2thetaijCase3b1},\ref{eq:JCPI1I2Case3b1}) can be made for $m=\frac n2$ (corresponding to $m=10$ for the present case). 
In particular, two approximate sum rules are found, see eq.~(\ref{eq:sumrulesCase3b1}). In the following, we investigate how these are affected by the presence of the ISS heavy sterile states. We proceed in an analogous way as done for the (approximate) sum rules found for the other cases. Our results are displayed in fig.~\ref{fig:Case3b1_Sigma45} for two different values of the Yukawa coupling, $y_0=0.1$ and $y_0=0.5$, and can be compared to the analytical estimates for the relative deviations 
$\Delta\Sigma_4$ and $\Delta\Sigma_5$, see eqs.~(\ref{eq:sumrule1Case3b1nonuni},\ref{eq:sumrule2Case3b1nonuni},\ref{eq:estimateDSigma45y01})
in section~\ref{sec4}. We note that the results have been obtained for the choice $s=4$ ($\cos3 \, \phi_s \approx -0.31$). This choice has been made since it leads to a value of the atmospheric mixing angle which agrees best with current experimental data~\cite{NuFIT50}. Furthermore, we remark that we have replaced $\theta_0$ by $\theta_S$ in the second approximate sum rule in eq.~(\ref{eq:sumrulesCase3b1})
 which, however, turns out to be very close to $\theta_0\approx 1.83$. 
 As can be seen in fig.~\ref{fig:Case3b1_Sigma45}, for $y_0 \sim 0.5$ and $M_0 \sim 500$~GeV we find a deviation of about $-3 \%$ with respect to the results obtained in the model-independent approach. We thus confirm the analytical expectation (see eq.~\eqref{eq:estimateDSigma45y01}), which was obtained for $y_0 = 1$ and $M_0 = 1000$~GeV (leading to the same value of $\eta_0$, cf. eq.~(\ref{eq:etaopt1})).
For large $M_0$ the displayed ratios may not lead to exactly one, since the two sum rules only hold approximately.

For the choice of $m = \frac{n}{2} = 10$,
we can also check the (approximate) validity of the statements made for the sines of the CP phases and for the lower bound on the absolute value of the sine of the CP phase $\delta$, as observed in the  
model-independent scenario (compare to point $e)$ in section~\ref{sec34}). Indeed, these hold, up to the expected deviations due to the effects of non-unitarity of $\widetilde{U}_\nu$; moreover,  the equality of the sines of the two Majorana phases $\alpha$ and 
$\beta$ still holds exactly (see first equality in eq.~(\ref{eq:sinaplhabetaCase3b1})).

For the choice $m=\frac n2=10$ and additionally $s=\frac n2=10$, one expects from the model-independent scenario that the atmospheric mixing angle and the Dirac phase are maximal, while both Majorana phases are trivial.
This also holds to a very good degree for option 1 of the $(3,3)$ ISS framework, for values of $M_0 \gtrsim 500$~GeV and $y_0 \lesssim 1$. 
In general, in all occasions in which a trivial CP phase is expected in the model-independent scenario, the same is obtained for option 1 of the $(3,3)$ ISS framework.

For Case 3 b.1) we numerically confirm that the symmetry transformations, given in eq.~(\ref{eq:symmtrafosCase3}) under point $i)$ for Case 3 a), also hold. 

\vspace{0.3in}
In summary, we find that the effects of non-unitarity (of the PMNS mixing matrix,  $\widetilde{U}_\nu$) on the lepton mixing parameters and on the (approximate) sum rules relating them, turn out to be below the $1\%$ level, once experimental limits on the quantities $\eta_{\alpha\beta}$
are taken into account, see section~\ref{sec50}. Consequently, the results obtained for option 1 of the $(3,3)$ ISS framework are very similar to those obtained in the model-independent scenario~\cite{HMM}. 
In particular, the dependence of the CP phases on the group theory parameters (especially those determining the CP transformation $X$) and the 
vanishing of a CP phase for certain choices of group theory parameters, are not affected.

\mathversion{bold}
\section{Results for neutrinoless double beta decay}
\mathversion{normal}
\label{sec6}

In the following, we briefly comment on $0\nu\beta\beta$ decay prospects for option 1 of the $(3,3)$ ISS framework.
First, we recall that  in the presence of light neutrinos and of heavy sterile states,  the effective mass $m_{ee}$, accessible in $0\nu\beta\beta$ decay experiments, is given by~\cite{Blennow:2010th}
\begin{equation}
\label{eq:0nubb_HSS}
m_{ee} \simeq \sum_{i=1}^{3+n_s} \, \mathcal U_{e i}^2 \, p^2 \, \frac{m_{ i}}{p^2-m_{i}^2} \simeq \sum_{i=1}^3 \, \mathcal U^2_{e i} \, m_i + \sum_{k=4}^{3+n_s} \mathcal U_{e k}^2 \, p^2 \, \frac{m_{k}}{p^2-m_{k}^2} \; ,
\end{equation}
where $n_s$ denotes the number of heavy sterile states, in our case $n_s=6$, and the virtual momentum $p^2$ is estimated as $p^2 \simeq -(100 \, \mathrm{MeV})^2$. 
For $i=1,2,3$, the mixing matrix elements 
$\mathcal U_{e i}$ coincide with the elements of the first row of the matrix $\widetilde{U}_\nu$ (and hence $\widetilde{U}_{\mathrm{PMNS}}$); for $k=4,..., 3+n_s=4, ..., 9$,  
in our case $\mathcal U_{e k}$ are approximately given by 
\begin{equation}
\label{eq:matchingUek}
\mathcal U_{e k} \approx -i \, \left( \frac{y_0 \, v}{2 \, M_0} \right) \, (U_S)_{1 \, k-3} \;\; \mbox{for} \;\; k=4,..., 6 \;\; \mbox{and} \;\; \mathcal U_{e k} \approx \left( \frac{y_0 \, v}{2 \, M_0} \right) \, (U_S)_{1 \, k-6} \;\; \mbox{for} \;\; k=7,..., 9
\end{equation}
according to the expression for $S$ presented in eq.~(\ref{eq:matSopt1}). 
For $i=1,2,3$ $m_{i}$  correspond to the light neutrino masses; we recall that, according to eq.~\eqref{eq:heavyMopt1} for option 1 of the $(3,3)$ ISS framework, the masses of the heavy sterile states
$m_{k}$ (with $k=4,..., 3+n_s=4, ..., 9$) are approximately degenerate
\begin{equation}
\label{eq:mnkapprox}
m_{k} \approx M_0 \,.
\end{equation}
Thus, we have 
\begin{equation}
\label{eq:0nubb_HSS_alone}
m_{ee} \simeq \sum_{i=1}^3 \, \mathcal U^2_{e i} \, m_i + \left( \frac{p^2 \, M_0}{p^2-M_0^2} \right) \, \left( \frac{y_0^2 \, v^2}{4 \, M_0^2} \right) \, \left( - \sum_{k=1}^3 (U_S)_{1 k}^2 + \sum_{k=1}^3 (U_S)_{1 k}^2 \right) = \sum_{i=1}^3 \, 
\mathcal U^2_{e i} \, m_i \; ,
\end{equation}
implying that the contribution of the heavy sterile states to $m_{ee}$ is very suppressed due to their pseudo-Dirac nature. Consequently, we expect that the results for $m_{ee}$
are very similar to those obtained in the model-independent scenario, as studied for example in~\cite{Hagedorn:2016lva}.
 \begin{figure}[t!]
\begin{center}
\parbox{3in}{\hspace{-0.3in}\includegraphics*[scale=0.55]{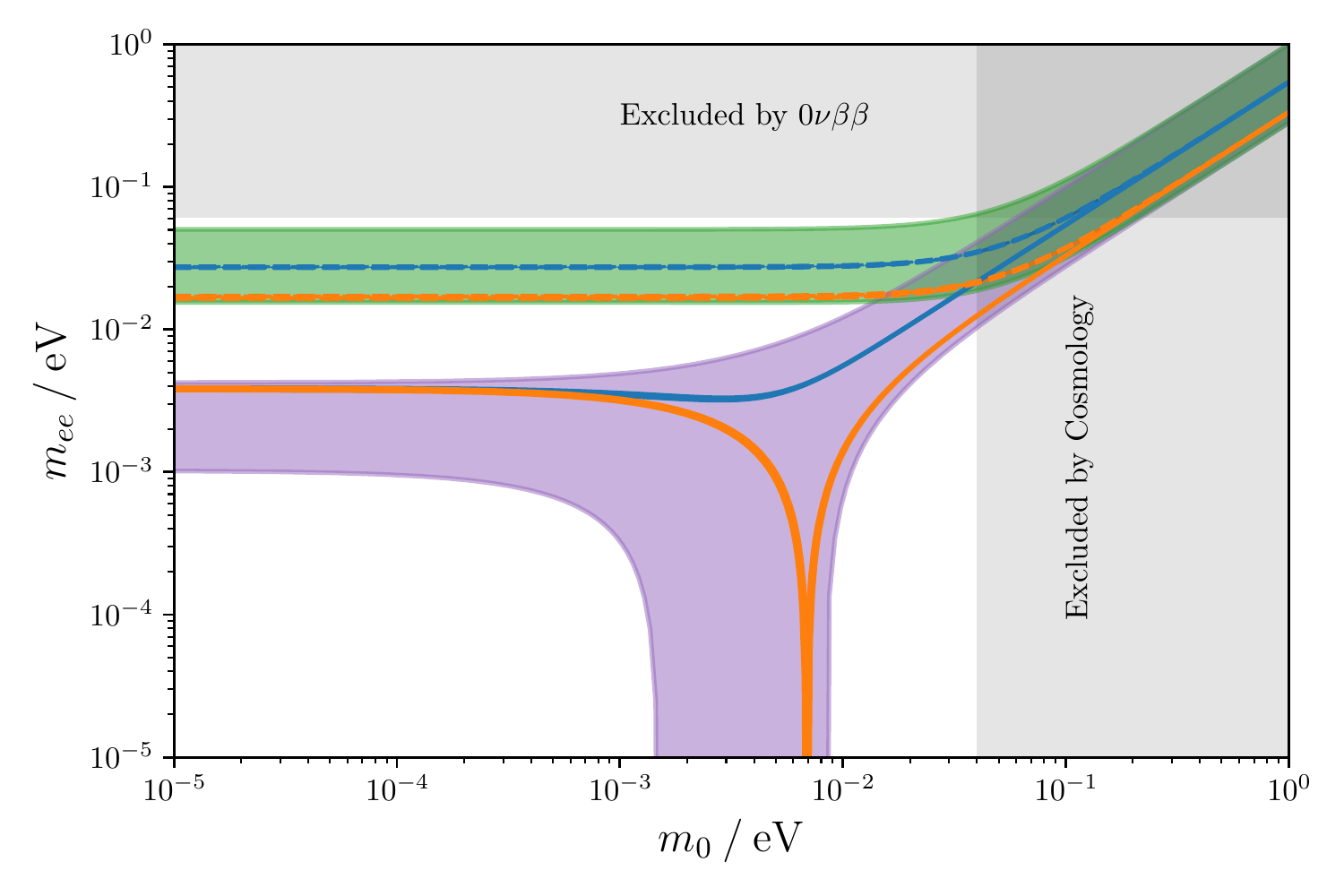}}
\hspace{-0.1in}
\parbox{3in}{\includegraphics*[scale=0.55]{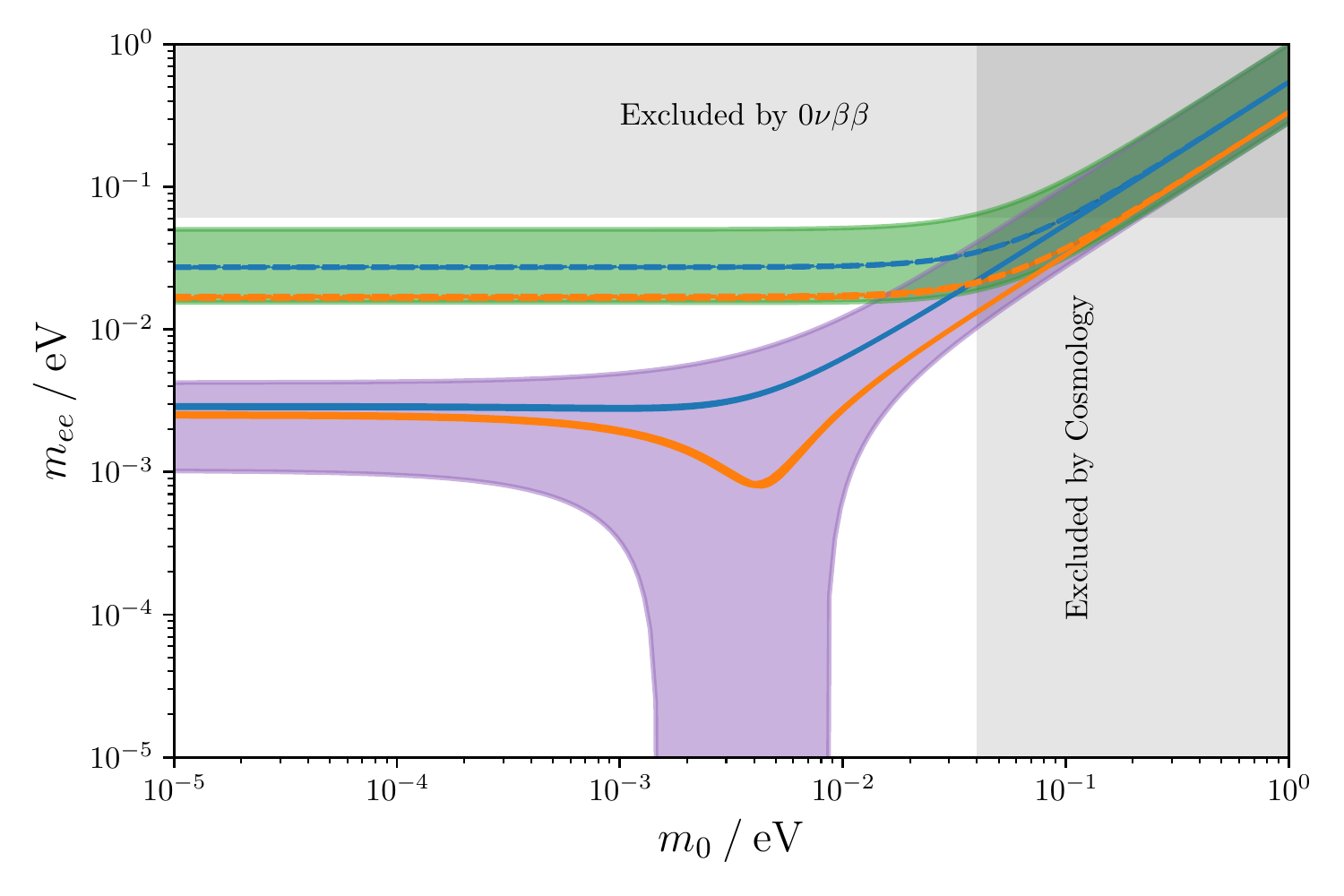}}
\end{center}
\caption{{\small {\bf Results for \mathversion{bold}{$0\nu\beta\beta$} decay for Case 3 b.1)}. Effective mass $m_{ee}$ as a function of the lightest neutrino mass $m_0$ (both in eV)
obtained for option 1 of the $(3,3)$ ISS framework. 
Leading to the results in the left plot, we take $n=20$, $m=10$ and $s=9$ (blue curves) and $s=10$ (orange curves), while the right plot is for $n=20$, $m=11$ and $s=9$ (blue curves) and $s=10$ (orange curves). In both plots  we fix $M_0=1000$~GeV and $y_0=0.1$.
Solid (dashed) curves correspond to a NO (IO)  light neutrino mass spectrum.
The purple (green) shaded area arises upon variation of the lepton mixing parameters and mass squared differences within the experimentally preferred
$3 \, \sigma$ ranges for NO (IO)~\cite{NuFIT50}. 
Values of $m_0$ disfavoured by the cosmological bound on the sum of the light neutrino masses~\cite{Planck2018} are identified by a vertical grey band, while regimes of
$m_{ee}$ already disfavoured by searches
for $0\nu\beta\beta$ decay~\cite{KamlandZen} are indicated by a horizontal grey band.
\label{fig:Case3b1_0nubb}}}
\end{figure}

For completeness, we show two plots for $m_{ee}$ in fig.~\ref{fig:Case3b1_0nubb}, where we have set $y_0=0.1$ and $M_0=1000$~GeV.
These plots were obtained for Case 3 b.1), and  we have chosen $n=20$ like in section~\ref{sec54}.
In the left plot of fig.~\ref{fig:Case3b1_0nubb} we fix $m=10$, while in the right one $m=11$.
In both plots we display results for two values of $s$, $s=9$ (blue) and $s=10$ (orange). 
Solid (dashed) curves correspond to a NO (IO) light neutrino mass spectrum.
We remind that for $s=10$ both Majorana phases turn out to be trivial, thus allowing for the strong 
cancellation observed in association with the orange solid curve in the left plot. The thickness of the curves is determined by the variation
of the mass squared differences in their experimentally preferred $3 \, \sigma$ ranges~\cite{NuFIT50}, see eqs.~(\ref{eq:dm2expNO},\ref{eq:dm2expIO}) in appendix~\ref{app3}.
The purple (green) shaded area arises upon variation of the lepton mixing parameters  and of the mass squared differences within the experimentally preferred
$3 \, \sigma$ ranges for NO (IO). 
The upper bound on the lightest neutrino mass 
$m_0$ arises from the cosmological bound on the sum of the light neutrino masses~\cite{Planck2018},
see eq.~(\ref{eq:sumnuexp}) in appendix~\ref{app3}. 
In fig.~\ref{fig:Case3b1_0nubb} we have depicted the experimental limit on $m_{ee}$ obtained by the KamLAND-Zen Collaboration (using the isotope $^{136} \mathrm{Xe}$)~\cite{KamlandZen}, 
\begin{equation}
\label{eq:meeKamLANDZen}
m_{ee} < (61 \div 165) \, \mathrm{meV} \; ,
\end{equation}
with the above range resulting from different 
theoretical estimates of the nuclear matrix elements. Similar limits have been obtained by other collaborations, for distinct choices of isotopes: 
$m_{ee} < (78 \div 239) \, \mathrm{meV}$ also for $^{136} \mathrm{Xe}$, by EXO-200~\cite{EXO200};
$m_{ee} < (79 \div 180) \, \mathrm{meV}$ for $^{76}\mathrm{Ge}$, as derived by GERDA~\cite{GERDA};
$m_{ee} < (200 \div 433) \, \mathrm{meV}$ also for $^{76} \mathrm{Ge}$ by the Majorana Demonstrator~\cite{MAJORANADemo};  
$m_{ee} < (75 \div 350) \, \mathrm{meV}$ for $^{130} \mathrm{Te}$, obtained by CUORE~\cite{CUORE}.
With further improvement of the experimental limits, certain combinations of group theory parameters in the different cases could be disfavoured, at least, if the light neutrino mass spectrum is assumed to follow IO.

\section{Impact for charged lepton flavour violation}\label{sec:clfv} 
We now proceed to discuss the impact of endowing the (3,3) ISS
realisation with flavour and CP symmetries concerning cLFV
observables, such as radiative and three-body lepton decays,
and neutrinoless $\mu-e$ conversion in matter. 
  
Before addressing the cLFV rates,
it is important to recall that in a regime of sufficiently small
$\mu_i$, the heavy Majorana states are approximately mass-degenerate
in pairs, and have opposite CP-parity, thus effectively leading to the
formation of pseudo-Dirac pairs, whose phases are closely related by (see eq.~\eqref{eq:matSopt1})
\begin{eqnarray}
    \mathcal{U}_{\alpha j } &=& \mathcal{U}_{\alpha j+3}\, e^{i
      (\varphi_{\alpha j} - \varphi_{\alpha j+3})}\:, \quad \text{with} \quad  
    \varphi_{\alpha j} - \varphi_{\alpha j+3} = -\pi/2\,, \label{eq:cLFV:PD:phase}
\end{eqnarray}
in which $\mathcal{U}_{\alpha j }$ are elements of the
unitary nine-by-nine matrix (cf. eq.~\eqref{eq:formBigU}), with $j = 4, 5, 6$ and $\alpha = e, \mu, \tau$. 
For option 1, not only is the mass splitting  extremely
small, typically $\mathcal{O}(1-100\:\mathrm{eV})$ but, as can be
seen from eq.~(\ref{eq:heavyMopt1}), the pseudo-Dirac pairs are
themselves degenerate in mass up to a very good approximation.
In view of the above, the loop functions entering the distinct
observables (see appendix~\ref{sec:loopfunctions}) can be taken
universal for the heavy states,  
$f(x_i) = f(x_0), \, \forall i = 4,5, ..., 9$ with $x_0 \simeq M_0^2/M_W^2$, where $M_W$ is the mass of the $W$-boson.

The full expressions for the cLFV rates arising in SM extensions via
$n_s$ heavy sterile states for the radiative and three-body decays are given by~\cite{Alonso:2012ji}  
\begin{equation}
    \mathrm{BR}(\ell_\beta\to \ell_\alpha \gamma) \,=
    \frac{\alpha_w^3\,
      s_w^2}{256\,\pi^2}\,\frac{m_{\beta}^4}{M_W^4}\,
\frac{m_{\beta}}{\Gamma_{\beta}}\, 
    \left|G_\gamma^{\beta \alpha} \right|^2\:, 
\end{equation}
\begin{eqnarray}
    \mathrm{BR}(\ell_\beta\to 3\ell_\alpha) &=&
    \frac{\alpha_w^4}{24576\,\pi^3}\,\frac{m_{\beta}^4}{M_W^4}\,
\frac{m_{\beta}}{\Gamma_{\beta}}\times\left\{2\left|\frac{1}{2}F_\text{box}^{\beta
      3\alpha} +F_Z^{\beta\alpha} - 2 s_w^2\,(F_Z^{\beta\alpha} -
    F_\gamma^{\beta\alpha})\right|^2 \right.  \nonumber\\ 
     &+& \left. 4 s_w^4\, |F_Z^{\beta\alpha} -
    F_\gamma^{\beta\alpha}|^2 + 16
    s_w^2\,\mathrm{Re}\left[(F_Z^{\beta\alpha} - \frac{1}{2}F_\text{box}^{\beta
        3\alpha})\,G_\gamma^{\beta \alpha
        \star}\right]\right.\nonumber\\ 
     &-&\left. 48 s_w^4\,\mathrm{Re}\left[(F_Z^{\beta\alpha} -
      F_\gamma^{\beta\alpha})\,G_\gamma^{\beta\alpha \star}\right] + 32
    s_w^4\,|G_\gamma^{\beta\alpha}|^2\left[\log\frac{m_{\beta}^2}{m_{\alpha}^2}
      - \frac{11}{4}\right] \right\}\,,  
\end{eqnarray}
in which $m_{\alpha}$ ($\Gamma_\alpha$) denotes the mass (total width)
of a charged lepton of flavour $\alpha$, $\alpha_w = g_w^2/4\pi$ the weak coupling, and $s_w$ the sine of the weak mixing angle.  
Concerning the conversion rate in nuclei, one has~\cite{Alonso:2012ji}   
\begin{equation}
    \mathrm{CR}(\mu - e,\,\mathrm{N}) = \frac{2 G_F^2\,\alpha_w^2\,
      m_\mu^5}{(4\pi)^2\,\Gamma_\text{capt.}}\left|4 V^{(p)}\left(2
    \widetilde F_u^{\mu e} + \widetilde F_d^{\mu e}\right) + 4 V^{(n)}\left(
    \widetilde F_u^{\mu e} + 2\widetilde F_d^{\mu e}\right)  + s_w^2
    \frac{G_\gamma^{\mu e}D}{2 e}\right|^2\,,  
\end{equation}
in which $D$, $V^{(p)}$ and $V^{(n)}$ are nuclear form factors whose
values can be found in Ref.~\cite{Kitano:2002mt} and $e$ is the unit electric charge. For a given nucleus
N, $\Gamma_\text{capt.}$ denotes the capture rate.   
The form factors present in the above equations are given by~\cite{Alonso:2012ji, Ilakovac:1994kj} 
\begin{eqnarray}
    G_\gamma^{\beta \alpha} &=& \sum_{i =1}^{3 + n_s}
    \mathcal{U}_{\alpha i}^{\phantom{\star}}\,\mathcal{U}_{\beta i}^\star\,
    G_\gamma(x_i)\:,\label{eq:cLFV:FF:Ggamma}\\     
    F_\gamma^{\beta \alpha} &=& \sum_{i =1}^{3 + n_s}
    \mathcal{U}_{\alpha i}^{\phantom{\star}}\,\mathcal{U}_{\beta i}^\star
    \,F_\gamma(x_i)\:,\label{eq:cLFV:FF:Fgamma}\\ 
    F_Z^{\beta \alpha} &=& \sum_{i,j =1}^{3 + n_s}
    \mathcal{U}_{\alpha i}^{\phantom{\star}}\,\mathcal{U}_{\beta j}^\star
    \left[\delta_{ij} \,F_Z(x_j) + 
    C_{ij}\, G_Z(x_i, x_j) + C_{ij}^\star \,H_Z(x_i,
    x_j)\right]\:, \label{eq:cLFV:FF:FZ}\\  
    F_\text{box}^{\beta 3 \alpha} &=&\sum_{i,j = 1}^{3+n_s}
    \mathcal{U}_{\alpha i}^{\phantom{\star}}\,\mathcal{U}_{\beta
      j}^\star\left[\mathcal{U}_{\alpha i}^{\phantom{\star}} \,\mathcal{U}_{\alpha
        j}^\star\, G_\text{box}(x_i, x_j) - 2 \,\mathcal{U}_{\alpha
        i}^\star \,\mathcal{U}_{\alpha j}^{\phantom{\star}}\, F_\text{Xbox}(x_i, x_j)
      \right]\:,\label{eq:cLFV:FF:Fbox}\\ 
    F_\text{box}^{\mu e uu} &=& \sum_{i = 1}^{3 + n_s}\sum_{q_d = d, s,
      b} \mathcal{U}_{e i}^{\phantom{\star}}\,\mathcal{U}_{\mu i}^\star\, V_{u q_d}^{\phantom{\star}}\,V_{u
      q_d}^\star \:F_\text{box}(x_i, x_{q_d})\,, 
    \label{eq:cLFV:FF:mueuu}\\
    F_\text{box}^{\mu e dd} &=& \sum_{i = 1}^{3 + n_s}\sum_{q_u = u, c,
      t} \mathcal{U}_{e i}^{\phantom{\star}}\,\mathcal{U}_{\mu i}^\star\, V_{q_u
      d}^{\phantom{\star}}\,V_{q_u d}^\star \:F_\text{Xbox}(x_i, x_{q_u})\,, 
    \label{eq:cLFV:FF:muedd}\\
    \widetilde F^{\mu e}_d &=& -\frac{1}{3}s_w^2 F_\gamma^{\mu e} - F_Z^{\mu e}\left(\frac{1}{4} - \frac{1}{3}s_w^2 \right) + \frac{1}{4}F^{\mu e dd}_\text{box}\\
    \widetilde F^{\mu e}_u &=& \frac{2}{3}s_w^2 F_\gamma^{\mu e} + F_Z^{\mu e}\left(\frac{1}{4} - \frac{2}{3}s_w^2 \right) + \frac{1}{4}F^{\mu e uu}_\text{box}\\
  \text{with } \quad  x_i &=&\frac{m_{i}^2}{M_W^2}\:,\quad x_{q} =
  \frac{m_{q}^2}{M_W^2}\:,\quad C_{ij} = \sum_{\rho = 1}^3
  \mathcal{U}_{i\rho}^\dagger \,\mathcal{U}_{\rho j}^{\phantom{\dagger}}\:. 
\end{eqnarray}
In the above, $i,j=1,...,9$ denote the
neutral lepton mass eigenstates, $\alpha,\beta$ the leptonic flavours,
and $V$ the Cabibbo-Kobayashi-Maskawa quark mixing matrix.  

While the radiative decays ($\ell_\beta \to \ell_\alpha \gamma$) only
call upon $G_\gamma^{\beta \alpha}$, three-body decays ($\ell_\beta
\to 3\ell_\alpha $) depend\footnote{For simplicity, here we only
  focus on identical flavour final states for the three-body cLFV decays,
  although one expects similar results for $\ell_\beta \to \ell_\alpha \ell_\gamma
  \ell_\gamma$ decays.}    
on $F_\text{box}^{\beta 3 \alpha}$, $G_\gamma^{\beta \alpha}$, 
$F_\gamma^{\beta \alpha}$ and $F_Z^{\beta\alpha}$.  
Finally, $\mu-e$ conversion in nuclei involves the latter three
form factors (for $\alpha, \beta = e, \mu$), as well as additional
ones corresponding to box diagrams with an internal quark line,
$F_\text{box}^{\mu e qq}$.  

It is worth noticing that the combination $\sum_{i=4}^9
\mathcal{U}_{\alpha i}^{\phantom{\star}} \,\mathcal{U}_{\beta i}^\star$ can be recast in
terms of the unitarity violation of the PMNS mixing matrix, $\widetilde
U_\nu$. As usually done~\cite{Xing:2007zj},  
one can write  
\begin{equation}\label{eq:cLFV:UtildeAU}
\widetilde{U}_\nu \, = \, A\,U_0\,,
\end{equation}
in which $U_0$ is a unitary three-by-three matrix (see eq.~\eqref{eq:UPMNStildeeta})
and $A$ is a triangular matrix, 
\begin{equation}
A\, =\, 
\begin{pmatrix}
\alpha_{11} & 0& 0\\
\alpha_{21} & \alpha_{22} & 0\\
\alpha_{31} & \alpha_{32} & \alpha_{33}
\end{pmatrix}\,,
\end{equation}
and we define
\begin{equation}
    \mathcal A \equiv \, A \,A^\dagger = \widetilde{U}_\nu^{\phantom{\dagger}} \, \widetilde{U}_\nu^\dagger\,.
\end{equation}
Recalling the definition of the quantity $\eta$ (see
eqs.~(\ref{eq:UPMNStildeeta},\ref{eq:UPMNStildeeta2})), it is manifest
that one has  
\begin{equation}
\mathcal{A}_{\alpha \beta} \, =\, \delta_{\alpha \beta} - 2 \eta_{\alpha \beta}\,.
\end{equation}
Unitarity of the full nine-by-nine matrix $\mathcal{U}$ implies that 
\begin{equation}
\sum_{i=4}^9 \mathcal{U}_{\alpha i}^{\phantom{\star}}\,\mathcal{U}_{\beta i}^\star
=\delta_{\alpha\beta} - \sum_{i=1}^3 \mathcal{U}_{\alpha
  i}^{\phantom{\star}}\,\mathcal{U}_{\beta i}^\star =\delta_{\alpha\beta} - (\widetilde{U}_\nu^{\phantom{\dagger}}\, \widetilde{U}_\nu^\dagger)_{\alpha \beta} =\delta_{\alpha \beta} -
\mathcal{A}_{\alpha \beta} = 2\, \eta_{\alpha \beta}\,.
\end{equation}
For option 1 one has $\eta_{\alpha \beta} =0$, $\forall
\alpha\neq \beta$, so that  
$\eta$ and thus $\mathcal{A}$ (and also $A$) are diagonal. This is of
paramount importance for the cLFV observables, since - and as
discussed below - any contribution proportional to $\eta_{\alpha
  \beta}$ will vanish (for $\alpha\neq \beta$).

\mathversion{bold}
\subsection{Dipole terms - radiative decays $\mathbf{\ell_\beta \to
    \ell_\alpha \gamma}$}  
\mathversion{normal}
Since the contribution of the light (mostly active) neutrinos to the
dipole form factor can be neglected 
(the relevant limits of the loop functions can be found in
appendix~\ref{sec:loopfunctions}) one has
\begin{equation}
G_\gamma^{\beta \alpha} \, \simeq\, G_\gamma (x_0) 
\sum_{i=4}^9 \mathcal{U}_{\alpha i}\,\mathcal{U}_{\beta i}^\star\,,
\end{equation}
or, and in view of the above discussion, 
\begin{equation}
G_\gamma^{\beta \alpha} \, \simeq\, 
G_\gamma (x_0)\, \left(\delta_{\alpha \beta } -
\mathcal{A}_{\alpha \beta} \right) \,=\, 2\, G_\gamma (x_0)\,  
\eta_{\alpha \beta}\,.
\end{equation}
As an illustrative example, for radiative cLFV muon decays one has 
$G_\gamma^{\mu e} = -G_\gamma (x_0)\, \alpha_{11}\,\alpha_{21}^\star$.
For the present scenario, in which $\mathcal{A}$ and $\eta$
are diagonal, one thus finds $G_\gamma^{\beta \alpha} \, \simeq\,0\,$.

In line with the analytical discussion on lepton mixing carried in
section~\ref{sec22}, let us also emphasise that similar results can be
obtained relying on the approximate analytical expression for $S$, which for
option 1 is given at leading order in $\mu_S/M_{NS}$ 
by
$    S \,\simeq \,\frac{y_0\, v}{2\,M_0}\,(- i U_S,\, U_S)$ (see eq.~(\ref{eq:matSopt1})).
The form factor can be recast as 
\begin{align}
    G_\gamma^{\mu e} &\simeq G_\gamma(x_0)\left\{\sum_{i = 4}^6
    \mathcal{U}_{ei} \,\mathcal{U}_{\mu i}^\star +  \sum_{i = 7}^9
    \mathcal{U}_{ei}\, \mathcal{U}_{\mu i}^\star\right\} \nonumber\\ 
    &\propto G_\gamma(x_0)\left\{\sum_{i = 1}^3 (i U_S)_{ei}
    \,(iU_S)_{\mu i}^\star +  \sum_{i = 1}^3 (U_S)_{ei}\, (U_S)_{\mu
      i}^\star\right\} \,= \,2 G_\gamma(x_0) \left\{\sum_{i = 1}^3
    (U_S)_{ei} \,(U_S)_{\mu i}^\star \right\}  = 0\:,
\end{align}
due to the orthogonality of the ($\mu - e$) rows of $U_S$ (which we
recall to be a unitary matrix, determined by the group theory parameters and
the free angle $\theta_S$). 

\mathversion{bold}
\subsection{Photon and $Z$ penguin form factors} 
\mathversion{normal}

Relevant for both ${\ell_\beta \to 3 \ell_\alpha}$
and $\mu-e$ conversion, these
include several contributions (reflecting the fact that two neutral
fermions can propagate in the loop).

A reasoning analogous to the one conducted for 
$G_\gamma^{\beta \alpha}$ leads to 
$F_\gamma^{\beta \alpha} \, \simeq\, F_\gamma (x_0)\, 
\left( \delta_{\alpha \beta} -\mathcal{A}_{\alpha \beta} \right)\,=\,
2\,F_\gamma (x_0)\, \eta_{\alpha\beta}$, which thus vanishes for the
flavour violating decays.
Likewise, the first term on the right-hand side of eq.~(\ref{eq:cLFV:FF:FZ})
leads to the same result, 
\begin{equation}
\sum_{i,j =1}^{9}
\mathcal{U}_{\alpha i}\,\mathcal{U}_{\beta j}^\star \left[\delta_{ij}
  F_Z(x_i) \right] \simeq 
F_Z(x_0)\left( \delta_{\alpha \beta} -\mathcal{A}_{\alpha \beta}
\right)\,=\, 2F_Z(x_0)\, \eta_{\alpha\beta}\,=\, 0\, .  
\end{equation}

Both terms associated with $G_Z(x,y)$ and $H_Z(x,y)$ correspond
to two neutral leptons
propagating in the loop. Although the loop functions do tend to zero
for the case of very light internal fermions, the same does not occur for
$G_Z$ if at least one of the states is heavy, i.e. $G_Z (0,x_i)$ (see
appendix~\ref{sec:loopfunctions}).
Introducing the following limits for the loop functions,
\begin{equation}
{\bar G}_Z(x) \, =\, \lim_{x_i \gg 1} G_Z (0,x_i)\, ,
\quad \quad 
{\bar{\bar G}}_Z(x) \, =\, \lim_{x_i \approx x_j \gg 1} G_Z (x_i,x_j)
\,,
\end{equation}
respectively corresponding to ``heavy-light" and
``heavy-heavy" (combinations of) fermion propagators, one thus has
\begin{equation}
\sum_{i,j = 1}^{9}
\mathcal{U}_{\alpha i}\,\mathcal{U}_{\beta j}^\star 
\,C_{ij}\, G_Z(x_i, x_j)\simeq
{\bar G_Z(x_0)} \left[ 2 \mathcal{A}
  \,(\mathbb{1}-\mathcal{A})\right]_{\alpha \beta} 
+
{\bar{\bar G}}_Z(x_0) \left[ (\mathbb{1}-\mathcal{A})^2\right]_{\alpha \beta},
\end{equation}
or in terms of $\eta$,
${\bar G_Z(x_0)} \left[ 4(\mathbb{1}-2\eta)\,\eta\right]_{\alpha
      \beta} + 4 \,{\bar{\bar G}}_Z(x_0) \,
[\eta^2]_{\alpha\beta}\,=\,0$,
as previously argued.

For the $H_Z(x,y)$-associated terms, only the ``heavy-heavy" case (two
heavy sterile states in 
the loop) can potentially contribute in a non-negligible way.
However, the corresponding contribution also vanishes, 
as a consequence of the nature of the (degenerate) heavy states, which
as mentioned
form pseudo-Dirac pairs. 
Defining (see appendix~\ref{sec:loopfunctions})   
\begin{equation}
{\bar{\bar H}}_Z(x) \, =\, \lim_{x_i \approx x_j \gg 1} H_Z (x_i,x_j)\,,
\end{equation}
one then finds (taking into account eq.~(\ref{eq:cLFV:PD:phase}))
\begin{align}
&\sum_{i,j =1}^{9}
\mathcal{U}_{\alpha i}\,\mathcal{U}_{\beta j}^\star 
\,C_{ij}^\star\, H_Z(x_i, x_j) \simeq
{\bar{\bar H}}_Z(x_0) \sum_{i,j =4}^{9} 
\sum_{\rho =1}^{3} 
\mathcal{U}_{\alpha i}\,\mathcal{U}_{\rho i}\,
\mathcal{U}_{\beta j}^\star\,\mathcal{U}_{\rho j}^\star  \nonumber \\
& \simeq
{\bar{\bar H}}_Z(x_0) \sum_{\rho =1}^{3}
\left\{\left[
\sum_{i =4}^{6}  \left(\mathcal{U}_{\alpha i}\,
\mathcal{U}_{\rho i} +
e^{i \pi/2} \mathcal{U}_{\alpha i}\,
e^{i \pi/2} \mathcal{U}_{\rho i}
\right)\right] \Bigg[ 
\sum_{j =4}^{6} \left(\mathcal{U}_{\beta j}^\star\,
\mathcal{U}_{\rho j}^\star +
e^{-i \pi/2}\mathcal{U}_{\beta j}^\star\,
e^{- i \pi/2}\mathcal{U}_{\rho j}^\star
\right)
\Bigg]\right\}\,=\, 0\,,
\end{align}
which is a direct consequence of the pseudo-Dirac nature of the heavy states.

\subsection{Box diagrams}
Several form factors contribute to both the three-body decays
$\ell_\beta \to 3 \ell_\alpha$, and neutrinoless $\mu-e$
conversion. The first ($F^{\beta3\alpha}_\text{box}$) can be decomposed in two terms,
``box" and cross-box ``Xbox", respectively associated with the loop
functions $G_\text{box}$ and $F_\text{Xbox}$. Similar contributions
(single internal neutral lepton) are present for the latter ($F^{\mu e
qq}_\text{box}$).

Only diagrams with two heavy neutrinos are at the source of 
non-vanishing contributions to $G_\text{box}$; however, and
analogously to what occurred for the previously discussed
$H_Z(x,y)$-associated terms, the contributions vanish, 
due to having the heavy states forming, to an excellent
approximation, pseudo-Dirac pairs. 

A priori, one can have contributions to  the $F_\text{Xbox}$ form factors
from ``light-light" and ``heavy-heavy" fermion propagators in
the box. However, both turn out to be proportional to $\mathcal{A}_{\alpha \beta}$ and thus to $\eta_{\alpha \beta}$, and
are hence vanishing.

The additional form factors relevant for 
${\mu-e}$ conversion, $F^{\mu eqq}_\text{box}$ lead to contributions again
proportional to $\eta_{e \mu}$, thus also vanishing in the present scenario.

\mathversion{bold}
\subsection{cLFV for option 1 of the $(3,3)$ ISS with flavour and CP symmetry}
\mathversion{normal}
In the present scenario, no new
contributions to the different cLFV observables due to the exchange of
heavy states are expected.\footnote{Numerical evaluations confirm that
  the rates are typically $\mathcal{O}(10^{-50})$.} Such a
``stealth" realisation of the ISS - which in general can account for
significant contributions to the observables, well within    
experimental sensitivity - is due to two peculiar features of 
option 1. First and most importantly, recall that here $\mu_S$ is the unique source of
flavour violation in the sector of neutral states; this is in contrast with other
ISS realisations in which the Dirac neutrino Yukawa couplings (and possibly
$M_{NS}$) are non-trivial in flavour space.  
Moreover, notice that for option 1 of the flavour symmetry-endowed ISS the heavy mass spectrum is composed of three degenerate
pseudo-Dirac pairs (to an excellent approximation), which further
suppresses any new contribution.  

Thus, cLFV processes will not offer any additional source of insight
in what concerns the underlying discrete flavour symmetries nor the
mass scale of the heavy states; however 
the observation of at least one cLFV transition would
strongly disfavour the 
flavour symmetry-endowed ISS in its option 1, with strictly
diagonal and universal $M_{NS}$ and $m_D$ in flavour space. 

\section{Summary and outlook}
\label{summ}
We have considered an inverse seesaw mechanism with $3+3$ heavy
sterile states, endowed with a flavour symmetry $G_f=\Delta (3 \,
n^2)$ or $G_f=\Delta (6 \, n^2)$ 
and a CP symmetry. The
peculiar breaking of the flavour and CP symmetry to different residual
symmetries $G_\ell$ in the charged lepton 
sector and $G_\nu$ in the sector of the neutral states, is 
the key to rendering this scenario predictive (and possibly testable). 
In the inverse seesaw mechanism, several terms in the Lagrangian 
determine the mass spectrum of the neutral states, in association with  
three matrices, $m_D$, $M_{NS}$ and $\mu_S$. 
Several realisations of the residual symmetry $G_\nu$ are possible, and here we have focused on one of the three minimal options, which we have
called ``option 1''. 
In this option only the Majorana mass matrix $\mu_S$ breaks $G_f$ and
CP to $G_\nu$, while $m_D$ and $M_{NS}$ preserve $G_f$ and CP.
In the sector of the neutral states, lepton number and lepton flavour violation are thus
both encoded in $\mu_S$.  
Left-handed lepton doublets and the $3+3$ heavy sterile states are assigned to the same triplet 
${\bf 3}$ of $G_f$, whereas right-handed
charged leptons are in singlets. 

In~\cite{HMM} mixing patterns arising from the breaking of $G_f$ and CP to $G_\ell$
and $G_\nu$ have been analysed and four of them have been identified as particularly interesting for leptons. We have studied examples of lepton mixing for each of the different mixing patterns, Case 1) through Case 3 b.1), both analytically and numerically.
For option 1, a significant consequence of the presence of the heavy
sterile states is that for certain regimes there is a sizeable
deviation from unitarity of the PMNS mixing matrix, and thus potential
conflict with the associated experimental bounds. 
This leads to stringent constraints on the Yukawa coupling $y_0$ and on the mass scale $M_0$, 
so that regimes of large $y_0$ and small $M_0$ are disfavoured.
In the viable regimes, the impact of the heavy sterile states on lepton mixing turns out to be
 small:
deviations typically below $1\%$ are found upon comparison of the results of the $(3,3)$ ISS framework to those derived in the model-independent scenario.
We have also discussed the potential impact of this ISS framework for
several observables. An interesting implication of option 1 here
discussed is that the heavy sterile states are degenerate to a very
good approximation, and combine to form three pseudo-Dirac pairs. 
As a consequence, the results for neutrinoless double
beta decay are hardly modified, compared to results obtained in the model-independent scenario.
We have also addressed in detail charged lepton flavour violating
processes: in sharp contrast to what generally occurs for 
inverse seesaw models (see, e.g.~\cite{Abada:2015oba,Abada:2014kba}), 
the cLFV rates are highly
suppressed, similar to what occurs in the
Standard Model with three light (Dirac) neutrinos. 
This is a consequence of having strictly flavour-diagonal and flavour-universal deviations from unitarity of the PMNS mixing matrix (and also due to a very high degree of 
degeneracy in the heavy mass spectrum).

Throughout this work we have assumed that the desired breaking of the flavour and CP symmetries can be realised,  and that the appropriate residual symmetries are preserved by the different mass matrices.
As has been shown in the literature,
it is possible to achieve the breaking of flavour (and CP) 
in different ways, e.g.~spontaneously, if flavour (and CP) symmetry breaking fields acquire non-vanishing vacuum expectation values, in supersymmetric theories (see for instance~\cite{JKCH2}), or explicitly via boundary conditions
in a model with an extra dimension (see e.g.~\cite{MSCH1,MSCH2}). The predictive power of concrete models 
is usually higher than the one of the model-independent approach: for example, by choosing a certain set of flavour (and CP) symmetry breaking fields, the ordering of the light neutrino mass  spectrum
can be predicted, and by extending the flavour (and CP) symmetry to the flavour sector of the new particles, as for instance supersymmetric particles or Kaluza-Klein states, many flavour observables can be constrained and correlated. 
It is thus interesting to consider the construction of 
such models. 

It is well-known that in concrete models
corrections to the desired breaking of flavour (and CP) can arise. This can for instance be the case if flavour (and CP) symmetry breaking fields, whose vacuum expectation values preserve the residual symmetry $G_\ell$, couple 
at a higher order to the neutral states as well. We have not discussed such corrections in our analysis, but we can briefly comment on their expected impact on lepton mixing as well as
predictions for branching ratios of different charged lepton flavour violating processes. Considering, for example, that corrections invariant under $G_\ell$ contribute to the mass matrices $m_D$ and $M_{NS}$,
we expect that lepton mixing can still be correctly explained for corrections not larger than a few percent\footnote{See also~\cite{MSCH1, MSCH2} for a similar analysis in the context of a type-I seesaw 
mechanism, implemented in a model with a warped extra dimension and a flavour symmetry $G_f$.} and possibly by re-fitting the value of the free angle $\theta_S$. 
At the same time, the branching ratios of charged lepton flavour violating processes would still remain strongly suppressed, beyond the reach of current and future experiments.\footnote{Notice that corrections that are invariant under $G_\ell$ only contribute to the diagonal entries of $m_D$ and $M_{NS}$, so that even in the presence of the latter the matrix $\eta$ will still be diagonal (see eq.~\eqref{eq:eta}). Moreover, 
in the considered mass regime, the dominant loop functions have an asymptotic logarithmic behaviour (or are even constant, cf. appendix~\ref{sec:loopfunctions}), thus being insensitive to percent 
level changes in the mass splitting of the heavy states; 
this thus still leads to a strong Glashow–Iliopoulos–Maiani (GIM) cancellation in the cLFV rates.}

As mentioned, here we have focused on one of the three minimal options to
realise the residual symmetry $G_\nu$ in the sector of the neutral
states. It could be interesting to analyse 
 lepton mixing, as well as neutrinoless double beta decay, effects of
 non-unitarity of the PMNS mixing matrix $\widetilde{U}_\nu$, and charged lepton flavour violating processes for the
 other two options, called option 2 and option 3. 
Both these options
could potentially lead to larger effects in charged lepton flavour violating
processes. For option 2 the non-trivial flavour structure is encoded in the
Dirac neutrino mass matrix $m_D$ and thus strongly resembles ISS
constructions typically associated with sizeable predictions to
numerous leptonic observables.  
Furthermore, a non-trivial flavour structure in $M_{NS}$ for option 3 also leads to off-diagonal terms in $\eta$, thus potentially having a strong impact on cLFV processes.
Should this be the case, a study of possible correlations among the lepton mixing parameters and the different charged lepton flavour violating processes for the distinct cases (Case 1) through Case 3 b.1)) could be valuable and may even help testing the hypotheses of $G_f$, CP and the residual symmetries $G_\ell$ and
$G_\nu$.   
Going beyond the three minimal options, 
we can also consider options,
in which at least two of the three mass matrices $m_D$, $M_{NS}$ and
$\mu_S$ carry non-trivial flavour information.  

Further variants could be also envisaged. These could include versions of the inverse
seesaw mechanism, for instance with two right-handed neutrinos $N_i$ and two (three)
neutral states $S_j$~\cite{ISScategories}, or even a minimal radiative inverse seesaw mechanism~\cite{MRISS}.
The latter generates light neutrino masses at the one-loop level and could offer a dark matter candidate. This might be an appealing playground for a scenario with $G_f$ and CP since the same symmetries could govern the phenomenology of both the visible and the dark sector.

\section*{Acknowledgements}
C.H. acknowledges the warm hospitality of the theory group at the Laboratoire de Physique de Clermont, Universit\'e{} Clermont Auvergne, in the beginning of this project, and thanks Juan Herrero-Garc\'i{}a and Jacobo
L\'o{}pez-Pav\'o{}n for interesting comments.
C.H. has been partly 
supported by the European Union's Horizon 2020 research and innovation programme under the Marie Sk\l{}odowska-Curie grant agreement No. 754496 (FELLINI programme) 
and is supported by Spanish MINECO through the Ram\'o{}n y Cajal programme RYC2018-024529-I and by the national grant FPA2017-85985-P. 
J.K. and A.M.T. are grateful to C. Weiland for enlightening discussions
and comments.
This project has received support from the European Union's Horizon 2020 research and innovation programme under the Marie Sk\l{}odowska-Curie grant agreement No.~860881 (HIDDe$\nu$ network).
J.K., J.O., and A.M.T also acknowledge support from the IN2P3 (CNRS) Master Project, ``Flavour probes: lepton sector and beyond'' (16-PH-169).

\appendix

\mathversion{bold}
\section{Generators of $\Delta (3 \, n^2)$ and $\Delta (6 \, n^2)$ and CP transformations $X$}
\mathversion{normal}
\label{app1}
This appendix contains necessary information concerning the flavour groups $\Delta (3 \, n^2)$ and $\Delta (6 \, n^2)$. Since the former is a subgroup of the latter, we
 focus on $\Delta (6 \, n^2)$. The groups $\Delta (3 \, n^2)$ and $\Delta (6 \, n^2)$ are series of discrete symmetries for integer $n$. 
 For $n \geq 2$, 
 $\Delta (3 \, n^2)$ is non-abelian, while all groups $\Delta (6 \, n^2)$ have this property. The groups $\Delta (3 \, n^2)$ are isomorphic to the semi-direct product $(Z_n \times Z_n) \rtimes Z_3$ 
 and can be described in terms of three generators $a$, $c$ and $d$ that fulfil the following relations
\begin{equation}
\label{eq:gen1}
a^3=e \; , \;\; c^n=e \; , \;\; d^n=e \; , \;\;
a \, c \, a^{-1} = c^{-1} \, d^{-1} \; , \;\; a \, d \, a^{-1} = c \; , \;\; c \, d= d \, c
\end{equation} 
with $e$ being the neutral element of the group. For the groups $\Delta (6 \, n^2)$ that are isomorphic to $(Z_n \times Z_n) \rtimes S_3$, one adds the fourth generator $b$
to the set $\{ a, c, d\}$ which fulfils the relations
\begin{equation}
\label{eq:gen2}
b^2=e \; , \;\; (a \, b)^2=e \; , \;\; b \, c \, b^{-1} = d^{-1} \;\; \mbox{and} \;\; b \, d \, b^{-1} = c^{-1} \; .
\end{equation}
We note that all elements of the groups can be written in terms of these generators as
\begin{equation}
\label{eq:eleGf}
g= a^\alpha \, c^\gamma \, d^\delta \;\; \mbox{and} \;\; g= a^\alpha \, b^\beta \, c^\gamma \, d^\delta \;\; \mbox{with} \;\; \alpha=0,1,2, \; \beta=0, 1, \; 0 \leq \gamma, \delta \leq n-1 \; ,
\end{equation}
respectively.
For the analysis of lepton mixing we are interested in the generators in the irreducible faithful (complex) three-dimensional representation ${\bf 3}$ and in the (trivial) singlet ${\bf 1}$. 
For ${\bf 3}$
 we have 
\begin{eqnarray}
\label{eq:genrep3}
&&a ({\bf 3})  = \left( \begin{array}{ccc}
1 & 0 & 0\\
0 & \omega & 0\\
0 & 0 & \omega^2
\end{array}
\right) \; , \;\; b ({\bf 3}) = \left( \begin{array}{ccc}
1 & 0 & 0\\
0 & 0 & \omega^2 \\
0 & \omega & 0
\end{array}
\right) \; ,
\\ \nonumber
&&c ({\bf 3}) =\frac 13 \, \left( \begin{array}{ccc}
1 + 2 \, \cos \phi_n & 1- \cos \phi_n-\sqrt{3} \, \sin \phi_n & 1- \cos \phi_n+\sqrt{3} \, \sin \phi_n \\
1- \cos \phi_n+\sqrt{3} \, \sin \phi_n & 1 + 2 \, \cos \phi_n & 1- \cos \phi_n-\sqrt{3} \, \sin \phi_n\\
1- \cos \phi_n-\sqrt{3} \, \sin \phi_n & 1- \cos \phi_n+\sqrt{3} \, \sin \phi_n & 1 + 2 \, \cos \phi_n
\end{array}
\right)  
\end{eqnarray}
with $\omega= e^{\frac{2 \, \pi \, i}{3}}$ and $\phi_n= \frac{2 \, \pi}{n}$, while for {\bf 1} we have
\begin{equation}
\label{eq:genrep1}
a ({\bf 1}) = b ({\bf 1}) = c ({\bf 1}) =1 \; .
\end{equation}
We note that the generator $d$ can be obtained from the generators $a$ and $c$, since we find $d= a^2 \, c \, a$ from eq.~(\ref{eq:gen1}).

\vspace{0.1in}
\noindent For completeness, we list the set of used CP symmetries.
CP symmetries are associated with the automorphisms of the flavour group $G_f$. In particular, the automorphism
\begin{equation}
\label{eq:X0auto}
a \; \rightarrow \; a \; , \;\; c \; \rightarrow \; c^{-1} \; , \;\; d \; \rightarrow \; d^{-1} \;\; \mbox{and} \;\; b \; \rightarrow \; b
\end{equation}
for $G_f =\Delta (3 \, n^2)$ and $G_f=\Delta (6 \, n^2)$, respectively, corresponds to the CP transformation $X_0$ that is of the following form in the representations ${\bf 1}$ and ${\bf 3}$ 
\begin{equation}
\label{eq:X0}
X_0 ({\bf 1}) = 1 \;\; \mbox{and} \;\; X_0 ({\bf 3}) = \left( \begin{array}{ccc}
1 & 0 & 0\\
0 & 0 & 1\\
0 & 1 & 0
\end{array}
\right) 
\; .
\end{equation}
All other CP transformations $X$ of interest correspond to the composition of the automorphism in eq.~(\ref{eq:X0auto}) and a group transformation $g$. The CP transformation $X (\mathrm{{\bf r}})$ in the representation $\mathrm{{\bf r}}$ is of the form
\begin{equation}
\label{eq:Xr}
X (\mathrm{{\bf r}}) = g (\mathrm{{\bf r}}) \, X_0 (\mathrm{{\bf r}}) \;\; \mbox{with} \;\; g (\mathrm{{\bf r}}) = a (\mathrm{{\bf r}})^\alpha \, c(\mathrm{{\bf r}}) ^\gamma \, d(\mathrm{{\bf r}}) ^\delta \;\; \mbox{and} \;\; g (\mathrm{{\bf r}}) = a(\mathrm{{\bf r}})^\alpha \, b(\mathrm{{\bf r}})^\beta \, c(\mathrm{{\bf r}})^\gamma \, d(\mathrm{{\bf r}})^\delta
\end{equation}
for $G_f=\Delta (3 \, n^2)$ and $G_f=\Delta (6 \, n^2)$, respectively, as long as $X (\mathrm{{\bf r}})$ represents a symmetric matrix in flavour space, see eq.~(\ref{eq:Xrcond}). 
The CP symmetries and transformations relevant for the different cases, Case 1) through Case 3 b.1), were given in section~\ref{sec3}.

\section{Conventions of mixing angles, CP invariants and neutrino masses}
\label{app2}

We follow the conventions of the PDG in the parametrisation of a unitary mixing matrix ($W$) in terms of 
the lepton mixing angles and the Dirac phase $\delta$~\cite{PDG}
\begin{equation}
\label{eq:defW}
W=
\begin{pmatrix}
c_{12} c_{13} & s_{12} c_{13} & s_{13} e^{- i \delta} \\
-s_{12} c_{23} - c_{12} s_{23} s_{13} e^{i \delta} & c_{12} c_{23} - s_{12} s_{23} s_{13} e^{i \delta} & s_{23} c_{13} \\
s_{12} s_{23} - c_{12} c_{23} s_{13} e^{i \delta} & -c_{12} s_{23} - s_{12} c_{23} s_{13} e^{i \delta} & c_{23} c_{13}
\end{pmatrix}
\end{equation}
with $s_{ij}=\sin \theta_{ij}$ and $c_{ij}=\cos \theta_{ij}$,
while we define the Majorana phases $\alpha$ and $\beta$ through
\begin{equation}
\label{eq:defP}
P=\left( \begin{array}{ccc}
 1  & 0 & 0\\
 0  &  e^{i \alpha/2}  & 0\\
 0 & 0  & e^{i (\beta/2 + \delta)}
\end{array}
\right)
\end{equation}
so that 
\begin{equation}
\label{eq:UPMNSWP}
U_{\mathrm{PMNS}} = \left( \begin{array}{ccc}
U_{e1} & U_{e2} & U_{e3}\\
U_{\mu1} & U_{\mu2} & U_{\mu3}\\
U_{\tau1} & U_{\tau2} & U_{\tau3}
\end{array}
\right)
= W \, P
\end{equation}
with $0 \leq \theta_{ij} \leq \pi/2$ and $0 \leq \alpha, \beta, \delta \leq 2 \, \pi$. We extract the sine squares of the lepton mixing angles as follows
\begin{equation}
\label{eq:sin2thij}
\sin^2 \theta_{13} = |U_{e3}|^2 \; , \;\; \sin^2 \theta_{12} = \frac{|U_{e2}|^2}{1-|U_{e3}|^2} \; , \;\; \sin^2 \theta_{23} = \frac{|U_{\mu3}|^2}{1-|U_{e3}|^2} \, .
\end{equation}
The CP phases are most conveniently extracted with the help of the CP invariants $J_{\mathrm{CP}}$~\cite{jcp}, $I_1$ and $I_2$~\cite{Jenkins_Manohar_invariants}
\begin{equation}
\label{eq:JCP}
J_{\mathrm{CP}} = {\rm Im} \left[ U_{e1} U_{e3}^\star U_{\tau 1}^\star U_{\tau 3}  \right]
= \frac 18 \, \sin 2 \theta_{12} \, \sin 2 \theta_{23} \, \sin 2 \theta_{13} \, c_{13} \, \sin \delta 
\end{equation}
and 
\begin{equation}
\label{eq:I1I2}
I_1 = {\rm Im} [U_{e2}^2 (U_{e1}^\star)^2] = s^2_{12} \, c^2 _{12} \, c^4_{13} \, \sin \alpha \; , \;\;
I_2 = {\rm Im} [U_{e3}^2 (U_{e1}^\star)^2] = s^2 _{13} \, c^2 _{12} \, c^2_{13} \, \sin \beta \; .
\end{equation}
From these, $\sin\delta$, $\sin\alpha$ and $\sin\beta$ can be computed.

The light neutrino masses $m_i$, $i=1,2,3$, can be expressed in terms of the lightest neutrino mass $m_0$ and the two measured mass squared differences $\Delta m_{\mathrm{sol}}^2$
and $\Delta m_{\mathrm{atm}}^2$, which are defined as 
 \begin{equation}
 \label{eq:dm2def}
 \Delta m_{\mathrm{sol}}^2=m_2^2 -m_1^2 \;\;\; \mbox{and} \;\;\; \Delta m_{\mathrm{atm}}^2 = \left\{
 \begin{array}{cc}
m_3^2-m_1^2 & \mbox{for NO}\\
m_3^2-m_2^2 & \mbox{for IO}
 \end{array} \right. \; ,
 \end{equation}
depending on the light neutrino mass ordering, NO or IO. For NO, the light neutrino masses $m_i$ read 
\begin{equation}
\label{eq:massesNO}
m_1= m_0 \;\; , \;\;\; m_2= \sqrt{m_0^2 + \Delta m_{\mathrm{sol}}^2} \;\; , \;\;\; m_3= \sqrt{m_0^2 + \Delta m_{\mathrm{atm}}^2} \; ,
\end{equation}
while for IO we have
\begin{equation}
\label{eq:massesIO}
m_1= \sqrt{m_0^2 + |\Delta m_{\mathrm{atm}}^2| - \Delta m_{\mathrm{sol}}^2 } \, , \;\; m_2= \sqrt{m_0^2 + |\Delta m_{\mathrm{atm}}^2| } \, , 
\;\; m_3= m_0 \; .
\end{equation}
Furthermore, we define the sum of the light neutrino masses 
\begin{equation}
\label{eq:sumnu}
\Sigma_\nu = m_1 + m_2 + m_3 \; ,
\end{equation}
which is constrained by cosmological measurements.

\section{Data on lepton mixing parameters and neutrino masses}
\label{app3}

We use the latest global fit results from the NuFIT Collaboration, NuFIT~5.0 (July 2020)~\cite{NuFIT50} (without SK atmospheric data). For the lepton mixing angles we have\\
for NO
\begin{eqnarray}
&& \sin^2 \theta_{13} = 0.02221^{+0.00068} _{-0.00062} \;\;\;\;\;\;\;\;\,\, \mbox{and} \;\;\; 0.02034 \leq \sin^2 \theta_{13} \leq 0.02430 \; , \label{eq:anglesexpNO}
\\ \nonumber
&& \sin^2 \theta_{12} = 0.304^{+0.013} _{-0.012} \;\;\;\;\;\;\;\;\;\;\;\;\;\;\;\,\mbox{and} \;\;\;  0.269 \leq \sin^2 \theta_{12} \leq 0.343 \; ,
\\ \nonumber
&& \sin^2 \theta_{23} = 0.570^{+0.018} _{-0.024} \;\;\;\,\;\;\;\;\;\;\;\;\;\;\;\; \mbox{and} \;\;\; 0.407 \leq \sin^2 \theta_{23} \leq 0.618 \; ,
\end{eqnarray}
and for IO
\begin{eqnarray}
&& \sin^2 \theta_{13} = 0.02240^{+0.00062} _{-0.00062} \;\;\;\;\;\;\;\;\,\, \mbox{and} \;\;\; 0.02053 \leq \sin^2 \theta_{13} \leq 0.02436 \; , \label{eq:anglesexpIO}
\\ \nonumber
&& \sin^2 \theta_{12} = 0.304^{+0.013} _{-0.012} \;\;\;\;\;\;\;\;\;\;\;\;\;\;\;\,\mbox{and} \;\;\;  0.269 \leq \sin^2 \theta_{12} \leq 0.343 \; ,
\\ \nonumber
&& \sin^2 \theta_{23} = 0.575^{+0.017} _{-0.021} \;\;\;\,\;\;\;\;\;\;\;\;\;\;\;\; \mbox{and} \;\;\; 0.411 \leq \sin^2 \theta_{23} \leq 0.621 \; .
\end{eqnarray}
The experimental constraint on the Dirac phase $\delta$ reads\\
for NO
\begin{equation} \label{eq:deltaexpNO}
\delta =\left(195^{+51}_{-25}\right)^\circ \;\;\; \mbox{and} \;\;\; 107^\circ \leq \delta \leq 403^\circ \; , 
\end{equation}
and for IO
\begin{equation}
\delta =\left(286^{+27}_{-32}\right)^\circ \;\;\; \mbox{and} \;\;\; 192^\circ \leq \delta \leq 360^\circ \;\;\; \mbox{at the} \;\; 3\,\sigma \;\; \mbox{level} . \label{eq:deltaexpIO}
\end{equation}
For NO the experimental results for the mass squared differences are
\begin{eqnarray}
\label{eq:dm2expNO}
&&\Delta m_{\mathrm{sol}}^2=\left( 7.42^{+0.21} _{-0.20} \right) \times 10^{-5} \, \mathrm{eV}^2 \;\;\;\;\;\;\;\;\; \mbox{and} \;\;\;  6.82 \leq  \frac{\Delta m_{\mathrm{sol}}^2}{10^{-5} \, \mathrm{eV}^2} \leq 8.04 \; ,
\\ \nonumber
&&\Delta m_{\mathrm{atm}}^2=\left( 2.514^{+0.028} _{-0.027} \right) \times 10^{-3} \, \mathrm{eV}^2 \;\;\;\;\, \mbox{and} \;\;\; 2.431 \leq  \frac{\Delta m_{\mathrm{atm}}^2}{10^{-3} \, \mathrm{eV}^2} \leq 2.598 \; ,
\end{eqnarray}
and for IO
\begin{eqnarray}
\label{eq:dm2expIO}
&&\Delta m_{\mathrm{sol}}^2=\left( 7.42^{+0.21} _{-0.20} \right) \times 10^{-5} \, \mathrm{eV}^2 \;\;\;\;\;\;\;\;\;\;\,\, \mbox{and} \;\;\;  6.82 \leq  \frac{\Delta m_{\mathrm{sol}}^2}{10^{-5} \, \mathrm{eV}^2} \leq 8.04 \; ,
\\ \nonumber
&&\Delta m_{\mathrm{atm}}^2=\left(-2.497^{+0.028} _{-0.028} \right) \times 10^{-3} \, \mathrm{eV}^2 \;\;\;\,\, \mbox{and} \;\;\;  -2.583 \leq  \frac{\Delta m_{\mathrm{atm}}^2}{10^{-3} \, \mathrm{eV}^2} \leq -2.412 \; .
\end{eqnarray}
(Note that NO is currently slightly preferred over IO by experimental data with $\Delta\chi^2=\chi^2_{\mathrm{IO}}-\chi^2_{\mathrm{NO}}=2.7$.)

As limit on the sum of the light neutrino masses, we use the one given by the Planck Collaboration in~\cite{Planck2018},
\begin{equation}
\label{eq:sumnuexp}
\Sigma_\nu \leq 0.12 \; \mathrm{eV} \;\; \mbox{corresponding to} \;\; m_0 \lesssim 0.04 \, \mathrm{eV}
\end{equation}
for the lightest neutrino mass. The experimental limits from KATRIN~\cite{KATRIN} and from the searches for $0\nu\beta\beta$ decay~\cite{KamlandZen,EXO200,GERDA,MAJORANADemo,CUORE} 
do not pose stronger constraints on the lightest neutrino mass $m_0$ than the ones obtained from cosmology.

\section{Numerical treatment and fit procedure}\label{app:numerics}
\label{APPFIT}
Predicting the CP phases for the different choices of the CP symmetry
requires identifying the values of the free angle $\theta_S$
which lead to a set of lepton mixing parameters in agreement with
experimental data. 
The free angle $\theta_S$ is fit by maximising the joint likelihood of the
predictions for $\sin^2\theta_{ij}$. 
As mentioned in appendix \ref{app3}, we refer to the combination of
experimental data (latest update on global fits) provided by NuFIT
5.0~\cite{NuFIT50}. 
In order to fit $\theta_S$ with high accuracy, and especially to take
into account the ``double well'' structure in $\sin^2\theta_{23}$, we
interpolate the numerical $\Delta\chi^2$ values (available
at~\cite{NuFIT50}), and linearly extrapolate beyond the provided ranges to
ensure a smooth behaviour for arbitrary input values. 
The interpolated $\chi^2$ functions are then transformed into 
probability distributions so that a global
joint likelihood of all relevant parameters ($\sin^2\theta_{ij}$ and,
in the case of the (3,3) ISS framework, also $\Delta m_{ij}^2$) can be
constructed. 
To ensure that the cosmological bound on the sum of light neutrino
masses is respected, a half-normal distribution (as a gaussian upper
limit) is further included.
We first fit predictions for $U_S$ and therefore $\theta_S$
on the data for $\sin^2\theta_{ij}$,  
from which we proceed to consider the model-dependent, i.e. (3,3) ISS
framework. 
This is done by maximising the joint likelihood function using the
{\tt migrad} algorithm of the {\tt iminuit} library~\cite{iminuit}. 
Local maximum likelihood estimators lying outside of the global
$3\,\sigma$ region around the (experimental) best-fit point are
rejected. 

Due to the peculiar structure and almost degenerate heavy states in
the ISS mass matrix, a large numerical precision ($\sim 100$ digits)
is needed for a reliable matrix diagonalisation. This is achieved
using the {\tt mpmath} python library~\cite{mpmath} and algorithms within.  
To study the effects of the heavy sterile states on the
predictions for lepton mixing parameters, we use
``{\it effective}'' mixing angles (and phases) which we define as in appendix~\ref{app2} (eqs.~(\ref{eq:sin2thij},\ref{eq:JCP},\ref{eq:I1I2}), with $U_\text{PMNS}$ replaced by $\widetilde{U}_\nu$). 
The free angle $\theta_S$ then needs to be re-fitted, using the
results obtained within the model-independent approach as starting
values for the fit, thus allowing to study deviations from those
predictions. 

Keeping the lightest neutrino mass $m_0$ fixed - and thus the
lightest Majorana mass - 
($\mu_1$ or $\mu_3$ depending on the ordering of the light neutrino mass spectrum),  
the remaining two Majorana masses $\mu_i$ are treated as free
parameters to be determined by a fit to $\sin^2\theta_{ij}$ and
$\Delta m_{ij}^2$ data. 
The starting values for $\mu_i$ are determined by
inverting the leading order expression given in eq.~\eqref{eq:mnuLO}
and a modified Casas-Ibarra parametrisation~\cite{phasesnorelation}  
\begin{equation}
    \mu_S \,\simeq \,M_{NS}^T \,m_D^{-1} \,U^\star_S \, m^\text{diag}_\nu
    \,U^\dagger_S \,{m_D^T}^{-1} \,M_{NS}\,, 
\end{equation}
where $U_S$, the matrix which diagonalises $\mu_S$ and at leading order also the light neutrino mass matrix, is determined by the flavour symmetry $G_f$ and CP and the residual symmetry $G_\nu$.
For choices of $y_0$ and $M_0$ in conflict with bounds on the
unitarity of $\widetilde U_\nu$, sizeable departures from the model-independent results are observed, as described in the main body of the paper. 

\section{Loop functions}
\label{sec:loopfunctions}

The loop functions and their relevant limits\footnote{Note that in
  Ref.~\cite{Ilakovac:1994kj} the loop function $F_\text{Xbox}$ is
  named $F_\text{box}$ and has an opposite global sign compared to
  Ref.~\cite{Alonso:2012ji}, which also reflects in the form factor
  $F_\text{box}^{\beta 3\alpha}$.} are taken from
Refs.~\cite{Alonso:2012ji,Ilakovac:1994kj}. 
The photon dipole and anapole functions, as well as some relevant
limits, are given by
\begin{eqnarray}
    F_\gamma(x) &=& \frac{7 x^3 - x^2 - 12x}{12(1-x)^3} - \frac{x^4 -
      10x^3 + 12x^2}{6(1-x)^4}\log x\,,\nonumber\\ 
    F_\gamma(x) &\xrightarrow[x\gg1]{}& -\frac{7}{12} -
    \frac{1}{6}\log x\,,\nonumber\\ 
    F_\gamma(0) &=& 0\,,\label{eqn:lfun:fgamma}\\
    G_\gamma(x) &=& -\frac{x(2x^2 + 5x - 1)}{4(1-x)^3} -
    \frac{3x^3}{2(1-x)^4}\log x\,,\nonumber\\ 
    G_\gamma(x) &\xrightarrow[x\gg1]{}& \frac{1}{2}\,,\nonumber\\
    G_\gamma(0) &=& 0\,.\label{eqn:lfun:ggamma}
\end{eqnarray}
The functions associated with the $Z$ penguins are given by a
two-point 
function
\begin{eqnarray}
    F_Z(x) &=& -\frac{5 x}{2(1 - x)} - \frac{5x^2}{2(1-x)^2}\log x\,,\nonumber\\
    F_Z(x) &\xrightarrow[x\gg 1]{}& \frac{5}{2} - \frac{5}{2}\log
    x\,,\nonumber\\ 
    F_Z(0) &=& 0\,,\label{eqn:lfun:fz}
\end{eqnarray}
and two three-point functions which are symmetric under interchange
of the arguments. 
\begin{eqnarray}
    G_Z(x,y) &=& -\frac{1}{2(x-y)}\left[\frac{x^2(1-y)}{1-x}\log x -
      \frac{y^2(1-x)}{1-y}\log y \right]\,,\nonumber\\ 
    G_Z(x, x) &=& -\frac{x}{2} - \frac{x\log x}{1-x}\,,\nonumber\\
    G_Z(0,x) &=& -\frac{x\log x}{2(1-x)}\,,\nonumber\\
    G_Z(0,x)&\xrightarrow[x\gg 1]{}& \frac{1}{2}\log x\,,\nonumber\\
    G_Z(0,0) &=& 0\,,\label{eqn:lfun:gz}\\
    H_Z(x,y) &=& \frac{\sqrt{xy}}{4(x-y)}\left[\frac{x^2 - 4x}{1 -
        x}\log x - \frac{y^2 - 4y}{1 - y}\log
      y\ \right]\,,\nonumber\\ 
    H_Z(x,x) &=& \frac{(3 - x)(1-x) - 3}{4(1-x)} - \frac{x^3 - 2x^2 +
      4x}{4(1-x)^2}\log x\,,\nonumber\\ 
    H_Z(0,x) &=& 0\,.\label{eqn:lfun:hz}
\end{eqnarray}
The (symmetric) box-loop-functions and their limits are given by
\begin{eqnarray}
    F_\text{box}(x,y) &=& \frac{1}{x-y}\left\{\left(4 +
    \frac{xy}{4}\right)\left[\frac{1}{1-x} + \frac{x^2}{(1-x)^2} \log
      x - \frac{1}{1-y} - \frac{y^2}{(1-y)^2}\log
      y\right]\right.\nonumber\\  
    &\phantom{=}& \left. -2xy\left[\frac{1}{1-x} + \frac{x}{(1-x)^2}
      \log x - \frac{1}{1-y} - \frac{y}{(1-y)^2}\log y
      \right]\right\}\,,\nonumber\\ 
    F_\text{box}(x,x) &=& -\frac{1}{4(1-x)^3}\left[x^4 - 16x^3 + 31x^2
      - 16 + 2x\left(3x^2 + 4x - 16\right)\log x\right]\,,\nonumber\\ 
    F_\text{box}(0,x) &=& \frac{4}{1 - x} + \frac{4x}{(1-x)^2}\log
    x\,,\nonumber\\ 
    F_\text{box}(0,x)&\xrightarrow[x\gg 1]{}& 0\,,\nonumber\\
    F_\text{box}(0,0) &=& 4\,,\label{eqn:lfun:fbox}\\
    F_\text{Xbox}(x,y) &=& -\frac{1}{x-y}\left\{\left(1 + \frac{xy}{4}
    \right)\left[\frac{1}{1-x} + \frac{x^2}{(1-x)^2} \log x -
      \frac{1}{1-y} - \frac{y^2}{(1-y)^2}\log
      y\right]\right.\nonumber\\  
    &\phantom{=}& \left. -2xy\left[\frac{1}{1-x} + \frac{x}{(1-x)^2}
      \log x - \frac{1}{1-y} - \frac{y}{(1-y)^2}\log y
      \right]\right\}\,,\nonumber\\ 
    F_\text{Xbox}(x,x) &=& \frac{x^4 - 16x^3 + 19x^2 - 4}{4(1-x)^3} +
    \frac{3x^3 + 4x^2 - 4x}{2(1-x)^3}\log x\,,\nonumber\\ 
    F_\text{Xbox}(0,x) &=& -\frac{1}{1-x} - \frac{x}{(1 - x)^2}\log
    x\,,\nonumber\\ 
    F_\text{Xbox}(0,x)&\xrightarrow[x\gg 1]{}& 0\,,\nonumber\\
    F_\text{Xbox}(0,0) &=& -1\,,\label{eqn:lfun:fxbox}\\
    G_\text{box}(x,y) &=& -\frac{\sqrt{xy}}{x-y}\left\{(4 +
    xy)\left[\frac{1}{1-x} + \frac{x}{(1-x)^2} \log x - \frac{1}{1-y}
      - \frac{y}{(1-y)^2}\log y\right]\right.\nonumber\\  
    &\phantom{=}& \left. -2\left[\frac{1}{1-x} + \frac{x^2}{(1-x)^2}
      \log x - \frac{1}{1-y} - \frac{y^2}{(1-y)^2}\log y
      \right]\right\}\,,\nonumber\\ 
    G_\text{box}(x,x) &=& \frac{2x^4 - 4x^3 + 8x^2 - 6x}{(1-x)^3} -
    \frac{x^4 + x^3 + 4x}{(1-x)^3}\log x\,,\nonumber\\ 
    G_\text{box}(0,x) &=& 0\,.\label{eqn:lfun:gbox}
\end{eqnarray}

 


\begin{thebibliography}{00}
 

  \bibitem{Gfreviews}
  H.~Ishimori, T.~Kobayashi, H.~Ohki, Y.~Shimizu, H.~Okada, M.~Tanimoto,
  Prog.\ Theor.\ Phys.\ Suppl.\  {\bf 183 } (2010)  1-163
  [arXiv:1003.3552 [hep-th]];
 S.~F.~King and C.~Luhn,
  Rept.\ Prog.\ Phys.\  {\bf 76} (2013) 056201
  [arXiv:1301.1340 [hep-ph]];
F.~Feruglio and A.~Romanino,
Rev. Mod. Phys. \textbf{93} (2021) no.1, 015007
[arXiv:1912.06028 [hep-ph]].

  \bibitem{Gfreview_math}
  W.~Grimus and P.~O.~Ludl,
  J.\ Phys.\ A {\bf 45} (2012) 233001
  [arXiv:1110.6376 [hep-ph]].
  
  
\bibitem{Mohapatra:1986bd}
R.~N.~Mohapatra and J.~W.~F.~Valle,
Phys. Rev. D \textbf{34} (1986), 1642.

\bibitem{Mohapatra:1986aw}
R.~N.~Mohapatra,
Phys. Rev. Lett. \textbf{56} (1986), 561-563.

\bibitem{Bernabeu:1987gr}
J.~Bernabeu, A.~Santamaria, J.~Vidal, A.~Mendez and J.~W.~F.~Valle,
Phys. Lett. B \textbf{187} (1987), 303-308.

\bibitem{GonzalezGarcia:1988rw}
M.~C.~Gonzalez-Garcia and J.~W.~F.~Valle,
Phys. Lett. B \textbf{216} (1989), 360-366.


 \bibitem{type1seesaw}
 P.~Minkowski,
Phys. Lett. B \textbf{67} (1977), 421-428;
 T.~Yanagida, in {\it Proceedings of the Workshop on the Unified Theory and the Baryon Number in the Universe}
 (O.~Sawada and A.~Sugamoto, eds.), KEK Tsukuba, Japan, 1979, p. 95; 
 S.~L.~Glashow, {\it The future of elementary particle physics}, in {\it Proceedings of the 1979 Carg\`ese Summer 
 Institute on Quarks and Leptons} (M.~L\'evy, J.-L.~Basdevant, D.~Speiser, J.~Weyers, R.~Gastmans, and M.~Jacob, eds.),
 Plenum Press, New York, 1980, pp. 687-713; 
 M.~Gell-Mann, P.~Ramond, and R.~Slansky, {\it Complex spinors and unified theories}, in {\it Supergravity} (P.~van~Nieuwenhuizen
 and D.~Z.~Freedman, eds.), North Holland, Amsterdam, 1979, p. 315;
 R.~N.~Mohapatra and G.~Senjanovic,
  Phys.\ Rev.\ Lett.\  {\bf 44} (1980) 912.
  
 \bibitem{type2seesaw}
M.~Magg and C.~Wetterich,
Phys. Lett. B \textbf{94} (1980), 61-64;
J.~Schechter and J.~W.~F.~Valle,
Phys. Rev. D \textbf{22} (1980), 2227;
T.~P.~Cheng and L.~F.~Li,
Phys. Rev. D \textbf{22} (1980), 2860;
G.~Lazarides, Q.~Shafi and C.~Wetterich,
Nucl. Phys. B \textbf{181} (1981), 287-300;
C.~Wetterich,
Nucl. Phys. B \textbf{187} (1981), 343-375;
R.~N.~Mohapatra and G.~Senjanovic,
Phys. Rev. D \textbf{23} (1981), 165.  

 \bibitem{type3seesaw}
R.~Foot, H.~Lew, X.~G.~He and G.~C.~Joshi,
Z. Phys. C \textbf{44} (1989), 441.

 \bibitem{radnureview}
Y.~Cai, J.~Herrero-Garc\'\i{}a, M.~A.~Schmidt, A.~Vicente and R.~R.~Volkas,
Front. in Phys. \textbf{5} (2017), 63
[arXiv:1706.08524 [hep-ph]].

    
  \bibitem{S4CPgeneral}
 F.~Feruglio, C.~Hagedorn and R.~Ziegler,
  JHEP {\bf 1307} (2013) 027
  [arXiv:1211.5560 [hep-ph]].
  
  

\bibitem{GfCPearly}
 G.~Ecker, W.~Grimus and H.~Neufeld,
  Nucl.\ Phys.\ B {\bf 247} (1984) 70;
G.~Ecker, W.~Grimus and H.~Neufeld,
  J.\ Phys.\ A {\bf 20} (1987) L807;
H.~Neufeld, W.~Grimus and G.~Ecker,
  Int.\ J.\ Mod.\ Phys.\ A {\bf 3} (1988) 603;
  W.~Grimus and M.~N.~Rebelo,
  Phys.\ Rept.\  {\bf 281} (1997) 239
  [arXiv:hep-ph/9506272];
  P.~F.~Harrison and W.~G.~Scott,
  Phys.\ Lett.\ B {\bf 535} (2002) 163
  [arXiv:hep-ph/0203209];
W.~Grimus and L.~Lavoura,
  Phys.\ Lett.\ B {\bf 579} (2004) 113
  [arXiv:hep-ph/0305309].

  
 \bibitem{GfCPothers}
 M.~Holthausen, M.~Lindner and M.~A.~Schmidt,
  JHEP {\bf 1304} (2013) 122
  [arXiv:1211.6953 [hep-ph]];
M.-C.~Chen, M.~Fallbacher, K.~T.~Mahanthappa, M.~Ratz and A.~Trautner,
  Nucl.\ Phys.\ B {\bf 883} (2014) 267
  [arXiv:1402.0507 [hep-ph]].
  
 \bibitem{Delta3n2}
 C.~Luhn, S.~Nasri and P.~Ramond,
  J.\ Math.\ Phys.\  {\bf 48} (2007) 073501
  [arXiv:hep-th/0701188].


 
 \bibitem{Delta6n2}
 J.~A.~Escobar and C.~Luhn,
  J.\ Math.\ Phys.\  {\bf 50} (2009) 013524
  [arXiv:0809.0639 [hep-th]].



 \bibitem{HMM}
C.~Hagedorn, A.~Meroni and E.~Molinaro,
  Nucl.\ Phys.\ B {\bf 891} (2015) 499
  [arXiv:1408.7118 [hep-ph]].
 
 \bibitem{DeltaCPothers}
 G.~J.~Ding, S.~F.~King and T.~Neder,
  JHEP {\bf 1412} (2014) 007
  [arXiv:1409.8005 [hep-ph]];
G.~J.~Ding and S.~F.~King,
  Phys.\ Rev.\ D {\bf 93} (2016) 025013
  [arXiv:1510.03188 [hep-ph]].
  


 
 \bibitem{D6n2CPZ2Z2}
S.~F.~King and T.~Neder,
  Phys.\ Lett.\ B {\bf 736} (2014) 308
  [arXiv:1403.1758 [hep-ph]].


 \bibitem{smallDeltaCPmodels}
 G.~J.~Ding, S.~F.~King, C.~Luhn and A.~J.~Stuart,
  JHEP {\bf 1305} (2013) 084
  [arXiv:1303.6180 [hep-ph]];
  F.~Feruglio, C.~Hagedorn and R.~Ziegler,
  Eur.\ Phys.\ J.\ C {\bf 74} (2014) 2753
  [arXiv:1303.7178 [hep-ph]];
  G.~J.~Ding, S.~F.~King and A.~J.~Stuart,
  JHEP {\bf 1312} (2013) 006
  [arXiv:1307.4212 [hep-ph]];
  C.~C.~Li and G.~J.~Ding,
  Nucl.\ Phys.\ B {\bf 881} (2014) 206
  [arXiv:1312.4401 [hep-ph]];
C.~C.~Li and G.~J.~Ding,
  JHEP {\bf 1508} (2015) 017
  [arXiv:1408.0785 [hep-ph]];
  G.~J.~Ding and Y.~L.~Zhou,
  Chin.\ Phys.\ C {\bf 39} (2015) 2,  021001
  [arXiv:1312.5222 [hep-ph]];
 G.~J.~Ding and Y.~L.~Zhou,
  JHEP {\bf 1406} (2014) 023
  [arXiv:1404.0592 [hep-ph]];
   G.~J.~Ding and S.~F.~King,
  Phys.\ Rev.\ D {\bf 89} (2014) 9,  093020
  [arXiv:1403.5846 [hep-ph]]. 




\bibitem{tHooft:1979rat}
G.~'t Hooft,
NATO Sci. Ser. B \textbf{59} (1980), 135-157.

\bibitem{Abada:2014vea}
A.~Abada and M.~Lucente,
Nucl. Phys. B \textbf{885} (2014), 651-678
[arXiv:1401.1507 [hep-ph]].

\bibitem{Arganda:2014dta}
E.~Arganda, M.~J.~Herrero, X.~Marcano and C.~Weiland,
Phys. Rev. D \textbf{91} (2015) no.1, 015001
[arXiv:1405.4300 [hep-ph]].

\bibitem{Abada:2014kba}
A.~Abada, M.~E.~Krauss, W.~Porod, F.~Staub, A.~Vicente and C.~Weiland,
JHEP \textbf{11} (2014), 048
[arXiv:1408.0138 [hep-ph]].

\bibitem{Abada:2014cca}
A.~Abada, V.~De Romeri, S.~Monteil, J.~Orloff and A.~M.~Teixeira,
JHEP \textbf{04} (2015), 051
[arXiv:1412.6322 [hep-ph]].

\bibitem{Arganda:2015naa}
E.~Arganda, M.~J.~Herrero, X.~Marcano and C.~Weiland,
Phys. Rev. D \textbf{93} (2016) no.5, 055010
[arXiv:1508.04623 [hep-ph]].

\bibitem{Abada:2015oba}
A.~Abada, V.~De Romeri and A.~M.~Teixeira,
JHEP \textbf{02} (2016), 083
[arXiv:1510.06657 [hep-ph]].

\bibitem{DeRomeri:2016gum}
V.~De Romeri, M.~J.~Herrero, X.~Marcano and F.~Scarcella,
Phys. Rev. D \textbf{95} (2017) no.7, 075028
[arXiv:1607.05257 [hep-ph]].

\bibitem{Abada:2016awd}
A.~Abada and T.~Toma,
JHEP \textbf{08} (2016), 079
[arXiv:1605.07643 [hep-ph]].

\bibitem{Abada:2014nwa}
A.~Abada, V.~De Romeri and A.~M.~Teixeira,
JHEP \textbf{09} (2014), 074
[arXiv:1406.6978 [hep-ph]].

\bibitem{Abada:2018qok}
A.~Abada, \'A.~Hern\'andez-Cabezudo and X.~Marcano,
JHEP \textbf{01} (2019), 041
[arXiv:1807.01331 [hep-ph]].

\bibitem{Baglio:2016bop}
J.~Baglio and C.~Weiland,
JHEP \textbf{04} (2017), 038
[arXiv:1612.06403 [hep-ph]].

 \bibitem{Gfmodels}
 E.~Ma,
  Phys.\ Rev.\ D {\bf 70} (2004) 031901
  [arXiv:hep-ph/0404199];
 W.~Grimus and L.~Lavoura,
  JHEP {\bf 0508}, 013 (2005)
  [arXiv:hep-ph/0504153];
  I.~de Medeiros Varzielas, S.~F.~King and G.~G.~Ross,
Phys. Lett. B \textbf{644} (2007), 153-157
[arXiv:hep-ph/0512313];
  G.~Altarelli and F.~Feruglio,
  Nucl.\ Phys.\ B {\bf 741} (2006) 215
  [ariXiv:hep-ph/0512103];
 X.~-G.~He, Y.~-Y.~Keum and R.~R.~Volkas,
  JHEP {\bf 0604} (2006) 039
  [arXiv:hep-ph/0601001];  
  Y.~Lin,
  Nucl.\ Phys.\ B {\bf 813} (2009) 91
  [arXiv:0804.2867 [hep-ph]].
  
\bibitem{ISSsymmetries}
M.~Hirsch, S.~Morisi and J.~W.~F.~Valle,
Phys. Lett. B \textbf{679} (2009), 454-459
[arXiv:0905.3056 [hep-ph]];
D.~Ibanez, S.~Morisi and J.~W.~F.~Valle,
Phys. Rev. D \textbf{80} (2009), 053015
[arXiv:0907.3109 [hep-ph]];
L.~Dorame, S.~Morisi, E.~Peinado, J.~W.~F.~Valle and A.~D.~Rojas,
Phys. Rev. D \textbf{86} (2012), 056001
[arXiv:1203.0155 [hep-ph]];
A.~E.~C\'arcamo Hern\'andez and H.~N.~Long,
J. Phys. G \textbf{45} (2018) no.4, 045001
[arXiv:1705.05246 [hep-ph]];
D.~Borah and B.~Karmakar,
Phys. Lett. B \textbf{780} (2018), 461-470
[arXiv:1712.06407 [hep-ph]];
A.~E.~C\'arcamo Hern\'andez and S.~F.~King,
Nucl. Phys. B \textbf{953} (2020), 114950
[arXiv:1903.02565 [hep-ph]];
T.~Nomura, H.~Okada and S.~Patra,
Nucl. Phys. B \textbf{967} (2021), 115395
[arXiv:1912.00379 [hep-ph]];
T.~P.~Nguyen, T.~T.~Thuc, D.~T.~Si, T.~T.~Hong and L.~T.~Hue,
arXiv:2011.12181 [hep-ph];
H.~B.~Camara, R.~G.~Felipe and F.~R.~Joaquim,
JHEP \textbf{05} (2021), 021
[arXiv:2012.04557 [hep-ph]];
M.~R.~Devi and K.~Bora,
arXiv:2103.10065 [hep-ph];
X.~Zhang and S.~Zhou,
arXiv:2106.03433 [hep-ph].


\bibitem{choiceof3}
  S.~F.~King, T.~Neder and A.~J.~Stuart,
  Phys.\ Lett.\ B {\bf 726} (2013) 312
  [arXiv:1305.3200 [hep-ph]].

\bibitem{ISSsubleading}
H.~Hettmansperger, M.~Lindner and W.~Rodejohann,
JHEP \textbf{04} (2011), 123
[arXiv:1102.3432 [hep-ph]].

\bibitem{NuFIT50}
I.~Esteban, M.~C.~Gonzalez-Garcia, M.~Maltoni, T.~Schwetz and A.~Zhou,
JHEP \textbf{09} (2020), 178
[arXiv:2007.14792 [hep-ph]], 
website: \verb3http://www.nu-fit.org/3.

\bibitem{etaconstraints}
E.~Fernandez-Martinez, J.~Hernandez-Garcia and J.~Lopez-Pavon,
JHEP \textbf{08} (2016), 033
[arXiv:1605.08774 [hep-ph]].

\bibitem{Antusch:2016ejd}
S.~Antusch, E.~Cazzato and O.~Fischer,
Int. J. Mod. Phys. A \textbf{32} (2017) no.14, 1750078
[arXiv:1612.02728 [hep-ph]].

\bibitem{Blennow:2010th}
M.~Blennow, E.~Fernandez-Martinez, J.~Lopez-Pavon and J.~Menendez,
JHEP \textbf{07} (2010), 096
[arXiv:1005.3240 [hep-ph]].

 
 \bibitem{Hagedorn:2016lva}
C.~Hagedorn and E.~Molinaro,
Nucl. Phys. B \textbf{919} (2017), 404-469
[arXiv:1602.04206 [hep-ph]].


\bibitem{Planck2018}
N.~Aghanim \textit{et al.} [Planck],
Astron. Astrophys. \textbf{641} (2020), A6
[arXiv:1807.06209 [astro-ph.CO]].


\bibitem{KamlandZen} 
A.~Gando \textit{et al.} [KamLAND-Zen],
Phys. Rev. Lett. \textbf{117} (2016) no.8, 082503, Phys. Rev. Lett. \textbf{117} (2016) no.10, 109903 (addendum)
[arXiv:1605.02889 [hep-ex]].
\bibitem{EXO200} 
G.~Anton \textit{et al.} [EXO-200],
Phys. Rev. Lett. \textbf{123} (2019) no.16, 161802
[arXiv:1906.02723 [hep-ex]].
\bibitem{GERDA} 
M.~Agostini \textit{et al.} [GERDA],
Phys. Rev. Lett. \textbf{125} (2020), 252502
[arXiv:2009.06079 [nucl-ex]].
\bibitem{MAJORANADemo} 
S.~I.~Alvis \textit{et al.} [Majorana],
Phys. Rev. C \textbf{100} (2019) no.2, 025501
[arXiv:1902.02299 [nucl-ex]].
\bibitem{CUORE} 
D.~Q.~Adams \textit{et al.} [CUORE],
Phys. Rev. Lett. \textbf{124} (2020) no.12, 122501
[arXiv:1912.10966 [nucl-ex]].


\bibitem{Alonso:2012ji}
R.~Alonso, M.~Dhen, M.~B.~Gavela and T.~Hambye,
JHEP \textbf{01} (2013), 118
[arXiv:1209.2679 [hep-ph]].

\bibitem{Kitano:2002mt}
R.~Kitano, M.~Koike and Y.~Okada,
Phys. Rev. D \textbf{66} (2002), 096002
[erratum: Phys. Rev. D \textbf{76} (2007), 059902]
[arXiv:hep-ph/0203110].


\bibitem{Ilakovac:1994kj}
A.~Ilakovac and A.~Pilaftsis,
Nucl. Phys. B \textbf{437} (1995), 491
[arXiv:hep-ph/9403398].


\bibitem{Xing:2007zj}
Z.~z.~Xing,
Phys. Lett. B \textbf{660} (2008), 515-521
[arXiv:0709.2220 [hep-ph]].




\bibitem{JKCH2}
C.~Hagedorn and J.~K\"onig,
Nucl. Phys. B \textbf{953} (2020), 114953
[arXiv:1811.09262 [hep-ph]].

\bibitem{MSCH1}
C.~Hagedorn and M.~Serone,
JHEP \textbf{10} (2011), 083
[arXiv:1106.4021 [hep-ph]].
\bibitem{MSCH2}
C.~Hagedorn and M.~Serone,
JHEP \textbf{02} (2012), 077
[arXiv:1110.4612 [hep-ph]].

\bibitem{ISScategories}
A.~Abada and M.~Lucente,
Nucl. Phys. B \textbf{885} (2014), 651-678
[arXiv:1401.1507 [hep-ph]].

\bibitem{MRISS}
P.~S.~B.~Dev and A.~Pilaftsis,
Phys. Rev. D \textbf{86} (2012), 113001
[arXiv:1209.4051 [hep-ph]].


\bibitem{PDG}
P.A.~Zyla \textit{et al.} [Particle Data Group],
PTEP \textbf{2020} (2020) no.8, 083C01.

\bibitem{jcp}
  C.~Jarlskog,
  Phys.\ Rev.\ Lett.\  {\bf 55} (1985)  1039.
  
  \bibitem{Jenkins_Manohar_invariants}
  E.~E.~Jenkins and A.~V.~Manohar,
  Nucl.\ Phys.\ B {\bf 792} (2008) 187
  [arXiv:0706.4313 [hep-ph]].

\bibitem{KATRIN}
M.~Aker, A.~Beglarian, J.~Behrens, A.~Berlev, U.~Besserer, B.~Bieringer, F.~Block, B.~Bornschein, L.~Bornschein and M.~B\"ottcher, \textit{et al.}
``First direct neutrino-mass measurement with sub-eV sensitivity,''
arXiv:2105.08533 [hep-ex].

\bibitem{iminuit}
Hans Dembinski and Piti Ongmongkolkul {\it et al.},
``scikit-hep/iminuit: v1.5.1 (Version v1.5.1)", 
Zenodo, 
September 20, 2020, 
website: \verb3http://doi.org/10.5281/zenodo.4041167/3.

\bibitem{mpmath}
Fredrik Johansson {\it et al.},
``mpmath: a {P}ython library for arbitrary-precision floating-point arithmetic (version 1.1.0)",
December 11, 2018, 
website: \verb3http://mpmath.org/3.

\bibitem{phasesnorelation}
J.~A.~Casas and A.~Ibarra,
  Nucl.\ Phys.\ B {\bf 618} (2001) 171
  [arXiv:hep-ph/0103065].

 \end{thebibliography}
\end{document}